%% file: causal.tex
\documentclass{article}
\usepackage{amsmath,amsfonts,amsthm,amssymb}
\usepackage{graphicx,float,wrapfig}

\newtheorem{assumption}{Assumption}
\def\hI{\widehat I}
\def\hH{\widehat H}
\def\Pr{\mathbb P}
\def\reals{\mathbb R}
\def\V{\textrm{Var}}
\def\Cov{\textrm{Cov}}
\def\tf{\tilde{f}}
\def\tn{n}
\def\cxy{{\cal C}(X\rightarrow Y)}
\def\calc{{\cal C}}
\def\cX{\mathcal{X}}
\def\cY{\mathcal{Y}}
\def\hC{\widehat {\cal C}}
\def\Beta{{\rm Beta}}

\def\hUMI{{\rm \widehat{UMI}}}
\def\hCMI{{\rm \widehat{CMI}}}
\def\hI{\widehat I}
\def\hH{\widehat H}
\def\Pr{\mathbb P}
\def\reals{\mathbb R}
\def\V{\textrm{Var}}
\def\Cov{\textrm{Cov}}
\def\tf{\tilde{f}}
\def\tn{n}
\def\hC{\widehat {\cal C}}
\def\cX{{\cal X}}

\include{newcommand}
\begin{document}

\title{ Conditional Dependence via Shannon Capacity:\\ Axioms, Estimators and Applications\thanks{Parts of this manuscript has appeared in the International Conference on Machine Learning (ICML), 2016.}}
\author{Weihao Gao\thanks{Department of Electrical and Computer  Engineering, University of Illinois at Urbana-Champaign, email:\texttt{wgao9@illinois.edu}}, \;\;
Sreeram Kannan\thanks{Department of Electrical Engineering, University of Washington, email:\texttt{ksreeram@uw.edu}}, \;\;
Sewoong Oh\thanks{Department of Industrial and Enterprise Engineering, University of Illinois at Urbana-Champaign, email:\texttt{swoh@illinois.edu}}, \;\;
Pramod Viswanath\thanks{Department of Electrical and Computer  Engineering, University of Illinois at Urbana-Champaign, email:\texttt{pramodv@illinois.edu}}
}
\date{}
\maketitle

\begin{abstract}
We consider axiomatically the problem of estimating the strength of a conditional dependence relationship $P_{Y|X}$ from a random variables $X$ to a random variable $Y$. This has applications in determining the strength of a known causal relationship, where the strength depends only  on the conditional distribution of the effect given the cause (and not on the driving distribution of the cause). Shannon capacity, appropriately regularized, emerges as a natural measure under these axioms. We  examine the problem of calculating  Shannon capacity from the observed samples 
and propose a novel fixed-$k$ nearest neighbor estimator, and demonstrate its consistency. Finally, we demonstrate an application to single-cell flow-cytometry, where the proposed estimators  significantly reduce sample complexity.
\end{abstract}

\input{section1}

\input{section2}

\input{section3}

\input{section4}

\input{section5}

\input{section6}

\section*{Acknowledgements}
This work is supported in part by ARO W911NF1410220, NSF SaTC award CNS-1527754, NSF CISE award CCF-1553452 and a University of Washington startup grant.

\bibliography{causal_new}
\bibliographystyle{alpha}

\input{section7}

\end{document}

%% file: newcommand.tex
\newtheorem{defn}{Definition}

\newtheorem{note}{Remark}

\newcommand{\bc}{\begin{center}}
\newcommand{\ec}{\end{center}}
\newcommand{\bfl}{\begin{flushleft}}
\newcommand{\efl}{\end{flushleft}}

\newcommand{\beqa}{\begin{eqnarray}}
\newcommand{\eeqa}{\end{eqnarray}}

\newcommand{\beqan}{\begin{eqnarray*}}
\newcommand{\eeqan}{\end{eqnarray*}}
\newcommand{\beq}{\begin{equation}}
\newcommand{\eeq}{\end{equation}}
\newcommand{\beit}{\begin{itemize}}
\newcommand{\eeit}{\end{itemize}}

\newcommand{\E}{{\mathbb{E}}}

\newcommand{\mc}{\mathcal}

\newcommand{\bit}{\begin{itemize}}
\newcommand{\eit}{\end{itemize}}
\newcommand{\ben}{\begin{enumerate}}
\newcommand{\een}{\end{enumerate}}

\newcommand{\bdefn}{\begin{defn}}
\newcommand{\edefn}{\end{defn}}
\newcommand{\bnote}{\begin{note}}
\newcommand{\enote}{\end{note}}

\newcommand{\blem}{\begin{lemma}}
\newcommand{\elem}{\end{lemma}}
\newcommand{\bthm}{\begin{theorem}}
\newcommand{\ethm}{\end{theorem}}
\newcommand{\bpf}{\begin{proof}}
\newcommand{\epf}{\end{proof}}
\newcommand{\bcor}{\begin{corollary}}
\newcommand{\ecor}{\end{corollary}}
\newcommand{\bprop}{\begin{proposition}}
\newcommand{\eprop}{\end{proposition}}

\newtheorem{proposition}{Proposition}
\newtheorem{lemma}{Lemma}
\newtheorem{theorem}{Theorem}

\newtheorem{corollary}{Corollary}

%% file: section1.tex
\section{Introduction}
\label{sec:intro}

The axiomatic study of dependence measures on joint distributions between two random variables $X$ and $Y$ has a long history in statistics \cite{shannon, renyi1959measures, csiszar2008axiomatic}. In this paper, we study the relatively unexplored terrain of measures that depend only on the conditional distribution $P_{Y|X}$. We are motivated to study conditional dependence measures from a problem in causal strength estimation.  Causal learning is a basic problem in many areas of scientific learning, where one wants to uncover the cause-effect relationship usually using interventions or sometimes directly from observational data \cite{Pearl,tutorial,Mooijetal2015}. 


In this paper, we are interested in an even simpler question: given a causal relationship, how does one measure the strength of the relationship. This problem arises in many contexts, for example, one may know causal genetic pathways but only a subset of these maybe active in a particular tissue or organ - therefore, deducing how much influence each causal link exerts becomes necessary.

We focus on a simple model: consider a pair of random variables $(X,Y)$ with {\em known} causal direction $X \rightarrow Y$, and suppose that there are no confounders - we are interested in {\em quantifying} the causal influence $X$ has  on $Y$. We denote the causal influence quantity by $\cxy$. There are two philosophically distinct ways to model the quantity: the first one is {\em factual influence}, i.e., how much influence does $X$ exert on $Y$ under the current probability of the cause $X$. The second possible way, which one can term as {\em potential influence} measures how much influence $X$ can potentially exert on $Y$ - without cognizance to the present distribution of the cause. For example, consider a (hypothetical) city which has very few smokers, but smoking  inevitably leads to lung-cancer. In such a city, the factual influence of smoking on lung-cancer will be small but the potential influence is very high. Depending on the setting, one may prefer the former or the latter. In this paper, we are interested in the {\em potential influence} of a cause on its effect.

We want $\cxy$ to be invariant to scaling and one-one transformations of the variables $X,Y$. This naturally suggests information theoretic metrics as  plausible choices of $\cxy$, starting with the mutual information $I(X;Y) = D(P_{XY} || P_X P_Y)$, at least in the case of factual influence. This measures the information through the channel from $X\rightarrow Y$ as given by the prior $P_X$. Observe that this metric is {\em symmetric} with respect to the directions $X\rightarrow Y$ and $Y \rightarrow X$; this property is not always desirable. In fact, this measure is taken as a starting point to develop an axiomatic approach to studying causal strength on general graphs in \cite{Janzing13}.

In a recent work \cite{Krishnaswamy14}, potential causal influence is posited as a relevant metric to spot ``trends" in gene pathways. In the particular application considered there, rare biological states of gene $X$ in a given data may nevertheless correspond to important biological states (or become common under different biological conditions), and therefore it is important to have causal measures that are not sensitive to the cause distribution but only depend on the relationship between the cause and the effect. 
To quantify the potential influence of those rare $X$, the following approach is proposed.  Replace the observed distribution $P_X$ by a {\em uniform} distribution $U_X$ and calculate  the mutual information under the joint distribution $U_X P_{Y|X}$.
%
%
The resulting causal strength quantification is $\cxy = D(U_X P_{Y|X} || P_U P_Y)$, where $P_Y$ represents the distribution at the output of a channel $P_{Y|X}$ with input given by $U_X$. We call this quantification as Uniform Mutual Information (UMI) and pronounced ``you-me".  A key challenge is to compute this quantity from i.i.d.\ samples 
 in a statistical efficient manner, especially when the channel output is continuous valued (and potentially in high dimensions). This is the first focus point of this paper.

UMI is not invariant under bijective transformations (since a uniform distribution on $X$ is different from a uniform distribution on $X^3$) and is also sensitive to the estimated support size of $X$. Even more fundamentally, it is unclear why one would prefer the uniform prior to measure potential influence through the channel $P_{Y|X}$. Based on natural axioms of data processing and additivity, we motivate an alternative measure of causal strength: the {\em largest amount of information} that can be sent through the channel, namely the {\em Shannon capacity}. Formally $\cxy = \max_{Q_X} D(Q_X P_{Y|X} || Q_X P_Y)$, where $P_Y$ represents the distribution at the output of a channel $P_{Y|X}$ with input given by $Q_X$. We refer to such a quantification as Capacitated Mutual Information (CMI) and pronounced ``see-me". A key challenge is to compute this quantity from i.i.d.\ samples 
 in a statistical efficient manner, especially when the channel output is continuous valued (and potentially in high dimensions). This is the second focus point of this paper.
We make the following {\bf main contributions} in this paper. 
\begin{itemize}
\item {\em UMI Estimation}:  We construct a novel estimator to compute UMI from data  sampled i.i.d.\ from a distribution $P_{XY}$. The estimator brings together ideas from three disparate threads in statistical estimation theory: nearest-neighbor methods, a correlation boosting idea in the estimation of (standard) mutual information from samples \cite{Kra04}, and importance sampling. The estimator has  only a {\em single} hyper parameter (the number of nearest-neighbors considered, set to 4 or 5 in practice), uses an off-the-shelf kernel density estimator of only $P_X$,  and has strong connections to the entropy estimator of \cite{KL87}. Our main technical result is to show that the estimator is consistent (in probability) supposing that the Radon-Nikodym derivative $\frac{dP_U}{dP_X}$ is uniformly bounded over the support. In simulations, the estimator has very strong performance in  terms of sample complexity (compared to a baseline of the partition-based estimator in \cite{moddemeijer1989estimation}).

\item {\em CMI Estimation}: We build upon the estimator derived for UMI and construct an optimization problem 
that mimics the optimization problem inherent in computing the capacity directly from the conditional probability  distribution of the channel. Our main technical result is to show the consistency of this estimator, supposing that the Radon-Nikodym derivative $\frac{dP_Q}{dP_X}$ is uniformly bounded over the support, where $P_Q$ is the optimizing input to the channel. Simulation results show strong empirical performance, compared to a baseline of a partition-based method followed by discrete optimization.

\item {\em Application to gene pathway influence}: In \cite{Krishnaswamy14},  considered an important result in single-cell flow-cytometry data analysis, a causal strength metric (termed DREMI) is proposed  for measuring the causal influence of a gene -- this estimator is a specific way of implementing UMI along with a ``channel amplification" step, and DREMI was successfully used to spot gene-pathway trends. We show that our proposed  CMI and UMI estimators also exhibit the same performance as DREMI when supplied with the full dataset, while at the same time,  having significantly smaller sample complexity for the same performance. 

\end{itemize}

%% file: section2.tex
\section{An Axiomatic Approach}
\label{sec:axiom}

We formally model an influence measure on conditional probability distributions, by postulating five natural  axioms.
Let $X$ be drawn from an alphabet $\mc{X}$, and $Y$ from an alphabet $\mc{Y}$. Let the probability distribution of $Y$ given $X$ be given as $P_{Y|X}$. Let $\mc{P}$ be a family of conditional distributions; usually we will consider the case when $\mc{P}$ is the set of all possible conditional distributions. Then the influence measure $\cxy$ is a function of the conditional distribution to non-negative real numbers:  $ \calc:  \mc{P}(\mc{Y}|\mc{X}) \rightarrow \mathbb{R}^{+}$, and we can write $\cxy$ as $\calc(P_{Y|X})$.
We postulate that the function $\calc$  satisfies five axioms on $\mc{P}$,
and show that  CMI satisfies all five axioms:

\ben
\setcounter{enumi}{-1}
\item{\bf Independence}: The  measure  ${\cal C}( P_{Y|X})  = 0$  if and only if  $P_{Y=y|X=x}$  depends only on $y$.
\item {\bf Data Processing}:
If $X \rightarrow Y \rightarrow Z$ be a processing chain, i.e.,
$P_{Z=z|X=x} = \sum_{y \in \mc{Y}} P_{Z=z|Y=y} P_{Y=y|X=x}$, then the natural data processing inequalities should hold:
(a) $\calc(P_{Y|X}) \geq \calc(P_{Z|X})$; and (b)
$\calc(P_{Z|Y}) \geq \calc(P_{Z|X})$.

\item {\bf Additivity}:
For a parallel channel
$P_{Y_1,Y_2 | X_1, X_2}  := P_{Y_1 | X_1} P_{Y_2|X_2}$, we need
\beqa \calc(P_{Y_1,Y_2 | X_1, X_2} ) = \calc(P_{Y_1 | X_1}) +  \calc(P_{Y_2 | X_2}). \eeqa

\item {\bf Monotonicity}:
A causal relationship is strong if many possible values of $P_Y$ are achievable by varying the input probability distribution $P_X$. Thus if we consider $P_{Y|X}$ as a map from the probability simplex in $X$ to the probability simplex in $Y$, the larger the range of this map, the stronger should be the causal strength.

\ben
\item $\calc$ should only depend on the range of the map, $\textrm{Range} (P_{Y|X})$, the convex hull of the output distributions $P_{Y|X=x}$.

\item $\calc$ should be a monotonic function of the range of the map. If $P_{Y|X}$ and $Q_{Y|X}$ are such that,
$\textrm{Range} (P_{Y|X}) \subseteq \textrm{Range} (Q_{Y|X})$ then:  $\calc (P_{Y|X}) \leq \calc(Q_{Y|X})$.
\een

\item {\bf Maximum value}: The maximum value over all possible conditional distributions for a particular output alphabet $\mc{Y}$ should be achieved exactly when the relationship is fully causal, i.e., each $Y=y$ can be achieved by setting $X=x$ for some $x$.

\een

We begin our exploration of appropriate influence measures with the alphabets for $X$ and $Y$ being discrete. Let $I(P_{XY}) := D(P_{XY} || P_X P_Y)$ denote the mutual information with respect to the joint distribution $P_{XY}$.
Since we are looking at {\em potential} influence measures,   Shannon capacity, defined as the maximum over input probability distributions of the mutual information, is a natural choice:
\beqa \textrm{CMI}(P_{Y|X}) := \max_{P_X} I(P_X P_{Y|X}). \eeqa
Our first claim is the following:
\begin{proposition}
\textrm{CMI} satisfies all the axioms of causal influence.
\label{prop:cmi}
\end{proposition}

{\em Proof}: The proof is fairly straightforward.
\beit \item Clearly Axiom 0 holds, cf.\ Chapter 2 of \cite{CoverThomas}.
\item Axiom 1: Suppose $\textrm{CMI}(P_{Z|X})$ is achieved with $P_X^{*}$. Consider the joint distribution $P_X^{*} P_{Y|X} P_{Z|Y}$. Utilizing the data-processing inequality for mutual information, we get
\beqa \textrm{CMI}(P_{Y|X})  & = &  \max_{P_X} I(P_X P_{Y|X})  \geq  I(P_X^{*} P_{Y|X}) \nonumber\\
& \geq & I(P_X^{*} P_{Z|X}) = \textrm{CMI}(P_{Z|X}). \eeqa
Thus Axiom 1a is satisfied. Now consider Axiom 1b. With the same joint distribution, let $P_{Y}^{*}$ be the marginal of $Y$. Then  we have,
\beqa \textrm{CMI}(P_{Z|Y})  & = &  \max_{P_Y} I(P_Y P_{Z|Y})  \geq  I(P_Y^{*} P_{Z|Y})\nonumber \\
& \geq & I(P_X^{*} P_{Z|X}) = \textrm{CMI}(P_{Z|X}). \eeqa

\item Axiom 2: This is a standard result for Shannon capacity and  we refer the interested reader to Chapter 7 of \cite{CoverThomas}.

\item Axiom 3a: First we rewrite capacity equivalently as the information-centroid (see \cite{csiszar2004information}):
\beqa \textrm{CMI}(P_{Y|X}) & := & \max_{P_X} \min_{q_Y} D(P_{Y|X} \| q_Y | P_X) \nonumber \\
& = & \min_{q_Y}  \max_{P_X} D(P_{Y|X} \| q_Y | P_X) \nonumber \\
& = & \min_{q_Y}  \max_{x} D(P_{Y|X=x} \| q_Y) \label{eq:caP_alt}.
\eeqa
Here the conditional KL divergence $D(X \| Y | Z)$ is defined in the usual way:
\beqa
D(X \| Y | Z) = \sum_{z}  P_Z(z) \sum_{(x,y)} P_{X|Z}(x|z) \log \frac{P_{X|Z}(x|z)}{P_{Y|Z}(y|z)}.
\eeqa
This characterization allows us to make the  observation that the capacity is a function only of the convex hull of the probability distributions $P_{Y|X=x}$. Given a conditional probability distribution $P_{Y|X}$, we augment the input alphabet to have one more input symbol $x'$ such that $P_{Y|X=x'} = \sum_x \alpha_x P_{Y|X=x}$ is a convex combination of the other conditional distributions. We claim that the capacity of the new channel is unchanged: one direction is obvious, i.e., the new channel has capacity greater than or equal to the original channel, since adding a new symbol cannot decrease capacity. To show the other direction, we use \eqref{eq:caP_alt} and observe that, due to the convexity of KL divergence in its arguments, we get,
\beqa &&D(P_{Y|X=x'} \| q_Y )   =  D( \sum_x \alpha_x P_{Y|X=x} \| q_Y ) \nonumber \\
& \leq & \sum_x \alpha_x D(P_{Y|X=x} \| q_Y )
 \leq \max_x D(P_{Y|X=x} \| q_Y ). \nonumber\eeqa
%
Thus Shannon capacity is only a function of the convex hull of the range of the map $P_{Y|X}$, satisfying Axiom $3a$. This function is monotonic directly from  \eqref{eq:caP_alt}, thus satisfying Axiom $3b$.

\item Axiom 4: For fixed output alphabet $\mc{Y}$, it is clear that $\max_{\mc{X}, P_{Y|X}} \textrm{CMI} = \log |\mc{Y}|$. Now suppose for some conditional distribution $P_{Y|X}$ we have  $\textrm{CMI}(P_{Y|X}) = \log |\mc{Y}|$. This implies that, with the optimizing input distribution, $H(Y) - H(Y|X) = \log |\mc{Y}|$. This implies that $H(Y) = \log |\mc{Y}|$ and $H(Y|X) = 0$, thus $Y$ is a deterministic function of the essential support of $X$ and since $H(Y) = \log |\mc{Y}|$, it implies that $P_Y = U_Y$, the uniform distribution and the deterministic function is onto.

\eeit

{\bf Axiomatic View of UMI }:
Now consider an alternative metric: Uniform Mutual Information (UMI) which is defined as the mutual information with uniform input distribution,
\beqa \textrm{UMI}(P_{Y|X}): = 
 I(U_X P_{Y|X}), \eeqa
where $U_X$ is the uniform distribution on $\mathcal{X}$. This estimator is motivated by the recent work in \cite{Krishnaswamy14}. We investigate how this estimator fares in terms of the proposed axioms.
\begin{itemize}
\item UMI clearly satisfies Axiom $0$. It also satisfies Axioms $1a$.
Data-processing inequality for mutual information on the joint distribution $U_X P_{Y|X} P_{Z|Y}$ implies that $I(U_X P_{Y|X}) \geq I(U_X P_{Z|X})$, which is the same as $\textrm{UMI}(P_{Y|X}) \geq \textrm{UMI}(P_{Z|X})$.  Thus
$I(U_Y P_{Z|Y}) \geq I(U_X P_{Z|X})$.

\item UMI however does not satisfy Axiom $1b$ in general. However, if the transition matrices $P_{Y|X}$ and $P_{Z|Y}$ are both doubly stochastic, then a straightforward calculation shows that UMI satisfies Axiom $1b$ too.
\newcommand{\indep}{\rotatebox[origin=c]{90}{$\models$}}
\item UMI satisfies Axiom $2$ since the uniform distribution on $X_1,X_2$ naturally factors as $U_{X_1,X_2} = U_{X_1} U_{X_2}$ and we have $ \textrm{UMI}(P_{Y_1,Y_2 | X_1, X_2}) $
\beqa&=& I(U_{X_1,X_2}P_{Y_1,Y_2 | X_1, X_2}) \\
&=& I(U_{X_1} U_{X_2} P_{Y_1|X_1} P_{Y_2 | X_2})\\
& = & \textrm{UMI}(P_{Y_1|X_1}) +  \textrm{UMI}(P_{Y_2|X_2}).  \eeqa

\item UMI does not satisfy Axiom $3a$ since multiple repeated values of $P_{Y|X=x}$ does not alter the convex hull but alters the UMI value.
\item Interestingly, UMI does satisfy Axiom $4$ for the same reason as CMI.
\end{itemize}


\subsection{Real-valued alphabets}
For real-valued $X$, the Shannon mutual information is not finite  without additional regularizations.  This is also true of other measures of relation such as the  Renyi correlation \cite{renyi1959measures}, and in each case the measure is studied in the context of some form penalty term. Typically this is done via a cost constraint on the real-valued input parameters. In this context, one possibility is to consider
 the following norm-constrained optimization to ensure the causal effect is finite valued:
\beqa \textrm{CMI} (P_{Y|X}, a) := \max_{P_X: \mathbb{E}\|X\|_2^2 \leq a} I(P_X P_{Y|X}). \eeqa
In practice, $a$ is chosen from the empirical  moments of $X$ from samples: $a := \frac{1}{N} \sum_{i=1}^N \|X_i\|_2^2$ for samples $X_1,\ldots , X_N$. 
 This regularization turns to be the so-called power constraint on the input, common in treatments of additive noise communication channels.

%% file: section3.tex
\section{Estimators}
\label{sec:estimator}

 Although the definition of UMI and CMI seamlessly applies to both discrete and continuous random variables,
 the estimation becomes relatively straightforward when both $\cX$ and $\cY$ are discrete;
the estimation of the conditional distribution $P_{Y|X}$ and  the computation of UMI and CMI can be separated in a straightforward manner.
For this reason and also due to an application in genomic biology that we study,
we focus on the more challenging regime  $\cY$ is continuous.
Due to certain subtleties in the estimation process,
we provide separate estimators each customized for
each case of discrete and continuous $\cX$, respectively.

\subsection{Uniform Mutual Information}

The idea of applying UMI to infer the strength of conditional dependence was first proposed in \cite{Krishnaswamy14}.
Off-the-shelf 2-dimensional kernel density estimators (KDE) are used to first estimate the joint distribution $P_{XY}$ from given samples.
Subsequently,   the channel $P_{Y|X}$ is computed directly from the joint distribution, 
and then UMI is computed via either
numerical integration or sampling from $U_X$ and $P_{Y|X}$.
This approach suffers from known drawbacks of KDE, such as
sensitivity to the choice of the bandwidth and
 increased bias in higher  dimensional $X$ and $Y$.
However, a more critical challenge in
using KDE to estimate the joint distribution at all points (and not just at samples)
 is the {\em overkill} nature: we only need to compute a single functional (UMI) of the joint distribution, which could in principle be computed more efficiently directly from the samples.
 It is not at all clear how to {\em directly} estimate UMI.

Perhaps surprisingly,  we
bring together ideas from three topics in statistical estimation to introduce novel estimators that are also provably convergent.
Our estimator is  based on $(a)$ $k$-nearest neighbor estimators, e.g. \cite{KL87};
$(b)$ the correlation boosting idea of the estimator from \cite{Kra04}--which is widely adopted in practice  \cite{khan2007relative};
and $(c)$   the importance sampling techniques to adjust for the uniform prior for UMI.
We explain each idea below.

Consider a simpler task of computing the mutual information from samples;
several approaches exist for this estimation: 
\cite{paninski2003estimation,Kra04,wang2009divergence,pal2010estimation,sricharan2010empirical,poczos2012nonparametric,gao2014efficient,gao2015estimating,kandasamy2015nonparametric}.
Note that three applications of the entropy estimator,
such as those from  \cite{BDG97},  gives an estimate of the mutual information, i.e.
$\widehat{I}(X;Y)=\widehat{H}(X)+\widehat{H}(Y)-\widehat{H}(X,Y)$.
Each entropy term can be computed using, for example,
a KDE based approach, which suffers from the same challenges, as in UMI.
Alternatively, to bypass estimating   $P_{XY}$ at every point,
the differential entropy estimation can be done
via $k$ nearest neighbor ($k$NN) methods (pioneering work in \cite{KL87}).
This KL entropy estimator provides the first step in designing the UMI estimator.
However,
taking the route of estimating the mutual information via estimating the three differential entropies (two marginals and one joint), it is entirely unclear how to estimate two of these quantities (differential entropy of $Y$ and that of $(U,Y)$) directly from  samples.




Perhaps surprisingly,  an innovative approach undertaken in \cite{Kra04}
to improve upon three applications of KL estimators provides a solution.
The KSG estimator of \cite{Kra04}  is  based on $k$NN distance $\rho_{k,i}$
defined as the distance to the $k$-th nearest neighbor from $(X_i,Y_i)$ in  $\ell_\infty$  distance,
i.e. $\rho_{k,i} = \max\{ \| X_{j_k}-X_i \|_\infty  , \|Y_{j_k}-Y_i\|_\infty \}$ where $(X_{j_k},Y_{j_k})$ is the $k$-th nearest neighbor to
$(X_i,Y_i)$.
Precisely,  the KSG estimator is $\widehat{I}(X;Y) =$
\begin{eqnarray}
	\frac1N \sum_{i=1}^N \big( \psi(k) + \psi(N) -  \psi(n_{x,i}) - \psi(n_{y,i})\big)\;,
	\label{eq:Kraskov}
\end{eqnarray}
      where
    $\psi(x)$ is the digamma function, $\psi(x) =\Gamma'(x)/\Gamma(x)$
     (for large $x$, $\psi(x) \approx \log x - 1/(2x)$),
    and       the $k$NN statistics $n_{x,i}$ and $n_{y,i}$ are defined as
\begin{eqnarray}
    n_{x,i} &\equiv& \sum_{j \neq i} \mathbb{I}\{\|X_j - X_i\|_\infty <  \rho_{k,i} \}\;, \text{ and } \label{eq:defnx0}\\
    n_{y,i} &\equiv& \sum_{j \neq i}  \mathbb{I}\{\|Y_j - Y_i\|_\infty <  \rho_{k,i} \}. \label{eq:defny0}
\end{eqnarray}
Note that the number of nearest neighbors in $X$ and $Y$ are computed with respect to $\rho_{k,i}$
in the joint space $(X,Y)$. This innovative idea, not only gives a simple estimator, but also
has an advantage of canceling correlations in three entropy estimates,
giving an improved performance.
However, despite its popularity in practice due to its simplicity, no convergence result has been known until very recently (when \cite{gao2016demystifying} showed some consistency and rate of convergence properties).

Inspired by the innovative mutual information estimator in \eqref{eq:Kraskov},
we combine importance sampling techniques to adjust for the uniform prior for UMI, and propose a novel estimator.
On top of the provable  convergence, our estimator has only one hyper-parameter $k$ (besides the choice of bandwidth $h_N$ for
estimating the marginal distribution $P_X$ which is a significantly simpler task compared to estimating the joint),
which is the number  of nearest neighbors to consider; in practice $k$ is set to a  small integer such as 4 or 5.

\noindent
{\bf Continuous $\cX$.}
We propose a novel UMI estimator based on the Kraskov mutual information estimator.
For a conditional probability density $f_{Y|X}$,
we want to compute the uniform mutual information from $N$ i.i.d. samples   $(X_1,Y_1), \dots, (X_N,Y_N)$
that are generated from $f_{Y|X}f_X$ for some  prior on $X$.
Our  UMI estimator is based on $k$ nearest neighbor ($k$NN) statistics. Given a choice of $k\in{\mathbb Z^+}$ and $N$ samples,
\begin{align}
\hUMI\equiv
	 \frac1N \sum_{i=1}^N w_i \,\Big( \psi(k)    + \log \frac{N c_{d_x} c_{d_y}}{c_{d_x+d_y} \, n_{x,i}\, n_{y,i}} \Big) ,
    \label{def:umi}
\end{align}
      where 
      $\cX\subseteq\reals^{d_x}$, $\cY\subseteq\reals^{d_y}$,
      $c_{d} = \pi^{\frac{d}{2}}/\Gamma(\frac{d}{2}+1)$ is the volume of $d$-dimensional unit ball,
    and        $w_i$ is the {\em self-normalized} importance sampling estimate \cite{cornuet2012adaptive} of $\frac{u(X_i)}{f(X_i)}$:
\begin{eqnarray}
w_i &\equiv& \frac{N/\tilde{f}(X_i)}{\sum_{j=1}^N \,\big( 1/\tilde{f}(X_j) \,\big)}\;,
\label{eq:defw}
\end{eqnarray}
where $\tf:\cX\to\reals$ is the estimate of  $f_X(x)$. We use the standard kernel density estimator with a bandwidth $h_N$:
\begin{eqnarray}
\tilde{f}(x) &\equiv& \frac{1}{N h_N^{d_x}} \sum_{i=1}^N K\Big(\frac{X_i-x}{h_N}\Big) \;.
\end{eqnarray}
We define  the $k$NN statistics $n_{x,i}$ and $n_{y,i}$ as follows.
For each sample $(X_i,Y_i)$,  calculate the Euclidean distance $\rho_{k,i}$ (as opposed to the $\ell_\infty$ distance proposed by \cite{Kra04})
 to the $k$-th nearest neighbor.
This determines the
(random) number of samples within $\rho_{k,i}$ in $\cX$: first $n_{x,i}$ is defined as the same as in \eqref{eq:defnx0}, but with Euclidean distance; second we have
 a {\em weighted} number of samples within $\rho_{k,i}$ in $\cY$ as
\begin{eqnarray}
    n_{y,i} &\equiv& \sum_{j \neq i} w_j \mathbb{I}\{\|Y_j - Y_i\|<  \rho_{k,i} \}. \label{eq:defny}
\end{eqnarray}

Compared to \eqref{eq:Kraskov}, we first exchange log function for the digamma functions of $N$, $n_{x,i}$, and $n_{y,i}$.
This step (especially for $n_{x,i}$, and $n_{y,i}$) is crucial for proving convergence.
We use ideas from importance sampling and introduce new variables $w_i$'s
that capture the correction for the mismatch in the prior.
The constants $c_{d_x}$, $c_{d_y}$, and $c_{d_x+d_y}$ correct for the volume measured in $\ell_2$.

\bigskip
\noindent{\bf Discrete $\cX$.}
Similarly, for a discrete random variable $X$,
 the joint probability density function is denoted by  $f(x, y) = p_X(x) f_{Y|X}(y|x)$. 
 We propose a UMI estimator, and overload the same notation for this discrete case.
\begin{align}
   \hUMI \equiv \frac1N \sum_{i=1}^N w_{X_i} \,\Big( \psi(k) + \log \frac{N}{n_{X_i}\,n_{y_i}}  \,\Big) \;,
    \label{def:mixed_umi}
\end{align}
            where 
      $n_{X_i}=|\{j\in[N]: j \neq i, X_j=X_i\}|$ is the number of samples $j$ such that $X_j = X_i$,
      $w_{X_i}$ is the self-normalizing estimate of $1/(|\mathcal{X}|p_X(X_i))$ defined as
\begin{eqnarray}
w_{x} &\equiv& \frac{N}{|\mathcal{X}| n_x}\;,
\label{eq:mixed_defw}
\end{eqnarray}
and       $n_{y,i}$ is the weighted $k$NN statistics defined as follows.
For each sample $(X_i,Y_i)$,  let the distance to the $k$-th nearest neighbor be $\rho_{k,i}$,
where those samples that have the same $X$ value as $X_i$ is considered and the
Euclidean distance is measured in  $\cY$.
We define the {\em weighted} number of samples within $\rho_{k,i}$ in $\cY$ as
\begin{eqnarray}
    n_{y,i} &\equiv& \sum_{j \neq i} w_{X_j} \mathbb{I}\{\|Y_j - Y_i\|<  \rho_{k,i} \}. \label{eq:mixed_defny}
\end{eqnarray}

\subsection{Capacitated Mutual Information}

Given standard estimators for mutual information and entropy,
it is not at all clear how to {\em directly} estimate CMI where $f_X$ is changed to the (unknown) optimal input distribution.
However, combining the mutual information estimator in \eqref{eq:Kraskov} with
importance sampling techniques,
we design a novel estimator as a solution to an optimize over the space of the weights.
Our estimator has only one hyper-parameter $k$, the number  of nearest neighbors to consider.

\noindent{\bf Continuous $\cX$.}
For a conditional distribution $f_{Y|X}$, we compute an  estimate of CMI
from i.i.d. samples   $(X_1,Y_1), \dots, (X_N,Y_N)$
 generated from $f_{Y|X}f_X$ for some  prior on $X$.
We introduce a novel CMI estimator
that is based on our UMI estimator.
Given a choice of $k\in{\mathbb Z^+}$ and $N$ samples, the estimated CMI is the solution of the following constrained optimization:
\begin{align*}
    \hCMI =  \max_{w \in T_{a,N}} \frac1N \sum_{i=1}^N w_i \,\Big( \psi(k) + \log (\frac{N c_{d_x}c_{d_y}}{c_{d_x+d_y} n_{x,i} n_{y,i}}) \Big) \;,
    \label{def:cmi}
\end{align*}
      where $d_x$, $d_y$,
      $n_{x,i}$, $n_{y,i}$ and $c_d$ are defined in the same as in~\eqref{def:umi}.
      We optimize over  $w_1, \dots, w_N$ under
      the second moment constraint, i.e. $T_{a,N}=\{ w\in \reals^N | w_i\geq 0,\forall i\in[N], (1/N)\sum_{i=1}^N w_i = 1,  (1/N) \sum_{i=1}^N w_i \|X_i\|^2 \leq a^2\}$. Observe that no KDE of $P_X$ is needed for CMI estimation, making it particularly simple and robust.

\noindent{\bf Discrete $\cX$.}
Similarly, we define the CMI estimate $\hCMI$ as the solution of the following constrained optimization:
\begin{eqnarray*}
    & \hCMI = \underset{w \in T_{\Delta}}{\max} & \frac1N \sum_{i=1}^N w_{X_i} \,\Big( \psi(k) + \log (\frac{N}{ n_{x,i}n_{y,i}})\Big) \,\\
    \label{def:mixed_cmi}
\end{eqnarray*}
      where
      $n_{x,i}$ and $n_{y,i}$ are defined  in ~\eqref{def:mixed_umi}.
      $T_{\Delta}$ is the set of quantized version of an interval $[C_1,C_2]$ with step size $\Delta$, i.e.
$ 
T_{\Delta} = \big\{ w \in \{C_1 + m_i \Delta\}^{|\cX|} \big| (1/N) \sum_{x=1}^{|\cX|} w_x\in[1-{|\cX|}\Delta,1+{|\cX|}\Delta] , \text{ and }  m_i \in \{0, 1, \dots, \lceil (C_2-C_1)/C_1\rceil\} \text{ for all } i\, \big\}.
$ 
Such a quantization is crucial in proving consistence in Theorem \ref{thm:convergence_mixed_cmi}.

%% file: section4.tex
\section{Convergence Guarantees}
\label{sec:guarantee}

We show  both the proposed UMI and CMI estimators are  consistent under typical assumptions on the distribution. While consistency of estimators in the large sample limit is generally only a (basic) first step in understanding their properties, this is not so for fixed-$k$ nearest neighbor based estimators. As far as we know, the only estimator  based on fixed-$k$ nearest neighbors that is known to be consistent is the entropy estimator of \cite{KL87}, and the convergence rate is only known for the univariate case \cite{tsybakov1996root} (and that too under significant assumptions on the univariate density). Our result below for the consistency of the UMI estimator for discrete alphabet
marks another instance where consistency of fixed-$k$ nearest neighbor based estimators is established.

{\bf Uniform Mutual Information}: As our estimators use the
off-the-shelf kernel density estimator of $P_X$ \cite{DP84,SJ91} and also the
ides from the nearest-neighbor methods  \cite{KL87},
we make assumptions on the conditional density $f_{Y|X}$ that are
typical in these literature.
One extra assumption we make for UMI is that the
Radon-Nikodym derivative $\frac{dP_U}{dP_X}$ is uniformly  bounded over the support.
This is necessary for controlling the importance-sampling estimates of $w_i$'s.
We refer to the Assumption \ref{assumption:umi} in the supplementary material for a precise description.

\begin{theorem}
    Under the Assumption \ref{assumption:umi} in the supplementary material,
    the UMI estimator  converges to
    the true value in probability, i.e. for all $\varepsilon>0$ and all $\delta>0$,
    \begin{eqnarray}
        \lim_{N\to\infty} \Pr\big(\, \big|\hUMI  -  {\rm UMI}(f_{Y|X})\big|>\varepsilon\,\big) = 0\;,
        \label{eq:convergence_umi}
    \end{eqnarray}
    if  $k>\max\{d_y/d_x,d_x/d_y\}$ for continuous $X$ and
    $(\log N)^{(1+\delta)d_y} <k< \sqrt{N}/(5\log N)$ for discrete $X$.
    \label{thm:convergence_umi}
\end{theorem}

 In practice, we regularize the $k$NN distance $\rho_{k,i}$  in case it is much smaller than the expected distance of order $N^{-1/(d_x+d_y)}$.
 For continuous $\cX$, we require $k$ to be larger than the ratio of the dimensions, which is a finite constant.
 For discrete $\cX$, however, the effective dimension of $\cX$ is zero, which makes the ratio $d_y/d_x$ unbounded.
 Hence, for concentration of measure to hold, we need $k^{1/d_y}$ scaling at least logarithmically in the number of samples $N$.

{\bf Capacitated Mutual Information}: We make analogous assumptions which are described precisely in Assumption \ref{assumption:mixed_cmi} in the supplementary material.
The following theorem establishes consistency of our estimator when $\cX$ is discrete and we quantize $\cY$.
Our analysis  requires uniform convergence over all possible choices of the weights $w$, making the quantization step inevitable; improvements on this technical condition are natural future steps.

\begin{theorem}
    Under the Assumption \ref{assumption:mixed_cmi} in the supplementary material,
    the CMI estimator converges in probability to the true value up to the resolution of the quantization,
    i.e.     if $k > (\log{N})^{(1+\delta)d_y}$ for some $\delta > 0$, and $k < \sqrt{N}/(5\log{N})$,
    for all $\varepsilon>0$ and $\Delta>0$ and $s(\Delta)=O(\Delta)$
    \begin{eqnarray*}
        \lim_{N\to\infty} \Pr\big(\, \big|\hCMI - {\rm CMI}(f_{Y|X}) \big|>\varepsilon +  s(\Delta) \,\big) = 0.
        \label{eq:convergence_mixed_cmi}
    \end{eqnarray*}
    \label{thm:convergence_mixed_cmi}
\end{theorem}

%% file: section5.tex
\section{Numerical Experiments}
\label{experiments}

\begin{figure}[h]
	\begin{center}
	\includegraphics[width=.3\textwidth]{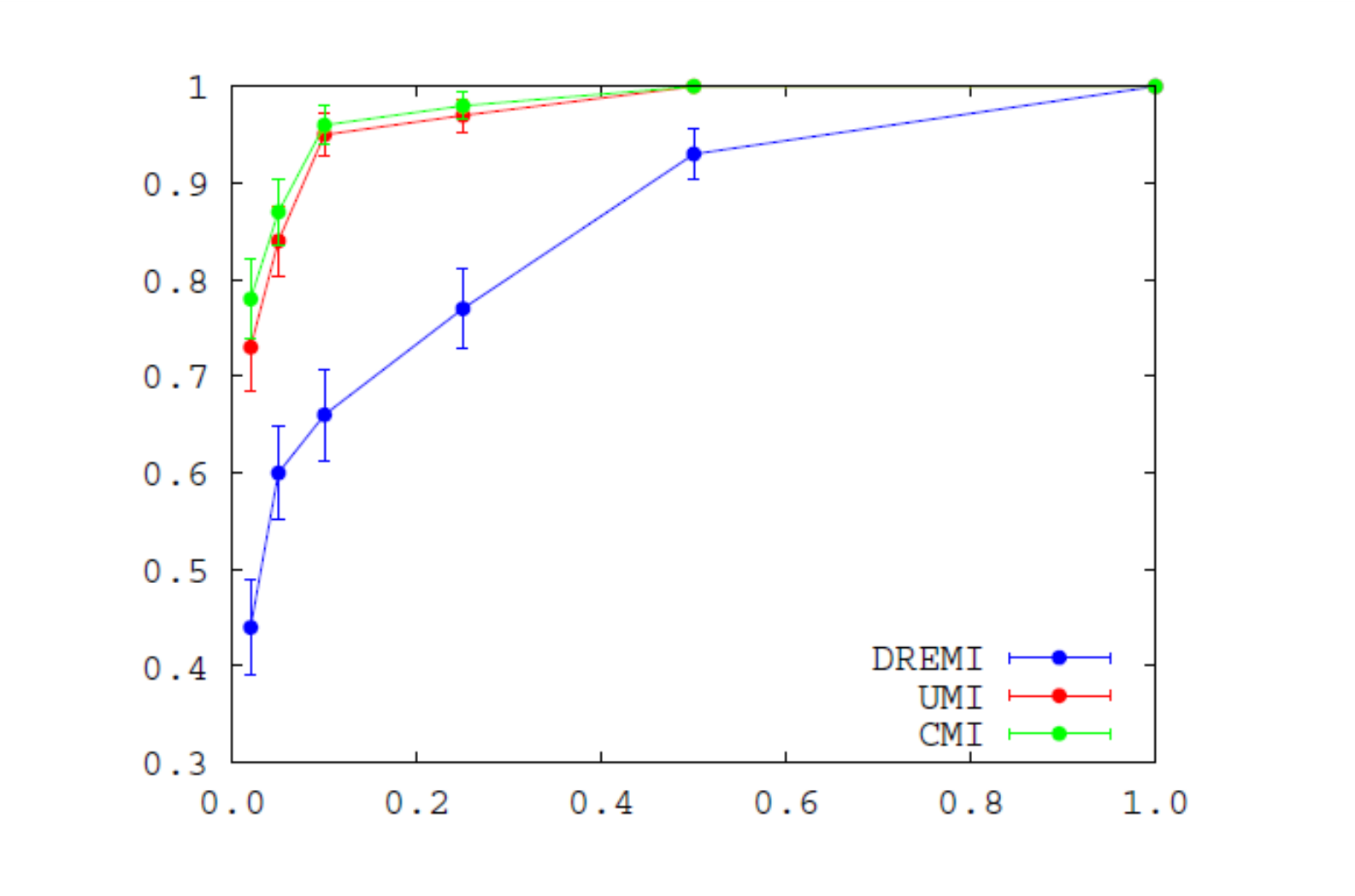}
	\includegraphics[width=.3\textwidth]{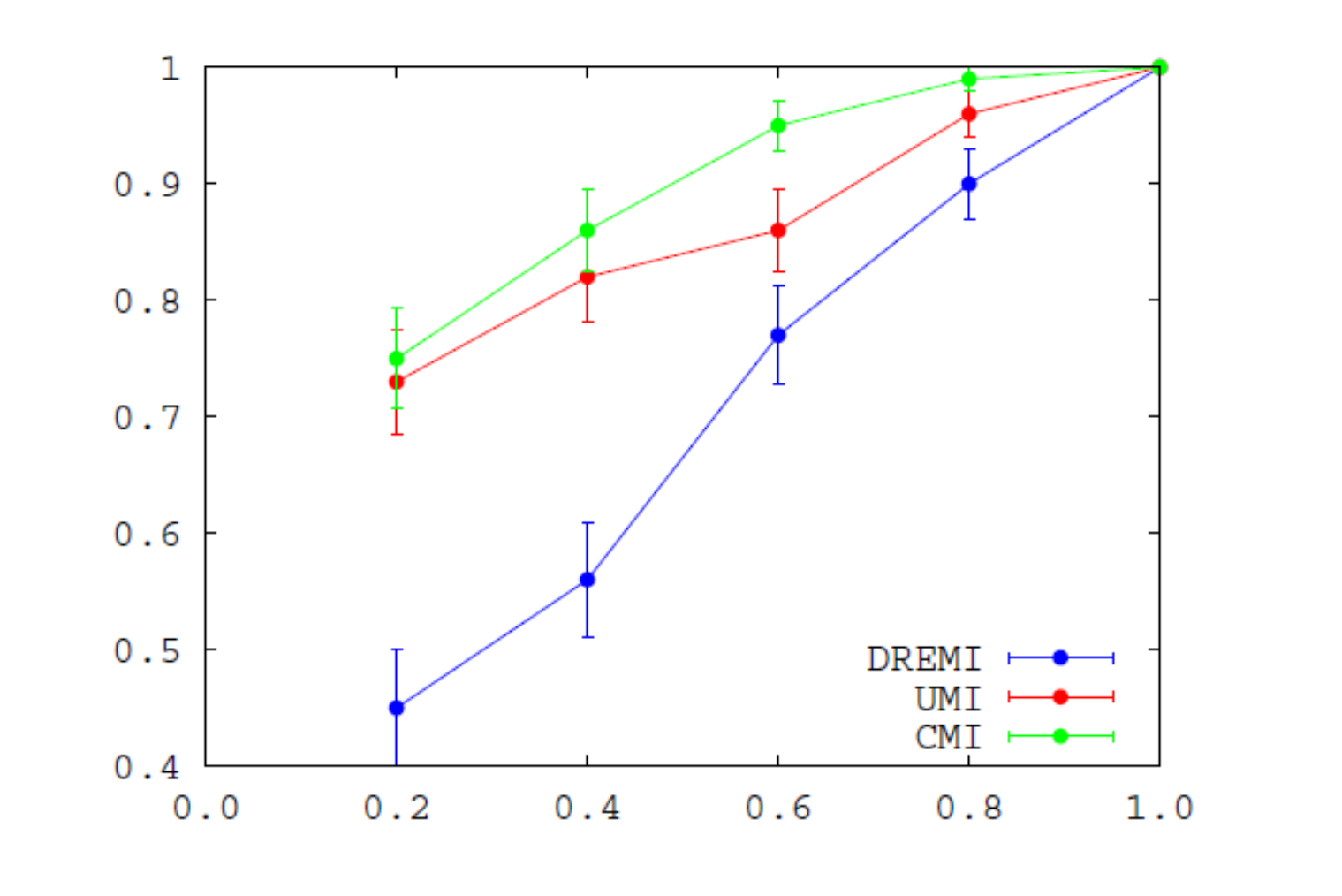}
	\put(-285,15){\rotatebox{90}{\tiny probability of success}}
	\put(-94,-2){\tiny resampling rate}
	\put(-220,-2){\tiny resampling rate}
	\end{center}
	\caption{CMI and UMI estimators significantly improve over DREMI in capturing the biological trend in flow-cytometry data: the figures above refer to the same setting as Figure 6 of   \cite{Krishnaswamy14}. }
	\label{fig:smitha1}
\end{figure}

\subsection{Gene Causal Strength from Single Cell Data}
We briefly describe the setup of \cite{Krishnaswamy14} to motivate our numerical experiments.  Consider a simple genetic pathway: a cascade of genes having expression values $X,Y,Z$ which interact linearly, i.e., $X \rightarrow Y \rightarrow Z$.  A key question of interest in this case is how the signaling in the pathway varies in different conditions of intervention. Let $T$ denote the time after the intervention (for example, after giving a certain drug). Then we may want to compare the strength of the causal relationship between two genes at different times after the intervention. In the experiments, usually samples are taken at very few time points, so $T$ has very small cardinality (for example, before the drug, $10$ minutes after the drug and $50$ minutes after the drug), but at each given time point, many cells are interrogated so we get samples from the distribution $P_{X,Y,Z;T=t} = P(Y | X; T=t) P(Z | Y; T=t)$. For each value of $T=t$, we observe $N_t$ i.i.d. samples $(X_i,Y_i,Z_i)$, for $i=1,2,...,N_t$ sampled from $P_{X,Y,Z;T=t}$. These samples are obtained using a technique called single-cell mass flow cytometry, see \cite{Krishnaswamy14} for details. We are interested in obtaining a causal measure ${\cal{C}}(X \rightarrow Y; T=t) = {\cal{C}}(P(Y | X; T=t))$ and another measure ${\cal C}(Y \rightarrow Z; T=t) = {\cal C}(P(Z|Y; T=t))$ for each time point $t$. This measure serves as a high level summary of how signaling proceeds in the cascade as a function of time, and lets one compare the strengths of a given causal relationship at different points after intervention.

%
%
%


If the drug indeed activates the causal pathway, one may expect the causal relationship to follow a certain {\em trend}, i.e., at earlier $t$, the strength of  ${\cal C}(X \rightarrow Y; T=t) $ will be high and at  a later value of $t$, the strength of ${\cal C}(Y \rightarrow Z;T=t)$ will be high before the effect of the drug wears off, at which time we expect all the relationships to fall back to its low nominal value. Such an analysis is conducted in \cite{Krishnaswamy14} where the causal strength function ${\cal C}$ is evaluated via the so-called DREMI estimator (essentially a version of UMI estimation with a ``channel amplification" step and careful choice of hyper parameters therein -- no theoretical properties of this estimator were evaluated). In that paper, it is shown that, for two example pathways, DREMI recovers the correct trend, i.e., it correctly identifies the time at which each causal relationship is expected to peak as per prior biological knowledge. This demonstrates the utility of DREMI for causal strength inference in gene networks (see Figure~6 of \cite{Krishnaswamy14}). The authors there also  demonstrate that other metrics which depend on the whole joint distribution, such as mutual information, maximal information coefficient, and correlation do not capture the trend. As an aside, we note that a somewhat different set of ``trend spotting" estimators, primarily trying to find genes which demonstrate a monotonic trend over time from single-cell RNA-sequencing data, have been proposed very recently in \cite{mueller2015modeling}.

%
%
%

In this paper, we have studied  influence measures axiomatically and proposed the UMI and CMI measures. It is natural to apply our estimators  to {\em each time point} in the same setting as \cite{Krishnaswamy14} --  and look to understand two distinct issues in our  experiments with the flow-cytometry data. The first is whether  the proposed quantities of UMI and CMI are able to capture the  same biological trend as DREMI was able to. The second question relates to the sample complexity: how does the ability to recover the trend vary as a function of the sample complexity? To study this, we  subsample the original data from \cite{Krishnaswamy14} multiple times (100 in the experiments) at each subsampling ratio and compute the fraction of times we recover the true biological trend. This is plotted in Figure~\ref{fig:smitha1}. The figure demonstrates  that when the whole dataset is made available, UMI and CMI are able to spot the trend correctly (just as DREMI does). When fewer samples are available, UMI uniformly dominates DREMI and, in turn, CMI uniformly dominates UMI in terms of capturing the biological trend as a function of number of samples available. We believe that this strong empirical evidence lends credence to our approach. For completeness, we note that the datasets represented in Figure~\ref{fig:smitha1} refer to regular T-cells (left figure) and T-cells exposed with an antigen (right figure), for which we expect different biological trends, but both of which are correctly captured by our metrics.

\begin{figure*}[t]
	\begin{center}
	\includegraphics[width=.3\textwidth]{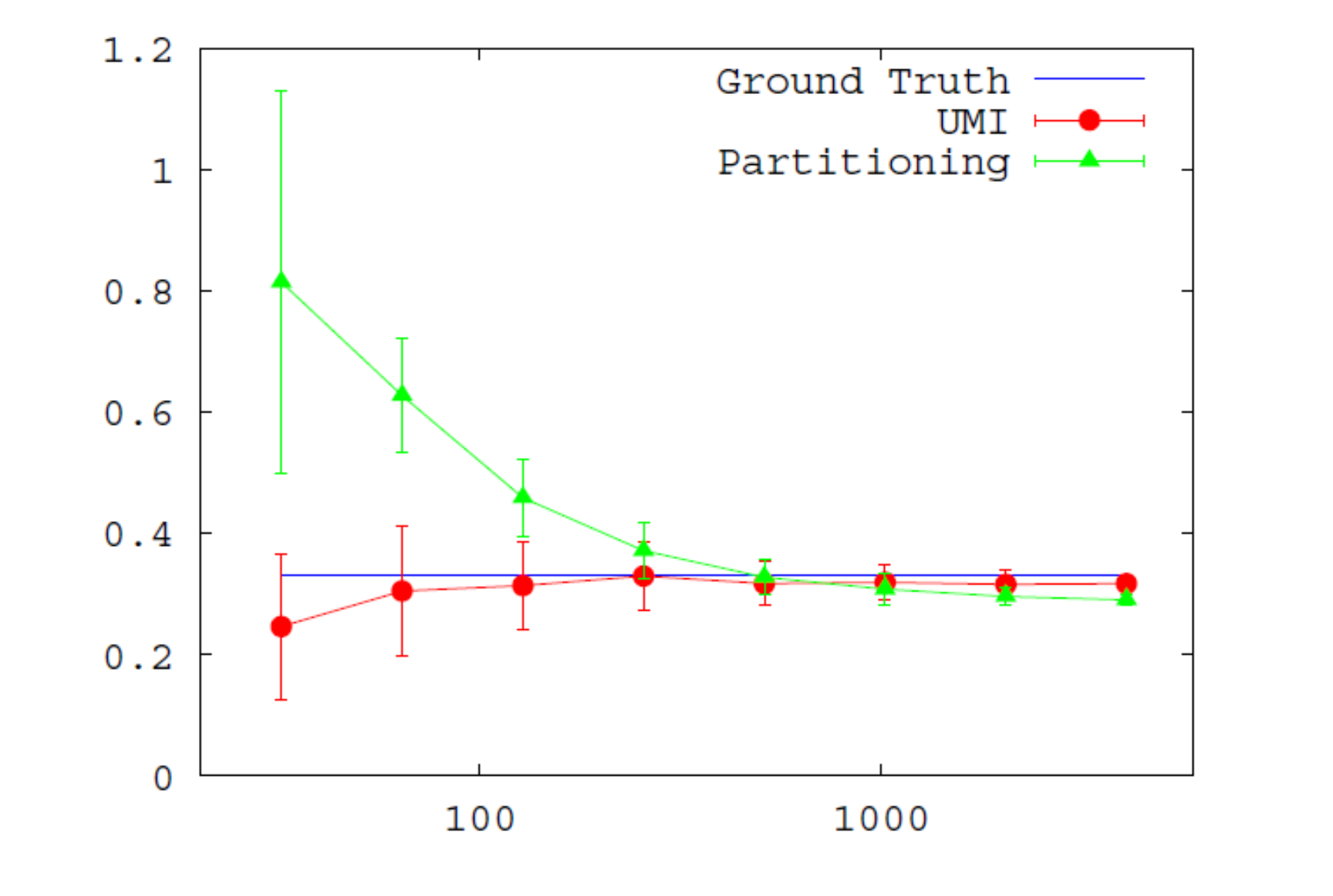}
	\put(-115,-2){number of samples $N$}
	\includegraphics[width=.3\textwidth]{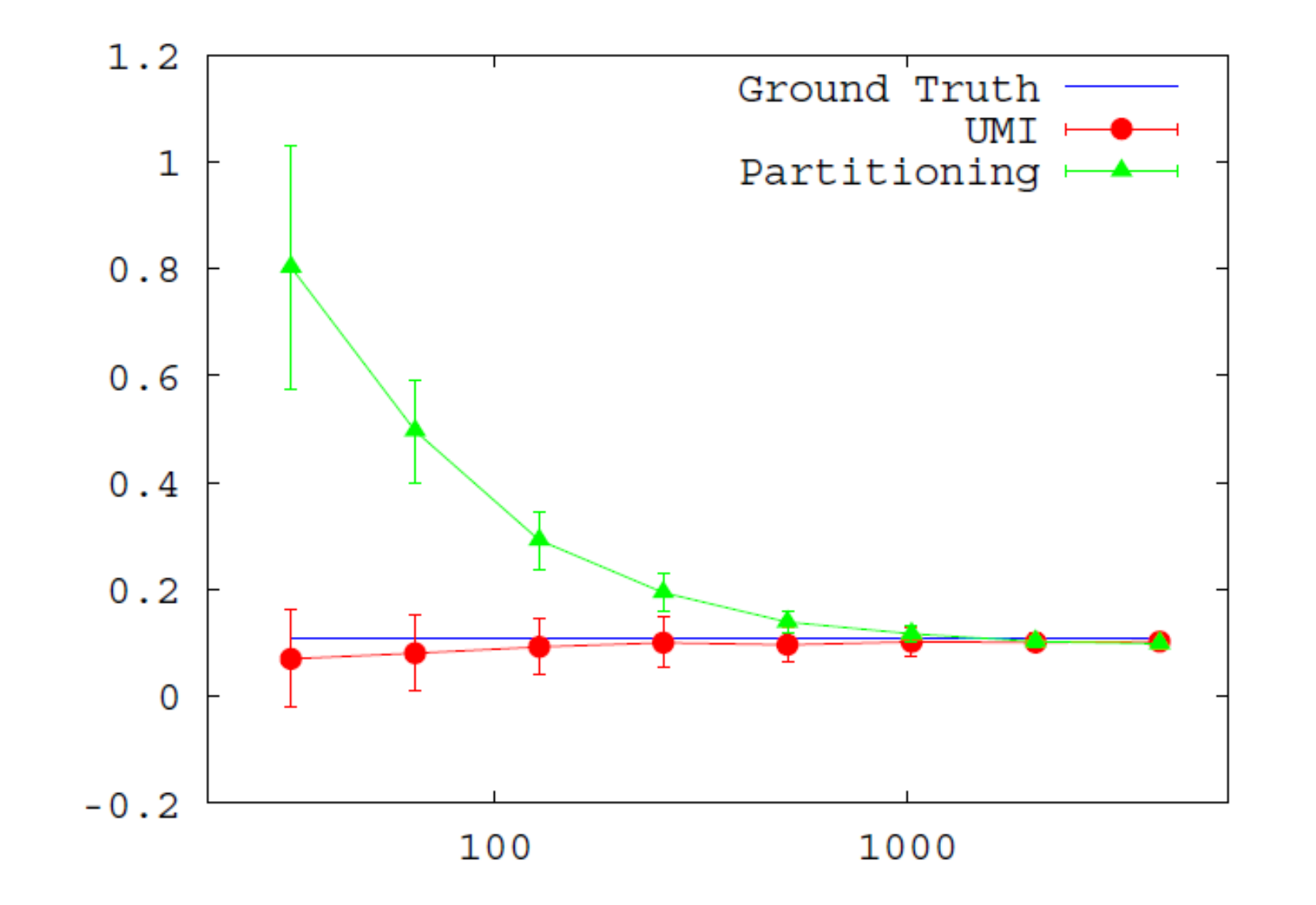}
	\put(-115,-2){number of samples $N$}
	\includegraphics[width=.3\textwidth]{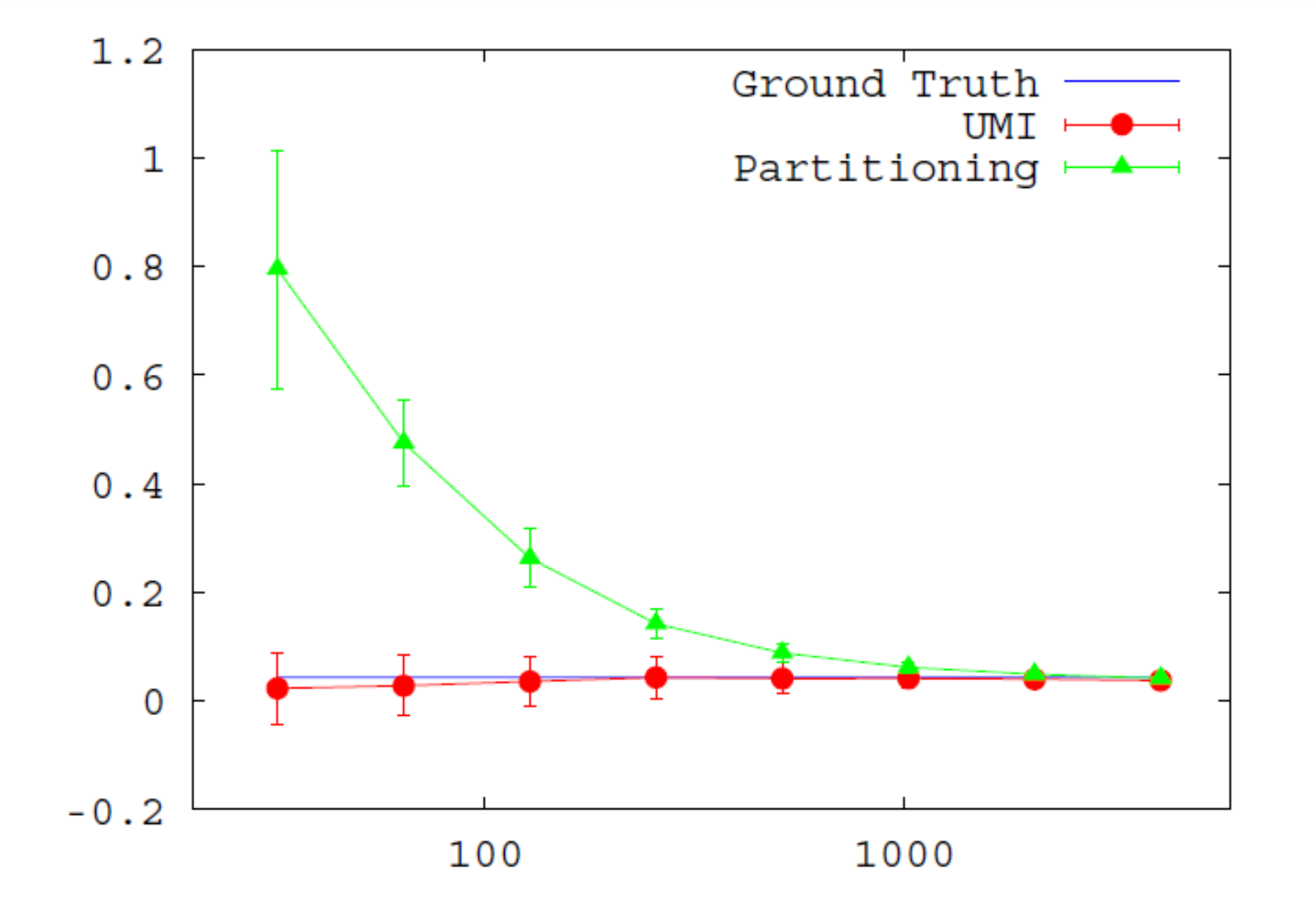}
	\put(-115,-2){number of samples $N$}
	\end{center}
	\caption{The proposed UMI estimator significantly outperforms partition based methods \cite{moddemeijer1989estimation} in sample complexity. Additive Gaussian channels are used with varying variances $\sigma^2$: $0.09$ (left), $0.36$ (middle), and $1.0$ (right). }
	\label{fig:umi}
\end{figure*}

\begin{figure*}[t]
	\begin{center}
	\includegraphics[width=.3\textwidth]{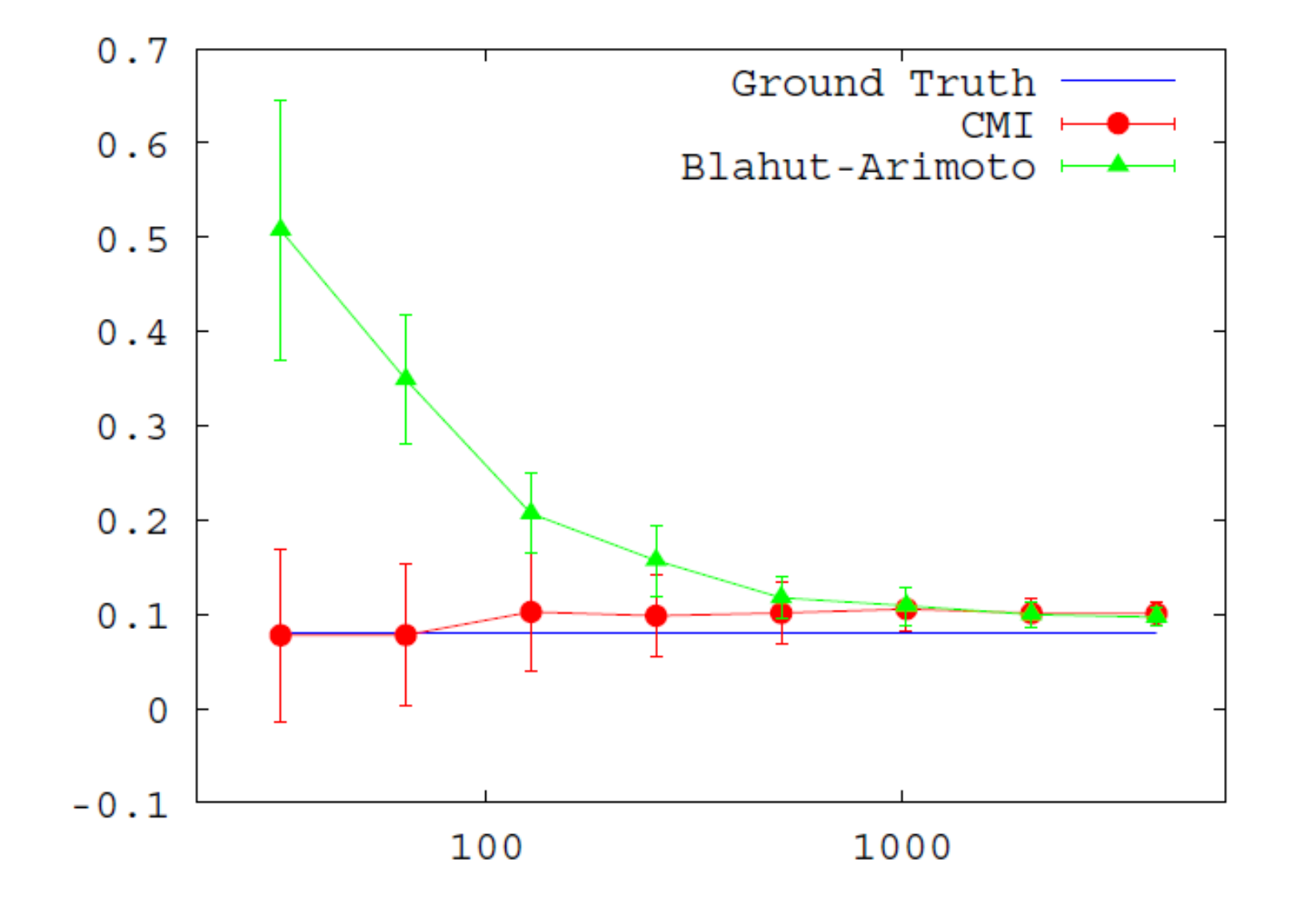}
	\put(-115,-2){number of samples $N$}
	\includegraphics[width=.3\textwidth]{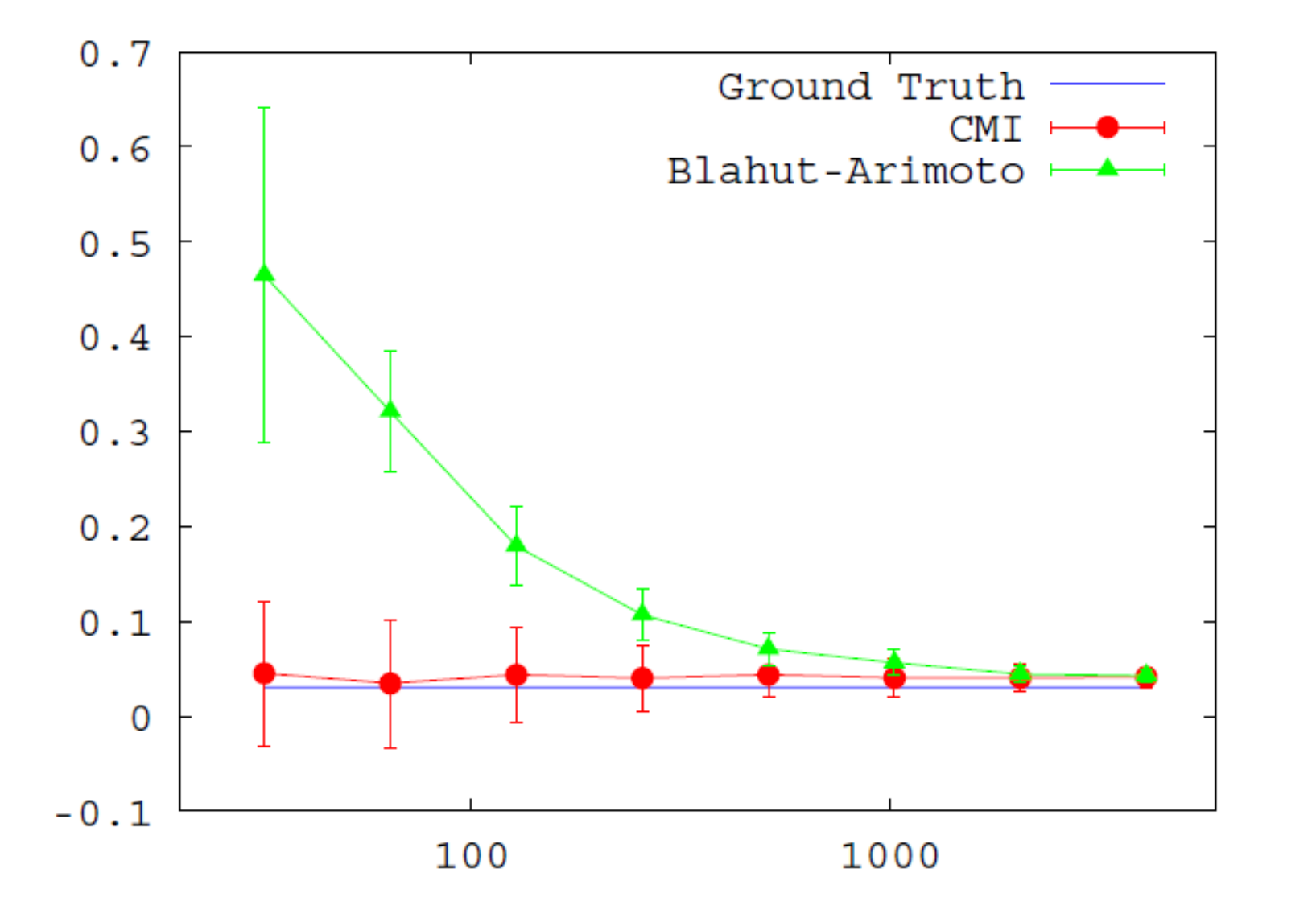}
	\put(-115,-2){number of samples $N$}
	\includegraphics[width=.3\textwidth]{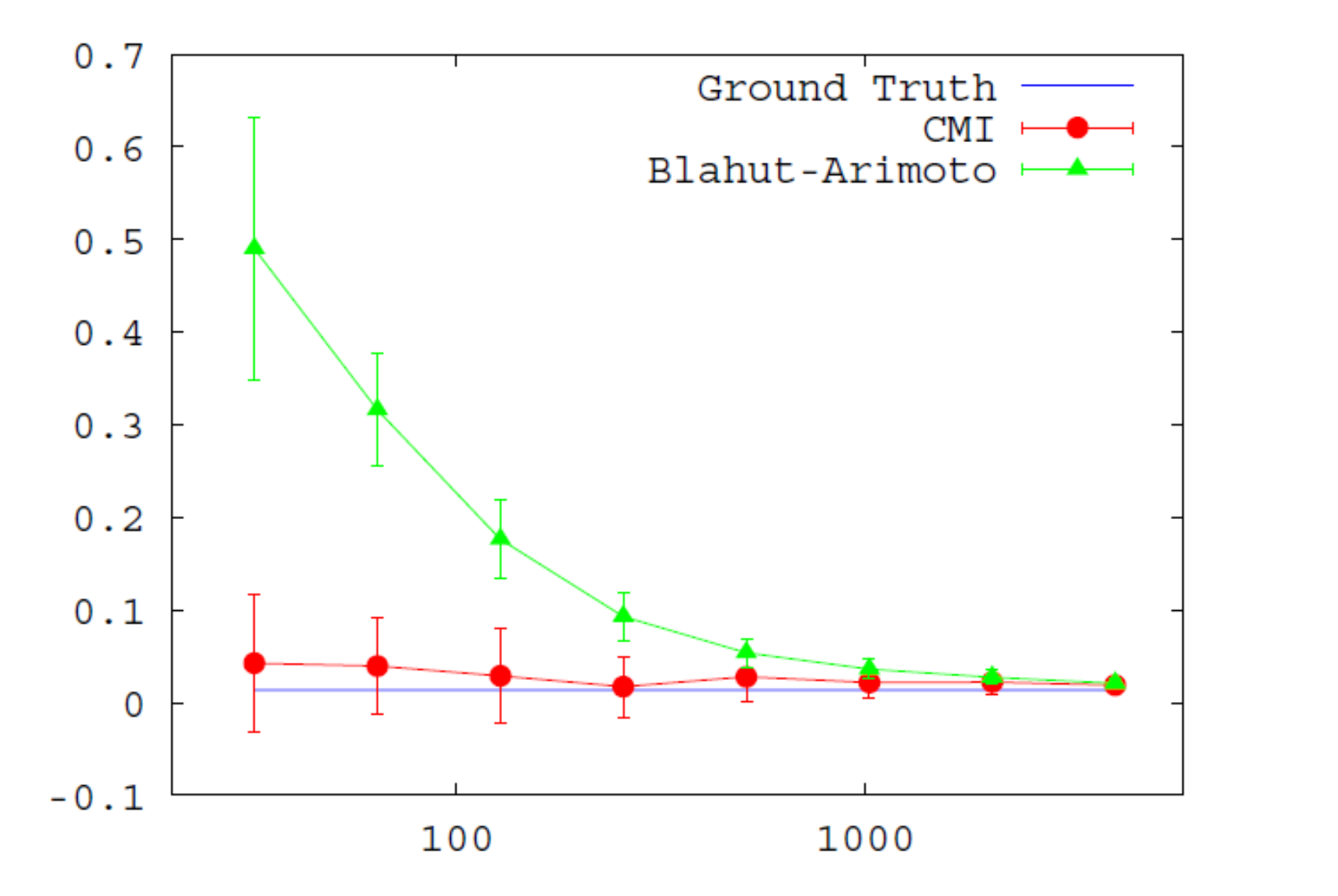}
	\put(-115,-2){number of samples $N$}
	\end{center}
	\caption{The proposed CMI estimator significantly outperforms partition based methods \cite{blahut1972computation,arimoto1972algorithm} in sample complexity.
	Additive Gaussian channels are used with varying variances $\sigma^2$: $0.36$ (left), $1.0$ (middle), and $2.25$ (right).}
	\label{fig:cmi}
\end{figure*}

\subsection{Synthetic data}

We demonstrate the accuracy of the proposed UMI  and CMI estimators on synthetic experiments.
We generate $N$ samples from $P_{XY}$ where $X$ is distributed as beta distribution $\Beta(1.5,1.5)$ and
$Y = X + N$, $N \sim \mathcal{N}(0, \sigma^2)$, independent of $X$.
We present three results with varying $\sigma^2 \in \{0.09,0.36,1.0\}$.
Figure \ref{fig:umi} shows the estimate of UMI, averaged over 100 instances.
This is compared to the ground truth and the state-of-the-art partition based estimators from \cite{moddemeijer1989estimation}.
 The ground truth has been computed via simulations with $8192$ samples from the desired distribution $P_{Y|X}U_X$ using
  Kraskov's mutual information estimator \cite{Kra04}.
 For CMI, we use exactly the same distribution $P_{XY}$ as in UMI,
but with  varying $\sigma^2 \in \{0.36,1.0,2.25\}$, which is illustrated in Figure \ref{fig:cmi}.
 Under the power constraint, the ground truth is given by $\frac{1}{2}\log(1+\frac{\sigma_X^2}{\sigma_{N}^2}) = \frac{1}{2}\log(1+1/16\sigma^2)$.
 The proposed CMI estimator is compared against
Blahut-Arimoto algorithm~\cite{blahut1972computation,arimoto1972algorithm} for computing discrete channel capacity, applied to quantized data.
Both figures illustrate that the proposed estimators significantly improves over the state-of-the-art partition based methods, in terms of sample complexity.

%% file: section6.tex
\section{Discussion}

In this paper we have proposed novel information theoretic  measures of potential influence of one variable on another, as well as provided novel estimators to compute the measures from i.i.d.\ samples. The technical innovation has been in proposing these estimators, by combining separate threads of ideas in statistics (including importance sampling and nearest-neighbor methods). The consistency proofs suggest that a similar analysis  the very popular estimator of  (traditional) mutual information in \cite{Kra04} can be conducted successfully; such work has been recently conducted in \cite{gao2016demystifying}. Several other issues in statistical estimation theory intersect with our current work and we  discuss some of these topics below.

(a) The main technical results of this paper have been weak consistency of the proposed estimators. Proving stronger consistency guarantees and rates of convergence would be natural improvements, albeit challenging ones. Rates of convergence in the nearest-neighbor methods are barely known in the literature even for traditional information theoretic quantities: for instance,  \cite{tsybakov1996root} derives a $\sqrt{N}$ consistency for the single dimensional case of differential entropy estimation (under strong assumptions on the underlying pdf), leaving higher dimensional scenarios open, and which recently have been successfully addressed in \cite{gao2016demystifying}.

(b) There is a natural generalization of our estimators when the alphabet $\cY$ is high dimensional, using the $k$NN approach (just as in the differential entropy estimator of \cite{KL87} or in the mutual information estimator of \cite{Kra04}). However, very recent works \cite{gao2014efficient,gao2015estimating,lombardi2016nonparametric}
 have shown that boundary biases common in high dimensional scenarios is much better handled  using local parametric methods (as in \cite{loader1996local,hjort1996locally}). Adapting these approaches to the estimators for UMI and CMI is an interesting direction of future research.

(c)  We have considered both the case of discrete and (single dimensional) continuous alphabet $\cX$. The scenario of high dimensional $\cX$ is significantly more challenging for CMI estimation: this is because of the (vastly) expanded space of distributions over which the optimization can be performed. Also challenging is to consider  application specific regularization of the inputs in this scenario.

(d) While the focus of this paper has been on quantifying potential causal influence, a related question involves testing the {\em direction} of causality for a pair of random variables. This is a widely studied topic with a long lineage \cite{Pearl} but also of strong topical interest \cite{Janzing13,JanSteShaSch15, Mooijetal2015, ShaJanSchBes15}. A natural inclination is to explore the efficacy of UMI and CMI measures to test for direction of causality -- especially in the context of the benchmark data sets collected in \cite{Mooijetal2015}.  Our results are as follows: UMI has a 45\% probability to predict the correct direction. CMI gives 53\% probability. Directly comparing the marginal entropy $H(X)$ and $H(Y)$ by the estimator in \cite{KL87} also only provides 45\% accuracy. While in ~\cite{Mooijetal2015}, different entropy estimators (with appropriate hyper parameter choices) were applied to get an accuracy up to 60\%-70\%. Further research is needed to shed conclusive light, although we point out that the benchmark data sets in \cite{Mooijetal2015} have substantial confounding factors that make causal direction hard to measure in the first place.

(e) The axiomatic derivation of potential causal influence naturally suggests CMI as an appropriate measure. We are also able to show  that a more general quantity -- the so-called R\'{e}nyi capacity --  also meets the axioms. For any $\lambda > 0, \lambda \neq 1$, define R\'{e}nyi entropy as
\beqa
H_\lambda(P) := \frac{1}{\lambda  - 1} \log {\mathbb E}_P [ (dP)^\lambda]
\eeqa
and
R\'{e}nyi divergence as: 
\beqa
D_{\lambda}(P \| Q) := \frac{1}{\lambda  - 1} \log {\mathbb E}_Q  \left [ \left ( \frac{dP}{dQ}\right )^\lambda \right ].
\eeqa
Now define the {\em asymmetric} information measure \cite{csiszar1995generalized}:
\beqa
K_{\lambda}(P_X P_{Y|X}) := \inf_{Q_Y} D_{\lambda}(P_{XY} \| P_X Q_Y),
\eeqa
which converges to the traditional mutual information when $\lambda \rightarrow 1$.
Now we can define the R\'{e}nyi capacity for any parameter $\lambda$ as, for any fixed conditional distribution $P_{Y|X}$:
\beqa
\textrm{CMI}_{\lambda} := \sup_{P_X} \inf_{Q_Y} D_{\lambda}(P_{XY} \| P_X Q_Y),
\eeqa
Observe that as $\lambda \rightarrow 1$, we have $\textrm{CMI}_{\lambda} \rightarrow \textrm{CMI}$, the traditional Shannon capacity.  We observe the following.

\begin{proposition}
For any $\lambda > 0, \lambda \neq 1$ we have that $\textrm{CMI}_{\lambda}$ satisfies the axioms in Section~\ref{sec:axiom}.
\label{prop:renyi}
\end{proposition}
The proof is available in Appendix~\ref{sec:proof_renyi}.
In the light of this result, it would be interesting to design estimators for the more general family of R\'{e}nyi capacity measures and confirm their performance on empirical tasks such as the ones studied in \cite{Krishnaswamy14}. It would also be very interesting to understand the role of additional axioms that would lead to uniqueness of Shannon capacity (in the same spirit as entropy being uniquely characterized by somewhat similar axioms \cite{csiszar2008axiomatic}).

(f) Finally, a  comment on the optimization problem in CMI estimation: the optimization problem involving the $w_i$'s is not necessarily a concave program for a given sample realization, 
 although this program converges to that  of Shannon capacity computation involves maximizing mutual information, which is a concave function of the input probability distribution. Standard (stochastic) gradient decent is used in our experiments,  and we did not face any disparity in convergent values over the set of synthetic experiments we conducted.

%% file: section7.tex
\clearpage
\appendix
\section*{Appendix}
\section{Proof of the UMI estimator convergence in Theorem \ref{thm:convergence_umi} }

We present the proof of the theorem for two separate UMI estimators: first for continuous $X$ and next for discrete $X$.
We first state the formal assumptions under which  the theorem holds.
\begin{assumption}
\label{assumption:umi}

For continuous $\mathcal{X}$, define
\begin{eqnarray}
f_U(y) &\equiv& \int_x u(x) f_{Y|X}(y|x) dx,  \\
 f_U(x\big|y) &\equiv& \frac{ u(x) f_{Y|X}(y|x)}{f_U(y)} .
\end{eqnarray}
We make the following assumptions:
\begin{itemize}
  \item[$(a)$] $\int f_{X,Y}(x,y) \big( \log f(x,y) \big)^2 dx dy < \infty$.
  \item[$(b)$] There exists a finite constant $C$ such that the Hessian matrix of $H(f)$ and $H(f_U)$ exists and \\ $\max\{\|H(f)\|_2 ,\|H(f_U)\|_2 \}< C$ almost everywhere.
   \item[$(c)$] There exists a positive constant $C'$ such that the conditional pdfs satisfy $f_{Y|X}(y\big|x) < C'$ and $f_U(x\big|y)<C'$ almost everywhere.
   	\item[$(d)$] There exist positive  constants $C_1 < C_2$ such that the marginal pdf satisfy,
	almost everywhere,
   	$$ \frac{C_1}{\mu(\cX)} < f_X(x) < \frac{C_2}{\mu(\cX)} \;.$$
   	\item [$(e)$] The bandwidth $h_N$ of kernel density estimator is chosen as
	$h_N = \frac{1}{2}N^{-1/(2d_x+3)}$.
\end{itemize}

For discrete $\mathcal{X}$, define
\begin{eqnarray}
	f_u(y) &\equiv& \sum_{x \in \mathcal{X}} \frac{1}{|\mathcal{X}|} f_{Y|X}(y|x) .
\end{eqnarray}
We make the following assumptions:
\begin{itemize}
  \item[$(a)$] $\int f_{Y|X}(y|x) \big(\, \log f_{Y|X}(y|x) \,\big)^2 dy < \infty$, for all $x \in \mathcal{X}$.
  \item[$(b)$] There exists a finite constant $C$ such that the Hessian matrix of $H(f_{Y|X})$ exists and $\|H(f_{Y|X})\|_2 < C$ almost everywhere, for all $x \in \mathcal{X}$.
   \item[$(c)$] There exists a finite constant $C'$ such that the conditional pdf $f_{Y|X}(y\big|x) < C'$  almost everywhere, for all $x \in \mathcal{X}$.
   \item[$(d)$] There exists finite constants $C_1 < C_2$ such that the prior $p_X(x) > C_1/|\mathcal{X}|$ and $f_X(x) < C_2/|\mathcal{X}|$ almost everywhere.
\end{itemize}

\end{assumption}

\subsection{The case of continuous $\mathcal{X}$}

Given these assumptions, we define
\begin{eqnarray}
g(X_i, Y_i) &\equiv& \psi(k) + \log(N) + \log\Big(\frac{c_{d_x}c_{d_y}}{c_{d_x+d_y}}\Big) - \big( \log(n_{x,i}) + \log( \tn_{y,i}) \big)
\end{eqnarray}
such that $\hat{I}^{U}_{k,N}(X,Y) = \frac{1}{N} \sum_{i=1}^N \,w_i \,g(X_i, Y_i)$. Define each quantity with the true prior $f_X(x)$ as
\begin{eqnarray}
w'_i &\equiv& \frac{u(X_i)}{f_X(X_i)} \,,\\
n'_{y,i} &\equiv& \sum_{j \neq i} w'_j \mathbb{I} \{\|Y_j - Y_i\| < \rho_{k, i}\} \,,\\
g'(X_i, Y_i) &\equiv& \psi(k) + \log(N) + \log\Big(\frac{c_{d_x}c_{d_y}}{c_{d_x+d_y}}\Big) - \big( \log(n_{x,i}) + \log(n'_{y,i}) \big) \,.
\end{eqnarray}
With $U_X$ equal to the uniform distribution on the support of $X$,
we apply the triangle inequality to show that each term converges to zero in probability.
\begin{eqnarray}
 \big| \hat{I}^U_{k,N}(X,Y) - I^U(f_{Y|X}) \big|
    & = & \Big| \frac{1}{N}\sum_{i=1}^N w_i g(X_i, Y_i) - \iint u(x) f_{Y|X}(y|x) \log \frac{f_{Y|X}(y|x)}{\int u(x') f_{Y|X}(y|x') dx'} dx dy \Big| \notag\\
    &\leq & \frac{1}{N} \Big|\, \sum_{i=1}^N \big( w_i g(X_i,Y_i) - w'_i g'(X_i, Y_i) \big) \Big| + \label{eq:kde_error}\\
    &  & \Big|\, \frac{1}{N} \sum_{i=1}^N w'_i g'(X_i, Y_i) - \iint u(x) f_{Y|X}(y|x) \log \frac{f_{Y|X}(y|x)}{\int u(x') f_{Y|X}(y|x') dx'} dx dy \,\Big| .\label{eq:knn_error}
\end{eqnarray}
The first term~(\ref{eq:kde_error}) captures the error in the kernel density estimator and we have the following claim, whose proof is delegated to Appendix~\ref{app:proof-of-kde_error}.
\begin{lemma}
    The term in Equation~\eqref{eq:kde_error} converges to 0 as $N\rightarrow \infty$ in probability.
      \label{lem:eqkde_error}
	 \end{lemma}

The second term in the error \eqref{eq:knn_error} comes from the sample noise in density estimation.
Similar to the decomposition of mutual information, $I(X;Y)=H(X)+H(Y)-H(X,Y)$, we decompose our estimator into three terms:
\begin{eqnarray*}
    \frac1N \sum_{i=1}^N w'_i g'(X_i, Y_i) &=& \hH^U_{k,N}(X) + \hH^U_{k,N}(Y) - \hH^U_{k,N}(X,Y) - \sum_{i=1}^N \frac{w'_i}{N}(2\log(N-1)-\psi(N)-\log(N))\;,
\end{eqnarray*}
where
\begin{eqnarray}
    \hH^U_{k,N}(X,Y) &\equiv& \sum_{i=1}^N \frac{w'_i}{N}  \big(\, -\psi(k) + \psi(N) + \log{c_{d_x+d_y}} + (d_x+d_y) \log \rho_{k,i} \,\big) \;, \label{def:hHUxy}\\
    \hH^U_{k,N}(X) &\equiv& \sum_{i=1}^N  \frac{w'_i}{N} \big(\, -\log {n_{x,i}} + \log(N-1) + \log c_{d_x} + d_x \log \rho_{k,i} \,\big)\;, \label{def:hHUx}\\
    \hH^U_{k,N}(Y) &\equiv& \sum_{i=1}^N \frac{w'_i}{N} \big(\, -\log {\tn_{y,i}} + \log(N-1) + \log c_{d_y} + d_y \log \rho_{k,i}  \, \big). \label{def:hHUy}
\end{eqnarray}
Notice that$\sum_{i=1}^N \frac{w'_i}{N}(2\log(N-1)-\psi(N)-\log(N))$ goes to 0 as $N$ goes to infinity.
The desired claim follows directly from
the following two lemmas showing the convergence each entropy estimates to corresponding entropies under UMI.

\begin{lemma}
    \label{lem:HUxy}
    Under the hypotheses of Theorem \ref{thm:convergence_umi}, for all $\varepsilon > 0$
    \begin{eqnarray}
    \lim_{N \to \infty} \Pr\left(\, \Big|\hH^U_{k,N}(X,Y) - \big(\, -\iint u(x) f_{Y|X}(y|x) \log f_{Y|X}(y|x)u(x) dx dy \,\big) \Big| > \varepsilon\,\right) = 0 \;.
    \end{eqnarray}
\end{lemma}

\begin{lemma}
    \label{lem:HUx_HUy}
    Under the hypotheses of Theorem  \ref{thm:convergence_umi}, for all $\varepsilon > 0$
    \begin{eqnarray}
    \lim_{N \to \infty} \Pr\left(\, \Big|\hH^U_{k,N}(X) + \hH^U_{k,N}(Y) - \big(\, -\iint u(x) f_{Y|X}(y|x) \big(\log f_{X}(x) + f_U(y)\big) dx dy\,\big)\Big| > \varepsilon\,\right) = 0 \;,
    \end{eqnarray}
    where $f_U(y)=\int f_{Y|X}(y|x)u(x)dx$.
\end{lemma}

A crucial  technical idea in proving these lemmas is
the concept of importance sampling. For any function $h(X_i,Y_i)$,
the \textit{importance sampling}  estimate of $\E [h]$ is given by
\begin{eqnarray}
\tilde{h}_N = \frac{1}{N} \sum_{i=1}^N w'_i h(X_i,Y_i) \;,
\end{eqnarray}
where
$ w_i' = N {u(X_i)}/{f(X_i)}$.
 The following lemma gives the almost sure convergence of $\tilde{h}_n$.

\begin{lemma}[Theorem 9.1 in \cite{mcbook}]
    \label{lem:snis}
    Assume
    $\E [h] = \iint u(x) f_{Y|X}(y|x) h(x,y) dxdy$
    exists, then $$\Pr \big(\, \lim_{n \to \infty}\tilde{h}_n = \E [h] \,\big) = 1 \;.$$
\end{lemma}


\subsubsection{Proof of Lemma \ref{lem:KDE}}
Given $X_i = x$, denote $a_j = \frac{1}{h_N^{d_x}} K(\frac{X_j - x}{h_N})$ such that $\tilde{f}_X(X_i) = \frac{1}{N} \sum_{j=1}^N a_j$.  For sufficiently small $h_N$, the mean of $a_j$ is given by:
\begin{eqnarray}
\E[a_j] &=& \int_{z \in \mathbb{R}^{d_X}} \frac{1}{h_N^{d_x}} K(\frac{z-X_i}{h_N}) f_X(z) dz \,\notag\\
&=& \int_{u \in \mathbb{R}^{d_X}} K(u) f_X(x+h_N u) du \,\notag\\
&=& \int_{u \in \mathbb{R}^{d_X}} K(u) \big(\, f_X(x) + h_N u^T \nabla f_X(x) + h_N^2 u^T H(f_X)(x) u + o(h_N^2)) du \,\notag\\
&=& f_X(x) + h_N^2 C \int_{u \in \mathbb{R}^{d_X}} \|u\|^2 K(u) du + o(h_N^2) \;,
\end{eqnarray}
where we used the fact that the kernel is centered such that $\int K(u) a^Tu du = 0$.
For sufficiently small $h_N$, we obtain $|\E[a_j] - f_X(x)| < h_N$. Therefore,
\begin{eqnarray}
&& \Pr \Big(\, \big|\, \tilde{f}_X(X_i) - f_X(X_i) \,\big| > N^{-1/(2d_x+3)} \Big| X_i = x\Big) \,\notag\\
&\leq& \Pr \big(\, \big|\, \sum_{j=1}^N a_j - N f_X(X_i) \,\big| > N^{(2d_x+2)/(2d_x+3)} \Big| X_i = x\big) \,\notag\\
&\leq& \Pr \big(\, \big|\, \sum_{j \neq i} a_j - (N-1)\E[a_j] \,\big| > N^{(2d_x+2)/(2d_x+3)} - \big| a_i - N f_X(x) + (N-1) \E[a_j]\big| \Big| X_i = x\big) .
\end{eqnarray}
Since $a_i$ is bounded by $A/h_N^{d_x}$, by choosing $h_N = \frac{1}{2}N^{-\frac{1}{2d_x+3}}$, the right hand side is lower bounded by:
\begin{eqnarray}
&& N^{(2d_x+2)/(2d_x+3)} - \big| a_i - N f_X(x) + (N-1) \E[a_j]\big| \,\notag\\
&\geq& N^{(2d_x+2)/(2d_x+3)} - |a_i - f_X(x)| - (N-1)|\E[a_j] - f_X(x)|\,\notag\\
&\geq& N^{(2d_x+2)/(2d_x+3)} - A/h_N^{d_x} - Nh_N\,\notag\\
&\geq& \frac{1}{3} N^{(2d_x+2)/(2d_x+3)}.
\end{eqnarray}
Since for $j \neq i$, $a_j's$ are i.i.d and bounded by $A/h_N^{d_x}$, by Hoeffding's inequality, we obtain
\begin{eqnarray}
&& \Pr \big(\, \big|\, \sum_{j \neq i} a_j - (N-1)\E[a_j] \,\big| > N^{(2d_x+2)/(2d_x+3)} - \big| a_i - N f_X(x) + (N-1) \E[a_j]\big| \Big| X_i = x\big) \,\notag\\
&\leq& \Pr \big(\, \big|\, \sum_{j \neq i} a_j - (N-1)\E[a_j] \,\big| > \frac{1}{3} N^{(2d_x+2)/(2d_x+3)} \Big| X_i = x\big)\,\notag\\
&\leq& 2\exp\{-2\frac{(\frac{1}{3} N^{(2d_x+2)/(2d_x+3)})^2}{(N-1) \cdot A^2 h_N^{-2d_x}}\} \,\notag\\
&\leq& 2\exp\{-\frac{2}{9A^2(N-1)} (N^{(2d_x+2)/(2d_x+3)} h_N^{d_x})^2\}
\,\notag\\
&\leq& 2\exp\{-\frac{N^{1/(2d_x+3)}}{9A^2}\}.
\end{eqnarray}
Since this upper bound is independent of $x$, we can take expectation over $x$ to obtain the desired claim.

\subsubsection{Proof of Lemma \ref{lem:HUxy}}

Define
\begin{eqnarray}
\hat{f}_{X,Y}(X_i,Y_i) = \frac{\exp\{\psi(k) - \psi(N)\}}{c_{d_x+d_y} \rho_{k,i}^{d_x+d_y}}\;,
\end{eqnarray}
so that
\begin{eqnarray}
\hat{H}^U_{k,N}(X,Y) = -\sum_{i=1}^N \frac{w'_i}{N} \log \hat{f}_{X,Y}(X_i, Y_i) \;.
\end{eqnarray}
By Theorem 8 of~\cite{singh2003nearest}, we have
\begin{eqnarray}
\lim_{N \to \infty} \E \big[\log \hat{f}_{X,Y}(X_i,Y_i) \big| (X_i, Y_i) = (x,y) \big] &=& \log f_{X,Y}(x,y)\;.
\end{eqnarray}
Notice that $w'_i \log \hat{f}(X_i,Y_i)$ are identically distributed, therefore, by plugging in $w'_i = u(X_i)/f_X(X_i)$, we have
\begin{eqnarray}
&&\lim_{N \to \infty} \E \hat{H}^U_{k,N}(X,Y) \notag\\
&=& -\lim_{N \to \infty} \E [w'_i \log \hat{f}_{X,Y}(X_i,Y_i)] \notag\\
&=& - \lim_{N \to \infty} \iint \E \big[ \frac{u(X_i)}{f_X(X_i)} \log \hat{f}_{X,Y}(X_i,Y_i) \big| (X_i, Y_i) = (x,y)\big] f_{X,Y}(x,y) dx dy  \notag \\
&=& - \lim_{N \to \infty} \iint u(x) f_{Y|X}(y|x)\E \big[\log \hat{f}_{X,Y}(X_i,Y_i) \big| (X_i, Y_i) = (x,y)\big]  dx dy.
\end{eqnarray}
Now we want to show that
\begin{eqnarray}
&& \lim_{N \to \infty} \iint u(x) f_{Y|X}(y|x) \E \big[\log \hat{f}_{X,Y}(X_i,Y_i) \big| (X_i, Y_i) = (x,y)\big]  dx dy \notag \\
&=& \iint u(x) f_{Y|X}(y|x) \big(\, \lim_{N \to \infty} \E \big[\log \hat{f}_{X,Y}(X_i,Y_i) \big| (X_i, Y_i) = (x,y)\big] \,\big) dx dy \notag\\
&=& \iint u(x) f_{Y|X}(y|x) \log f_{X,Y}(x,y)dx dy \;,
\end{eqnarray}
which follows from the
reverse Fatou's lemma and the fact that
\begin{eqnarray}
&& \limsup_{N \to \infty} \iint \big|\, \frac{u(x)}{f(x)} \E \big[\log \hat{f}_{X,Y}(X_i,Y_i) \big| (X_i, Y_i) = (x,y)\big] \,\big|^{2} f(x,y) dx dy \,\notag\\
&\leq& C_1^{-2}  \limsup_{N \to \infty} \iint \big|\, \E \big[\log \hat{f}_{X,Y}(X_i,Y_i) \big| (X_i, Y_i) = (x,y)\big] \,\big|^{2} f(x,y) dx dy \,\notag\\
&\leq& C_1^{-2}  \iint \limsup_{N \to \infty} \big|\, \E \big[\log \hat{f}_{X,Y}(X_i,Y_i) \big| (X_i, Y_i) = (x,y)\big] \,\big|^{2} f(x,y) dx dy \,\notag\\
&\leq& C_1^{-2} \iint \limsup_{N \to \infty}  \big(\, \log f_{X,Y}(x,y) \,\big)^{2} f(x,y) dx dy \,\notag\\
&<& +\infty \;. \label{eq:Fatou1}
\end{eqnarray}
As explained in the main result section, we regularize the $k$NN distance such that
$\rho_{k,i}^{d_x+d_y} > C k/N$ for some positive constant $C$.
This ensures that $\log \hat{f}_{X,Y}(X_i,Y_i) < C'$ almost surely.
It follows that $\E[\log \hat{f}_{X,Y}(X_i,Y_i)|X_i=x,Y_i=y] < C'$ and one can apply reverse Fatou's lemma.
Similar interchange of limit has been used in
\cite{KL87,wang2009divergence} without the regularization; in this context
\cite{pal2010estimation} claims that this step is not justified
(although no counterexample is pointed out). But in our case
given the practical way the algorithm is implemented with the regularization, reverse Fatou's lemma is justified.
Therefore,
\begin{eqnarray}
\lim_{N \to \infty} \E \hat{H}^U_{k,N}(X,Y) = -\iint u(x) f_{Y|X}(y|x) \log f_{X,Y}(x,y)dx dy.\label{eq:mean_HU}
\end{eqnarray}
Moreover, by Theorem11 of~\cite{singh2003nearest}, we have:
\begin{eqnarray}
\lim_{N \to \infty} \V [\hat{f}_{X,Y}(X_i,Y_i)] &=& (\frac{\Gamma'(k)}{\Gamma(k)})' \, \V [\log f_{X,Y}(x,y)]\;,
\end{eqnarray}
and for any $j \neq i$:
\begin{eqnarray}
\lim_{N \to \infty} \Cov [\hat{f}_{X,Y}(X_i,Y_i), \hat{f}_{X,Y}(X_j,Y_j)] = 0\;.
\end{eqnarray}
By $w'_i \leq 1/C_1$ for all $i$ and the fact that $\hat{f}_{X,Y}(X_i, Y_i)$ are identically distributed, we have:
\begin{eqnarray}
&&  \V \big[ \hat{H}^U_{k,N}(X,Y) \big] \notag \\
&=& \sum_{i=1}^N \frac{(w'_i)^2}{N^2} \V [\hat{f}_{X,Y}(X_i,Y_i)] + \sum_{j \neq i} \frac{w'_i w'_j}{N^2} \Cov [\hat{f}_{X,Y}(X_i,Y_i), \hat{f}_{X,Y}(X_j, Y_j)] \notag \\
&\leq& \sum_{i=1}^N \frac{1}{C_1^2 N^2} \V [\hat{f}_{X,Y}(X_i,Y_i)] + \sum_{j \neq i} \frac{1}{C_1^2 N^2} \Cov [\hat{f}_{X,Y}(X_i,Y_i), \hat{f}_{X,Y}(X_j, Y_j)] \notag \\
&=& \frac{1}{C_1^2 N}((\frac{\Gamma'(k)}{\Gamma(k)})' \V [\log f_{X,Y}(x,y)]) + \frac{1}{C_1^2 N^2} {N \choose 2} \Cov [\hat{f}_{X,Y}(X_1,Y_1), \hat{f}_{X,Y}(X_2, Y_2)] \;.
\end{eqnarray}
Therefore,
\begin{eqnarray}
\lim_{N \to \infty} \V \big[ \hat{H}^U_{k,N}(X,Y) \big] = 0 \label{eq:var_HU}\;.
\end{eqnarray}
Combining~(\ref{eq:mean_HU}) and~(\ref{eq:var_HU}), we get:
\begin{eqnarray}
&&\lim_{N \to \infty} \E \Big[ \big(\hH^U_{k,N}(X,Y) - \big(\, -\iint u(x) f_{Y|X}(y|x) \log f_{X,Y}(x,y) dx dy \,\big) \big)^2 \Big] \notag\\
&=& \lim_{N \to \infty} \V \big[ \hat{H}^U_{k,N}(X,Y) \big] + \lim_{N \to \infty} \Big( \E \hat{H}^U_{k,N}(X,Y) - \big(\, -\iint u(x) f_{Y|X}(y|x) \log f_{X,Y}(x,y) dx dy \,\big) \Big)^2 \notag\\
&=& 0.
\end{eqnarray}
Therefore, $\hat{H}^U_{k,N}(X,Y)$ converges to its mean in $L^2$, and hence in probability, i.e.,
\begin{eqnarray}
    \lim_{N \to \infty} \Pr\Big(\, \big|\hH^U_{k,N}(X,Y) - \big(\, -\iint u(x) f_{Y|X}(y|x) \log f_{X,Y}(x,y) dx dy \,\big) \big| > \varepsilon\,\Big) = 0.
    \end{eqnarray}

\subsubsection{Proof of Lemma \ref{lem:HUx_HUy}}
Define
\begin{eqnarray}
\hat{f}_X(X_i) &\equiv& \frac{n_{x,i}}{(N-1)c_{d_x}\rho_{k,i}^{d_x}} \;, \\
\hat{f}_U(Y_i) &\equiv& \frac{\tn_{y,i}}{(N-1)c_{d_y}\rho_{k,i}^{d_y}} \;,
\end{eqnarray}
such that
\begin{eqnarray}
\hat{H}_{k,N}^U(X) + \hat{H}^U_{k,N}(Y) = -\sum_{i=1}^N \frac{w'_i}{N}  \big( \, \log \hat{f}_X(X_i) + \log \hat{f}_U(Y_i) \,\big).
\end{eqnarray}
By triangle inequality, we can write the formula in Lemma~\ref{lem:HUx_HUy} as:
\begin{eqnarray}
    && \big|\hH^U_{k,N}(X) + \hH^U_{k,N}(Y) - \big(\, -\iint u(x) f_{Y|X}(y|x) \big(\log f_{X}(x) + \log f_U(y)\big) dx dy\,\big)\big| \notag \\
    & = & \big|\sum_{i=1}^N \frac{w'_i}{N}  \big(\, \log \hat{f}_X(X_i) + \log \hat{f}_U(Y_i) \,\big) - \iint u(x) f_{Y|X}(y|x) \big(\log f_{X}(x) + \log f_U(y)\big) dx dy \big| \notag\\
    &　\leq & \big| \sum_{i=1}^N \frac{w'_i}{N} \big(\, \log f_X(X_i) + \log f_U(Y_i) \,\big) - \iint u(x) f_{Y|X}(y|x) \big(\log f_{X}(x) + \log f_U(y)\big) dx dy \big| \label{eq:snis_error}\\
    &+&  \sum_{i=1}^N \frac{w'_i}{N} \Big|\big(\, \log \hat{f}_X(X_i) + \log \hat{f}_U(Y_i) \,\big) - \big(\, \log f_X(X_i) + \log f_U(Y_i) \,\big) \Big| . \label{eq:estimation_error}
\end{eqnarray}

The first term~(\ref{eq:snis_error}) comes from  sampling. Recall that $w'_i = u(X_i)/f_X(X_i)$. Since the random variables $w'_i \big(\, \log f_X(X_i) + \log f_U(Y_i)\,\big)$ are i.i.d., therefore, by the strong law of large numbers,
\begin{eqnarray}
     \sum_{i=1}^N \frac{w'_i}{N} \big(\, \log f_X(X_i) + \log f_U(Y_i) \,\big) \to \E \Big(\, \frac{u(x)}{f_X(x)} \big(\, \log f_X(x) + \log f_U(y) \,\big) \,\Big)
\end{eqnarray}
almost surely. The mean is given by
\begin{eqnarray}
    \E \big(\, \frac{u(x)}{f_X(x)} (\log f_X(x) + \log f_U(y)) \,\big) &=& \iint \frac{u(x)}{f_X(x)} \big(\, \log f_X(x) + \log f_U(y) \,\big) f(x,y) dx dy \,\notag \\
    &=&  \iint u(x) f_{Y|X}(y|x) \big(\, \log f_X(x) + \log f_U(y) \,\big) dx dy.
\end{eqnarray}
Therefore,~\eqref{eq:snis_error} converges to 0 almost surely.
\\

The second term \eqref{eq:estimation_error} comes from density estimation. To simplfy the  notations, let $Z_i = (X_i, Y_i)$, $z = (x,y)$ and $f(z) = f(x,y)$. For any fixed $\varepsilon > 0$, by union bound, we obtain that
\begin{eqnarray}
    && \Pr\big(\,\sum_{i=1}^N \frac{w'_i}{N} \big|\big(\, \log \hat{f}_X(X_i) + \log \hat{f}_U(Y_i) \,\big) - \big(\, \log f_X(X_i) + \log f_U(Y_i) \,\big) \big| > \varepsilon\,\big) \notag\\
    & \leq & \Pr\big(\,\bigcup_{i=1}^N \big\{ \big|\big(\, \log \hat{f}_X(X_i) + \log \hat{f}_U(Y_i) \,\big) - \big(\, \log f_X(X_i) + \log f_U(Y_i) \,\big) \big| > \varepsilon/2 \big\} \,\big) + \Pr(\sum_{i=1}^N w'_i > 2N).
\end{eqnarray}
The second term converges to zero by Lemma~\ref{lem:snis}. The first term is bounded by:
\begin{eqnarray}
    && \Pr\big(\,\bigcup_{i=1}^N \big\{ \big|\big(\, \log \hat{f}_X(X_i) + \log \hat{f}_U(Y_i) \,\big) - \big(\, \log f_X(X_i) + \log f_U(Y_i) \,\big) \big| > \varepsilon/2 \big\} \,\big) \\
    & \leq & N \cdot \Pr\big(\,  \big|\big(\, \log \hat{f}_X(X_i) + \log \hat{f}_U(Y_i) \,\big) - \big(\, \log f_X(X_i) + \log f_U(Y_i) \,\big) \big| > \varepsilon/2\,\big) \notag\\
    & = & N \int \underbrace{\Pr\big(\,\big| \big(\, \log \hat{f}_X(X_i) + \log \hat{f}_U(Y_i) \,\big) - \big(\, \log f_X(X_i) + \log f_U(Y_i) \,\big) \Big| > \varepsilon/2\big|Z_i = z\,\big)}_{\leq I_1(z) + I_2(z) + I_3(z) + I_4(z)} f(z) dz
\end{eqnarray}
where
\begin{eqnarray}
     && I_1(z) = \Pr\big(\,\rho_{k,i} > \log{N} (N f(x,y)c_{d_x+d_y})^{-\frac{1}{d_x+d_y}}\big|Z_i = z\,\big) \label{eq:i1}\\
     && I_2(z) = \Pr\big(\,\rho_{k,i} < \max\big\{\, (\log{N})^2 (N f_X(x) c_{d_x})^{-\frac{1}{d_x}}, (\log{N})^2 (N f_U(y) c_{d_y})^{-\frac{1}{d_y}} \,\}\big|Z_i = z\,\big) \label{eq:i2}\\
     && I_3(z) = \int_{r=r_2}^{r_1} \Pr\big(\,\big|\log \hat{f}_X(X_i) - \log f_X(X_i) \big| > \varepsilon/4 \Big| \rho_{k,i} = r, Z_i = z \,\big) f_{\rho_{k,i}}(r) dr \label{eq:i3}\\
     && I_4(z) = \int_{r=r_2}^{r_1} \Pr\big(\,\big|\log \hat{f}_U(Y_i) - \log f_U(Y_i) \big| > \varepsilon/4 \Big| \rho_{k,i} = r, Z_i = z \,\big) f_{\rho_{k,i}}(r) dr \label{eq:i4}
\end{eqnarray}
where $f_{\rho_{k,i}}(r)$ is the pdf of $\rho_{k,i}$ given $Z_i = z$. Here $r_1 = \log{N} (N f(z)c_{d_x+d_y})^{-\frac{1}{d_x+d_y}}$ and
\beqa
r_2 = \max\big\{\, (\log{N})^2 (N f_X(x) c_{d_x})^{-\frac{1}{d_x}}, (\log{N})^2 (N f_U(y) c_{d_y})^{-\frac{1}{d_y}} \,\}.
 \eeqa
 $I_1(z)$ and $I_2(z)$ are the probability that the $k$-NN distance $\rho_{k,i}$ is large or small given $Z_i = z$. $I_3(z)$ and $I_4(z)$ gives the probability that the estimator deviates from the true value, given that $\rho_{k,i}$ is medium. We will consider the four terms separately.
\\

\textit{$I_1(z)$:} Let $B_Z(z, r) = \{Z: \|Z-z\| < r\}$ be the $(d_x+d_y)$-dimensional ball centered at $z$ with radius $r$. Since the Hessian matrix of $H(f)$ exists and $\|H(f)\|_2 < C$ almost everywhere, then for sufficiently small $r$, the probability mass within $B_Z(z,r)$ is given by
\begin{eqnarray}
&& \Pr\big(\,u \in B_Z(z,r)\,\big) = \int_{\|u-z\| \leq r} f(u) du \notag \\
& = & \int_{\|u-z\| \leq r} f(z) + (u-z)^T \nabla f(z) + (u-z)^T H(f)(z) (u-z) + o(\|u-z\|^2) du \notag \\
& \in & \big[\, f(z)c_{d_x+d_y} r^{d_x+d_y}(1 - C r^2)), f(z)c_{d_x+d_y} r^{d_x+d_y}(1 + C r^2))\,\big] .
\end{eqnarray}
Then for sufficiently large $N$, the probability mass within $B_Z(z,r_1)$ is lower bounded by
\begin{eqnarray}
    p_1 & \equiv & \Pr\big(\,u \in B_Z(z,\log{N} (N f(z)c_{d_x+d_y})^{-\frac{1}{d_x+d_y}})\,\big) \notag \\
    & \geq & f(z) c_{d_x+d_y} \big(\, \log{N} (N f(z)c_{d_x+d_y})^{-\frac{1}{d_x+d_y}} \,\big)^{d_x+d_y} \big(\, 1-C (\log{N} (N f(z)c_{d_x+d_y})^{-\frac{1}{d_x+d_y}})^2 \,\big) \notag \\
    & \geq & \frac{(\log{N})^{d_x+d_y}}{2N} .
\end{eqnarray}
$I_1(z)$ is the probability that at most $k$ samples fall in $B_Z(z, r_1)$, so it is upper bounded by
\begin{eqnarray}
I_1(z) & = & \Pr\big(\,\rho_{k,i} > \log{N} (N f(z)c_{d_x+d_y})^{-\frac{1}{d_x+d_y}}\big| Z_i = z\,\big) \notag \\
& = &\sum_{m=0}^{k-1} {N-1 \choose m} p_1^m (1-p_1)^{N-1-m} \notag \\
&\leq & \sum_{m=0}^{k-1} N^m (1-p_1)^{N-1-m} \notag \\
&\leq & k N^{k-1} (1-\frac{(\log{N})^{d_x+d_y}}{2N})^{N-k-1} \notag \\
&\leq & k N^{k-1} \exp\{-\frac{(\log{N})^{d_x+d_y}(N-k-1)}{2N}\} \notag \\
&\leq & k N^{k-1}\exp\{-\frac{(\log{N})^{d_x+d_y}}{4}\} \label{eq:ub_i1}
\end{eqnarray}
for any $d_x, d_y \geq 1$.
\\

\textbf{$I_2$:} Let $r_{2,1} \equiv (\log{N})^2 (N f_X(x) c_{d_x})^{-\frac{1}{d_x}}$. Then for sufficiently large $N$, the probability mass within $B_Z(z,r_{2,1})$ is given by:
\begin{eqnarray}
    p_{2,1} & \equiv &\Pr\big(\,u \in B_Z(z,(\log{N})^2 (N f_X(x) c_{d_x})^{-\frac{1}{d_x}})\,\big) \notag \\
    & \leq & f(z) c_{d_x+d_y} \big(\, (\log{N})^2 (N f_X(x)c_{d_x})^{-\frac{1}{d_x}} \,\big)^{d_x+d_y} \big(\, 1+C (\log{N} (N f_X(x)c_{d_x})^{-\frac{1}{d_x}})^2 \,\big) \notag \\
    & \leq & \frac{2f(z) c_{d_x+d_y}}{(f(x)c_{d_x})^{\frac{d_x+d_y}{d_x}}} (\log{N})^{2(d_x+d_y)} N^{-\frac{d_x+d_y}{d_x}} \notag \\
    & \leq & 2f_{Y|X}(y|x) \frac{c_{d_x+d_y}}{c_{d_x}} (\log{N})^{2(d_x+d_y)} N^{-\frac{d_x+d_y}{d_x}} \notag \\
    & \leq & 2C' \frac{c_{d_x+d_y}}{c_{d_x}} (\log{N})^{2(d_x+d_y)} N^{-\frac{d_x+d_y}{d_x}}
\end{eqnarray}
where the last equation comes from the assumption that $f_{Y|X}(y|x) < C'$. Similarly, let $r_{2,2} = (\log{N})^2 (N f_U(y) c_{d_y})^{-\frac{1}{d_y}}$, the probability of being in $B_Z(z,r_2)$ is
\begin{eqnarray}
    p_{2,2} & \equiv &\Pr\big(\,u \in B_Z(z,(\log{N})^2 (N f_U(y) c_{d_y})^{-\frac{1}{d_y}})\,\big) \notag \\
    & \leq & f(z) c_{d_x+d_y} \big(\, (\log{N})^2 (N f_U(y)c_{d_x})^{-\frac{1}{d_x}} \,\big)^{d_x+d_y} \big(\, 1+C (\log{N} (N f_U(y)c_{d_x})^{-\frac{1}{d_y}})^2 \,\big) \notag \\
    & \leq & \frac{2f(z) c_{d_x+d_y}}{(f_U(y)c_{d_y})^{\frac{d_x+d_y}{d_y}}} (\log{N})^{2(d_x+d_y)} N^{-\frac{d_x+d_y}{d_y}} \notag \\
    & \leq & 2 \frac{f(z)}{f_U(y)} \frac{c_{d_x+d_y}}{c_{d_y}} (\log{N})^{2(d_x+d_y)} N^{-\frac{d_x+d_y}{d_y}} \notag \\
    &\leq & 2 C_2\frac{f_U(x,y)}{f_U(y)} \frac{c_{d_x+d_y}}{c_{d_y}} (\log{N})^{2(d_x+d_y)} N^{-\frac{d_x+d_y}{d_y}} \notag \\
    & \leq & 2C_2 C' \frac{c_{d_x+d_y}}{c_{d_y}} (\log{N})^{2(d_x+d_y)} N^{-\frac{d_x+d_y}{d_y}}.
\end{eqnarray}

$I_2(z)$ is the probability that at least $k$ samples lie in $B_Z(z, \max\{r_{2,1},r_{2,2}\})$. It is upper bounded as follows:
\begin{eqnarray}
I_2(z) &= & \Pr\big(\,\rho_{k,i} < \max\big\{\, (\log{N})^2 (N f_X(x) c_{d_x})^{-\frac{1}{d_x}}, (\log{N})^2 (N f_U(y) c_{d_y})^{-\frac{1}{d_y}} \,\}\big | Z_i = z\,\big) \notag \\
& = &\sum_{m=k}^{N-1} {N-1 \choose m} \max\{p_{2,1},p_{2,2}\}^m (1-\{p_{2,1},p_{2,2}\})^{N-1-m} \notag \\
&\leq &\sum_{m=k}^{N-1} N^m \max\{p_{2,1},p_{2,2}\}^m \notag \\
&\leq &\sum_{m=k}^{N-1} (\, 2C'C_2 \frac{c_{d_x+d_y}}{\min\{c_{d_x},c_{d_y}\}} (\log{N})^{2(d_x+d_y)} N^{- \min\{\frac{d_y}{d_x}, \frac{d_x}{d_y}\}} \,\big)^m \notag \\
&\leq &(4C'C_2 \frac{c_{d_x+d_y}}{\min\{c_{d_x}, c_{d_y}\}})^k (\log{N})^{2k(d_x+d_y)} N^{-k \min\{\frac{d_y}{d_x}, \frac{d_x}{d_y}\}} \label{eq:ub_i2}
\end{eqnarray}
for sufficiently large $N$ such that $2C' \frac{c_{d_x+d_y}}{\min\{c_{d_x},c_{d_y}\}} (\log{N})^{2(d_x+d_y)} N^{-\min\{\frac{d_x+d_y}{d_x}, \frac{d_x+d_y}{d_y}\}} < 1/2$, the last inequality comes from sum of geometric series. This holds for any $d_x, d_y \geq 1$ and $k \geq 1$.
\\

\textbf{$I_3$:} Given that $Z_i = z = (x,y)$ and $\rho_{k,i} = r$. Recall that $\hat{f}_X(X_i) = \frac{n_{x,i}}{(N-1)c_{d_x}\rho_{k,i}^{d_x}} $, so we have
\begin{eqnarray}
    && \Pr\big(\,\big|\log \hat{f}_X(X_i) - \log f_X(X_i)\big| > \varepsilon/4 \big| \rho_{k,i} = r, Z_i = z\,\big) \notag \\
    &=&\Pr\big(\,\big|\log n_{x,i} - \log(N-1) - \log c_{d_x} - d_x \log \rho_{k,i} - \log f_X(x)\big| > \varepsilon/4 \big| \rho_{k,i} = r, Z_i = z\,\big) \notag \\
    &=& \Pr\big(\,\big|\log n_{x,i} - \log(N-1)c_{d_x}r^{d_x}f_X(x)\big| > \varepsilon/4 \big| \rho_{k,i} = r, Z_i = z\,\big) \notag \\
    &=& \Pr\big(\, n_{x,i} > (N-1)c_{d_x}r^{d_x}f_X(x)e^{\varepsilon/4} \big| \rho_{k,i} = r, Z_i = z\,\big) \notag\\
    &+& \Pr\big(\, n_{x,i} < (N-1)c_{d_x}r^{d_x}f_X(x)e^{-\varepsilon/4} \big| \rho_{k,i} = r, Z_i = z\,\big).
\end{eqnarray}

Given $Z_i = z$ and $\rho_{k,i} = r \in [r_{2,1}, r_1]$, the probability distribution of $n_{x,i}$  is given in the following lemma:

\begin{lemma}~\label{lemma:n_xi}
Given $Z_i = z = (x,y)$ and $\rho_{k,i} = r < r_N$ for some deterministic sequence of $r_N$ such that $\lim_{N \to \infty} r_N = 0$ and
for any positive $\varepsilon>0$, the number of neighbors    $n_{x,i}-k$ is distributed as $\sum_{l=k+1}^{N-1} U_l$, where $U_l$ are i.i.d Bernoulli random variables with mean $f_X(x) c_{d_x} r^{d_x} (1-\varepsilon/8) \leq \E[U_l] \leq f_X(x) c_{d_x} r^{d_x} (1+\varepsilon/8)$ for sufficiently large $N$.
\end{lemma}

Given lemma.~\ref{lemma:n_xi}, we obtain
\begin{eqnarray}
    &&\Pr\big(\,n_{x,i} > (N-1)c_{d_x}r^{d_x}f_X(x)e^{\varepsilon/4} \Big| \rho_{k,i} = r, Z_i = z\,\big) \notag\\
    &=& \Pr\big(\, \sum_{l=k+1}^{N-1} U_l > (N-1)c_{d_x}r^{d_x}f_X(x)e^{\varepsilon/4} - k \big) \notag\\
    &=& \Pr\big(\, \sum_{l=k+1}^{N-1} U_l - (N-k-1)\E[U_l] > (N-1)c_{d_x}r^{d_x}f_X(x)e^{\varepsilon/4} - k - (N-k-1)\E[U_l]\big) \label{eq:lb_nxi}\;,
\end{eqnarray}
and the right hand side in the probability is lower bounded by
\begin{eqnarray}
&& (N-1)c_{d_x}r^{d_x}f_X(x)e^{\varepsilon/4} - k -\E[U_l] \,\notag\\
&\geq& (N-1)c_{d_x}r^{d_x}f_X(x)e^{\varepsilon/4} - k - (N-k-1)f_X(x) c_{d_x} r^{d_x} (1+\varepsilon/8) \,\notag\\
&\geq& (N-k-1)c_{d_x}r^{d_x}f_X(x)(e^{\varepsilon/4}-1-\varepsilon/8) - k \,\notag\\
&\geq& (N-k-1)c_{d_x}r^{d_x}f_X(x)\varepsilon/16
\end{eqnarray}
for sufficiently large $N$ such that $ (N-k-1)c_{d_x}r^{d_x}f_X(x)(e^{\varepsilon/4}-1-\varepsilon/16) > k$. Since $U_l$ is bernoulli, we have $\E[U_l^2] = \E[U_l]$. Now applying Bernstein's inequality,~\eqref{eq:lb_nxi} is upper bounded by:
\begin{eqnarray}
&& \Pr\big(\, \sum_{l=k+1}^{N-1} U_l - (N-k-1)\E[U_l] > (N-1)c_{d_x}r^{d_x}f_X(x)e^{\varepsilon} - k - (N-k-1)\E[U_l]\big) \,\notag\\
&\leq& \Pr\big(\, \sum_{l=k+1}^{N-1} U_l - (N-k-1)\E[U_l] > (N-k-1)c_{d_x}r^{d_x}f_X(x)\varepsilon/16\big) \,\notag\\
&\leq& \exp\{-\frac{((N-k-1)c_{d_x}r^{d_x}f_X(x)\varepsilon/16)^2}{2\big(\, (N-k-1) \E[U_l^2] + \frac{1}{3} ((N-k-1)c_{d_x}r^{d_x}f_X(x)\varepsilon/16)\,\big)}\} \,\notag\\
&\leq& \exp\{-\frac{((N-k-1)c_{d_x}r^{d_x}f_X(x)\varepsilon/16)^2}{2\big(\, (N-k-1) c_{d_x}r^{d_x}f_X(x)(1+\varepsilon/8) + \frac{1}{3} ((N-k-1)c_{d_x}r^{d_x}f_X(x)\varepsilon/16)\,\big)}\} \,\notag\\
&=& \exp\{-\frac{\varepsilon^2}{512(1+7\varepsilon/48)}(N-k-1) c_{d_x}r^{d_x}f_X(x)\} \;.
\end{eqnarray}

Similarly, the tail bound on the other direction is given by:
\begin{eqnarray}
    &&\Pr\big(\,n_{x,i} < (N-1)c_{d_x}r^{d_x}f_X(x)e^{-\varepsilon/4} \Big| \rho_{k,i} = r, Z_i = z\,\big) \notag\\
    &=& \Pr\big(\, \sum_{l=k+1}^{N-1} U_l < (N-1)c_{d_x}r^{d_x}f_X(x)e^{-\varepsilon/4} - k \big) \notag\\
    &=& \Pr\big(\, \sum_{l=k+1}^{N-1} U_l - (N-k-1)\E[U_l] < (N-1)c_{d_x}r^{d_x}f_X(x)e^{-\varepsilon/4} - k - (N-k-1)\E[U_l]\big) \label{eq:ub_nxi}\;,
\end{eqnarray}
where the right hand side is negative and upper bounded by:
\begin{eqnarray}
&& (N-1)c_{d_x}r^{d_x}f_X(x)e^{-\varepsilon/4} - k -\E[U_l] \,\notag\\
&\leq& (N-1)c_{d_x}r^{d_x}f_X(x)e^{-\varepsilon/4} -k - (N-k-1)f_X(x) c_{d_x} r^{d_x} (1-\varepsilon/8) \,\notag\\
&\leq& (N-k-1)c_{d_x}r^{d_x}f_X(x)(e^{-\varepsilon/4}-1+\varepsilon/8)  \,\notag\\
&\leq& -(N-k-1)c_{d_x}r^{d_x}f_X(x)\varepsilon/16
\end{eqnarray}
for small enough $r$ such that $c_{d_x}r^{d_x}f_X(x)e^{-\varepsilon/4} \leq 1$ and small enough $\varepsilon$ that $e^{-\varepsilon/4}-1+3\varepsilon/16 < 0$. Similarly,~\eqref{eq:ub_nxi} is upper bounded by:
\begin{eqnarray}
&&\Pr\big(\, \sum_{l=k+1}^{N-1} U_l - (N-k-1)\E[U_l] < (N-1)c_{d_x}r^{d_x}f_X(x)e^{-\varepsilon/4} - k - (N-k-1)\E[U_l]\big) \,\notag\\
&\leq& \exp\{-\frac{\varepsilon^2}{512(1+7\varepsilon/48)}(N-k-1) c_{d_x}r^{d_x}f_X(x)\}.
\end{eqnarray}

Therefore, $I_3(z)$ is upper bounded by:
\begin{eqnarray}
    I_3(z) &=& \int_{r=r_2}^{r_1} \Pr\big(\,\big|\log \hat{f}_X(X_i) - \log f_X(X_i)\big| > \varepsilon\big| \rho_{k,i} = r, Z_i = z\,\big) f_{\rho_{k,i}}(r) dr \notag \\
    &\leq& \int_{r=(\log{N})^2 (N f_X(x) c_{d_x})^{-\frac{1}{d_x}}}^{\log{N} (N f(z)c_{d_x+d_y})^{-\frac{1}{d_x+d_y}}} \Pr\big(\,\big|\log \hat{f}_X(X_i) - \log f_X(X_i)\big| > \varepsilon\big| \rho_{k,i} = r, Z_i = z\,\big) f_{\rho_{k,i}}(r) dr \notag \\
    &\leq& \int_{r=(\log{N})^2 (N f_X(x) c_{d_x})^{-\frac{1}{d_x}}}^{\log{N} (N f(z)c_{d_x+d_y})^{-\frac{1}{d_x+d_y}}} 2\exp\{-\frac{\varepsilon^2}{512(1+7\varepsilon/48)}(N-k-1) c_{d_x}r^{d_x}f_X(x)\} f_{\rho_{k,i}}(r) dr \notag \\
    &\leq& 2\exp\{-\frac{\varepsilon^2}{1024}N c_{d_x}f_X(x)((\log{N})^2 (N f_X(x) c_{d_x})^{-\frac{1}{d_x}})^{d_x}\}  \notag \\
    &\leq& 2\exp\{-\frac{\varepsilon^2}{1024} (\log{N})^{2d_x}\} \label{eq:ub_i3}
\end{eqnarray}
for sufficiently large $N$ such that $(N-k-1)/(1+\frac{7}{48}\varepsilon) > N/2$.
\\

\textbf{$I_4$:} Given that $Z_i = z = (x,y)$ and $\rho_{k,i} = r$. Recall that $\hat{f}_U(Y_i) = \frac{\tn_{y,i}}{(N-1)c_{d_y}r^{d_y}}$, then we have
\begin{eqnarray}
    && \Pr\big(\,\big|\log \hat{f}_U(Y_i) - \log f_U(Y_i)\big| > \varepsilon/4\big| \rho_{k,i} = r, Z_i = z\,\big) \notag \\
    &=&\Pr\big(\,\big|\log \tn_{y,i} - \log(N-1) - \log c_{d_y} - d_y \log \rho_{k,i} - \log f_U(y)\big| > \varepsilon/4\big| \rho_{k,i} = r, Z_i = z\,\big) \notag \\
    &=& \Pr\big(\,\big|\log \tn_{y,i} - \log(N-1)c_{d_y}r^{d_y}f_Y(y)\big| > \varepsilon/4\big| \rho_{k,i} = r, Z_i = z\,\big) \notag \\
    &=& \Pr\big(\, \tn_{y,i} > (N-1)c_{d_y}r^{d_y}f_U(y)e^{\varepsilon/4}\big| \rho_{k,i} = r, Z_i = z\,\big) \notag\\
    &+& \Pr\big(\, \tn_{y,i} < (N-1)c_{d_y}r^{d_y}f_U(y)e^{-\varepsilon/4}\big| \rho_{k,i} = r, Z_i = z\,\big).
\end{eqnarray}
Recall that
\begin{eqnarray}
\tn_{y,i} = \sum_{j \neq i} w'_j \mathbb{I}\{\|Y_j - Y_i\| < \rho_{k,i}\} = \sum_{j \neq i} \frac{u(X_i)}{f_X(X_i)}\mathbb{I}\{\|Y_j - Y_i\| < \rho_{k,i}\}.
\end{eqnarray}
We write $\tn_{y,i} = m^{(1)}_{y,i} +m^{(2)}_{y,i}$, where \begin{eqnarray}
m^{(1)}_{y,i} &=& \sum_{j: \|Z_j - z\| < \rho_{k,i}} \frac{u(X_j)}{f_X(X_j)} \,\notag \\
m^{(2)}_{y,i} &=& \sum_{j: \|Z_j - z\| > \rho_{k,i}} \frac{u(X_j)}{f_X(X_j)} \mathbb{I}\{\|Y_j - Y_i\| < \rho_{k,i}\}.
\end{eqnarray}
Since $C_1/\mu(K) \leq f_X(X_j) \leq C_2/\mu(K)$, we have: $k/C_2 \leq m^{(1)}_{y,i} \leq k/C_1$. Given that $Z_i = z$ and $\rho_{k,i} = r \in [r_{2,2},r_1]$, the probability distribution of $m^{(2)}_{y,i}$ is given by the following lemma:

\begin{lemma}~\label{lemma:n_yi}
Given $Z_i = z = (x,y)$ and $\rho_{k,i} = r < r_N$ for some deterministic sequence of $r_N$ such that $\lim_{N \to \infty} r_N = 0$ and for a positive $\varepsilon>0$,  the distribution of $m^{(2)}_{y,i}$ is distributed  as $\sum_{l=k+1}^{N-1} V_l$ where $V_l$ are i.i.d random variables with $V_l \in [0,1/C_1]$ and mean $f_U(y) c_{d_y} r^{d_y} (1-\varepsilon/8) \leq \E[V_l] \leq f_U(y) c_{d_y} r^{d_y} (1+\varepsilon/8)$, for sufficiently large $N$.
\end{lemma}

Given Lemma~\ref{lemma:n_yi} and the fact that $m^{(1)}_{y,i} \geq k/C_2$, we obtain
\begin{eqnarray}
    &&\Pr\big(\,\tn_{y,i} > (N-1)c_{d_y}r^{d_y}f_U(y)e^{\varepsilon/4} \Big| \rho_{k,i} = r, Z_i = z\,\big) \notag\\
    &\leq& \Pr\big(\, \sum_{l=k+1}^{N-1} V_l > (N-1)c_{d_y}r^{d_y}f_U(y)e^{\varepsilon/4} - k/C_2 \big) \notag\\
    &=& \Pr\big(\, \sum_{l=k+1}^{N-1} V_l - (N-k-1)\E[V_l] > (N-1)c_{d_y}r^{d_y}f_U(y)e^{\varepsilon/4} - k/C_2 - (N-k-1)\E[V_l] \big); \label{eq:lb_nyi}
\end{eqnarray}
here the right hand side is lower bounded by
\begin{eqnarray}
&& (N-1)c_{d_y}r^{d_y}f_U(y)e^{\varepsilon/4} - k/C_2 -\E[V_l] \,\notag\\
&\geq& (N-1)c_{d_y}r^{d_y}f_U(y)e^{\varepsilon/4} - k/C_2 - (N-k-1)c_{d_y}r^{d_y}f_U(y) (1+\varepsilon/8) \,\notag\\
&\geq& (N-k-1)c_{d_y}r^{d_y}f_U(y)(e^{\varepsilon/4}-1-\varepsilon/8) - k/C_2 \,\notag\\
&\geq& (N-k-1)c_{d_y}r^{d_y}f_U(y)\varepsilon/16
\end{eqnarray}
for sufficiently large $N$ such that $ (N-k-1)c_{d_y}r^{d_y}f_U(y)(e^{\varepsilon/4}-1-\varepsilon/16) > k/C_2$. Since $V_l$ is upper bounded by $1/C_1$, so $\E[V_l^2] \leq \E[V_l]/C_1$. Now applying Bernstein's inequality,~\eqref{eq:lb_nyi} is upper bounded by:
\begin{eqnarray}
&& Pr\big(\, \sum_{l=k+1}^{N-1} V_l - (N-k-1)\E[V_l] > (N-1)c_{d_y}r^{d_y}f_U(y)e^{\varepsilon} - k - (N-k-1)\E[V_l]\big) \,\notag\\
&\leq& \Pr\big(\, \sum_{l=k+1}^{N-1} V_l - (N-k-1)\E[V_l] > (N-k-1)c_{d_y}r^{d_y}f_U(y)\varepsilon/16\big) \,\notag\\
&\leq& \exp\{-\frac{((N-k-1)c_{d_y}r^{d_y}f_U(y)\varepsilon/16)^2}{2\big(\, (N-k-1) \E[V_l^2] + \frac{1}{3C_1} ((N-k-1)c_{d_y}r^{d_y}f_U(y)\varepsilon/8)\,\big)}\} \,\notag\\
&\leq& \exp\{-\frac{((N-k-1)c_{d_y}r^{d_y}f_U(y)\varepsilon/16)^2}{2\big(\, (N-k-1) c_{d_y}r^{d_y}f_U(y)(1+\varepsilon/8)/C_1 + \frac{1}{3C_1} ((N-k-1)c_{d_y}r^{d_y}f_U(y)\varepsilon/16)\,\big)}\} \,\notag\\
&\leq& \exp\{-\frac{C_1\varepsilon^2}{512(1+7\varepsilon/48)}(N-k-1) c_{d_y}r^{d_y}f_U(y)\} \;.
\end{eqnarray}

Similarly, since $m^{(1)}_{y,i} < k/C_1$, the tail bound on the other way is given by:
\begin{eqnarray}
    &&\Pr\big(\,\tn_{y,i} < (N-1)c_{d_y}r^{d_y}f_U(y)e^{-\varepsilon/4} \Big| \rho_{k,i} = r, Z_i = z\,\big) \notag\\
    &\leq& \Pr\big(\, \sum_{l=k+1}^{N-1} V_l < (N-1)c_{d_y}r^{d_y}f_U(y)e^{-\varepsilon/4} - k/C_1 \big) \notag\\
    &=& \Pr\big(\, \sum_{l=k+1}^{N-1} V_l - (N-k-1)\E[V_l] < (N-1)c_{d_y}r^{d_y}f_U(y)e^{-\varepsilon/4} - k/C_1 - (N-k-1)\E[V_l]\big) \label{eq:ub_nyi}\;,
\end{eqnarray}
where the right hand side is negative and upper bounded by:
\begin{eqnarray}
&& (N-1)c_{d_y}r^{d_y}f_U(y)e^{-\varepsilon/4} - k/C_1 -\E[V_l] \,\notag\\
&\leq& (N-1)c_{d_y}r^{d_y}f_U(y)e^{-\varepsilon/4} - k/C_1 - (N-k-1)c_{d_y}r^{d_y}f_U(y) (1-\varepsilon/8) \,\notag\\
&\leq& (N-k-1)c_{d_y}r^{d_y}f_U(y)(e^{-\varepsilon/4}-1+\varepsilon/8)  \,\notag\\
&\leq& -(N-k-1)c_{d_y}r^{d_y}f_U(y)\varepsilon/16
\end{eqnarray}
for small enough $r$ such that $c_{d_y}r^{d_y}f_U(y)e^{-\varepsilon/4} \leq 1/C_1$ and small enough $\varepsilon$ that $e^{-\varepsilon/4}-1+3\varepsilon/16 < 0$. Similarly,~\eqref{eq:ub_nyi} is upper bounded by:
\begin{eqnarray}
&&\Pr\big(\, \sum_{l=k+1}^{N-1} V_l - (N-k-1)\E[V_l] < (N-1)c_{d_y}r^{d_y}f_U(y)e^{-\varepsilon/4} - k - (N-k-1)\E[V_l]\big) \,\notag\\
&\leq& \exp\{-\frac{C_1\varepsilon^2}{512(1+7\varepsilon/48)}(N-k-1) c_{d_y}r^{d_y}f_U(y)\}.
\end{eqnarray}
Therefore, $I_4(z)$ is upper bounded by:
\begin{eqnarray}
    I_4(z)&=& \int_{r=r_2}^{r_1} \Pr\big(\,\big|\log \hat{f}_U(Y_i) - \log f_U(Y_i)\big| > \varepsilon\big| \rho_{k,i} = r, Z_i = z\,\big) f_{\rho_{k,i}}(r) dr \notag \\
    &\leq& \int_{r=(\log{N})^2 (N f_U(y) c_{d_y})^{-\frac{1}{d_y}}}^{\log{N} (N f(z)c_{d_x+d_y})^{-\frac{1}{d_x+d_y}}} \Pr\big(\,\big|\log \hat{f}_U(Y_i) - \log f_U(Y_i)\big| > \varepsilon\big| \rho_{k,i} = r, Z_i = z\,\big) f_{\rho_{k,i}}(r) dr \notag \\
    &\leq& \int_{r=(\log{N})^2 (N f_U(y) c_{d_y})^{-\frac{1}{d_y}}}^{\log{N} (N f(z)c_{d_x+d_y})^{-\frac{1}{d_x+d_y}}} \exp\{-\frac{C_1 \varepsilon^2}{512(1+7\varepsilon/48)}(N-k-1) c_{d_y}r^{d_y}f_U(y)\} f_{\rho_{k,i}}(r) dr \notag \\
    &\leq& 2\exp\{- \frac{C_1 \varepsilon^2}{1024}N c_{d_y}f_U(y)((\log{N})^2 (N f_U(y) c_{d_y})^{-\frac{1}{d_y}})^{d_y}\}  \notag \\
    &\leq& 2\exp\{-\frac{C_1 \varepsilon^2}{1024} (\log{N})^{2d_y}\} \label{eq:ub_i4}
\end{eqnarray}
for sufficiently large $N$ such that $(N-k-1)/(1+7\varepsilon/48)) > N/2$.
\\

Now combining \eqref{eq:ub_i1}, \eqref{eq:ub_i2}, \eqref{eq:ub_i3} and \eqref{eq:ub_i4}, we obtain
\begin{eqnarray}
    && \Pr\big(\,\sum_{i=1}^N \frac{w'_i}{N} \big|\big(\, \log \hat{f}_X(X_i) + \log \hat{f}_U(Y_i) \,\big) - \big(\, \log f_X(X_i) + \log f_U(Y_i) \,\big) \big| > \varepsilon\,\big) \notag\\
    & \leq & N \iint (I_1(z) + I_2(z) + I_3(z) + I_4(z)) f(z) dz \notag\\
    & \leq & k N^{k}\exp\{-\frac{(\log{N})^{d_x+d_y}}{4}\} + (4C'C_2 \frac{c_{d_x+d_y}}{\min\{c_{d_x},c_{d_y}\}})^k (\log{N})^{2k(d_x+d_y)} N^{1-k \min\{\frac{d_y}{d_x}, \frac{d_y}{d_y}\}} \,\notag\\
    &+& 2N\exp\{-\frac{\varepsilon^2}{1024} (\log{N})^{2d_x}\} + 2N\exp\{-\frac{C_1^2 \varepsilon^2}{1024} (\log{N})^{2d_y}\}.
\end{eqnarray}
If $k > \max\{d_y/d_x,d_x/d_y\}$, we have $1 - k\min\{\frac{d_x+d_y}{d_x}, \frac{d_x+d_y}{d_y}\} < 0$. Then each of the four terms goes to 0 as $N \to \infty$ and we conclude:
\begin{eqnarray}
\lim_{N \to \infty} \Pr\big(\,\sum_{i=1}^N \frac{w'_i}{N} \big|\big(\, \log \hat{f}_X(X_i) + \log \hat{f}_U(Y_i) \,\big) - \big(\, \log f_X(X_i) + \log f_U(Y_i) \,\big) \big| > \varepsilon\,\big) = 0.
\end{eqnarray}
Therefore, by combining the convergence properties  of error from kernel density estimation, error from self-normalized importance sampling and error from density estimation, we obtain that $\hat{I}^U_{k,N}(X,Y)$ converges to $I^U(f_{Y|X})$ in probability.

\subsubsection{Proof of Lemma~\ref{lemma:n_xi}}
Given that $Z_i = z = (x,y)$ and $\rho_{k,i} = r $, let $\{1, 2, \dots, i-1, i+1, \dots, N\} = S \cup \{j\} \cup T$ be a partition of the indexes with $\big|S\big| = k-1$ and $\big|T\big| = N-k-1$.  Then define an event $\mathcal{A}_{S, j, T}$ associated to the partition as:
\begin{eqnarray}
    \mathcal{A}_{S, j, T} = \big\{\, \|Z_s - z\| < \|Z_j - z\|, \forall s \in S,  \textrm{ and } \|Z_t - z\| > \|Z_j - z\|,\forall t \in T \,\big\}.
\end{eqnarray}
Since $Z_j - z$ are i.i.d. random variables, each event $\mathcal{A}_{S, j, T}$ has identical probability. The number of such partitions is $\frac{(N-1)!}{(N-k-1)!(k-1)!}$, and thus $\Pr\big(\,\mathcal{A}_{S, j, T}\,\big) = \frac{(N-k-1)!(k-1)!}{(N-1)!}$. So the cdf of $n_{x, i}$ is given by:
\begin{eqnarray}
    &&\Pr\big(\,n_{x,i} \leq k+m\big|\rho_{k,i} = r, Z_i = z\,\big) \notag \\
    &= &\sum_{S,j,T} \Pr\big(\,\mathcal{A}_{S, j, T}\,\big) \Pr\big(\,n_{x,i} \leq k+m\big|\mathcal{A}_{S, j, T}, \rho_{k,i} = r, Z_i = z\,\big) \notag \\
    &= &\frac{(N-k-1)!(k-1)!}{(N-1)!} \sum_{S,j,T} \Pr\big(\,n_{x,i} \leq k+m\big|\mathcal{A}_{S, j, T}, \rho_{k,i} = r, Z_i = z\,\big).
\end{eqnarray}

Now condition on event $\mathcal{A}_{S,j,T}$ and $\rho_{k,i} = r$, namely $Z_j$ is the $k$-nearest neighbor with distance $r$, $S$ is the set of samples with distance smaller than $r$ and $T$ is the set of samples with distance greater than $r$. Recall that $n_{x,i}$ is the number of samples with $\|X_j-x\| < r$. For any index $s \in S \cup \{j\}$, $\|X_s-x\| < r$ is satisfied. Therefore, $n_{x,i} \leq k+m$ means that there are no more than $m$ samples in $T$ with $X$-distance smaller than $r$. Let $U_l = \mathbb{I}\{\|X_l - x\| < r \big| \|Z_l - z\| > r\}. $Therefore,
\begin{eqnarray}
    &&\Pr\big(\,n_{x,i} \leq k+m\big|\mathcal{A}_{S, j, T}, \rho_{k,i} = r, Z_i = z\,\big) \notag \\
    &=& \Pr\big(\,\sum_{t \in T} \mathbb{I}\{\|X_t - x\| < r\} \leq m\big|~\|Z_s - z\| < r, \forall s \in S, \|Z_j - z\| = r, \|Z_t - z\| > r,\forall t \in T, Z_i = z\,\big) \notag \\
    &=& \Pr\big(\,\sum_{t \in T} \mathbb{I}\{\|X_t - x\| < r\} \leq m\big|~ \|Z_t - z\| > r,\forall t \in T\,\big) \notag \\
    &=& \Pr \big(\, \sum_{l=k+1}^{N-1} U_l \leq m\,\big).
\end{eqnarray}

We can drop the conditions of $Z_s$'s for $s \not\in T$ since $Z_s$ and $X_t$ are independent. Therefore, given that $\|Z_t-z\| > r$ for all $t \in T$, the variables $\mathbb{I}\{\|X_t - x\| < r\}$ are i.i.d.\ and have the same distribution as $U_l$. Therefore, we have:
\begin{eqnarray}
    && \Pr\big(\,n_{x,i} \leq k+m\big|\rho_{k,i} = r, Z_i = z\,\big) \notag \\
    &=& \frac{(N-k-1)!(k-1)!}{(N-1)!} \sum_{S,j,T} \Pr\big(\,n_{x,i} \leq k+m\big|\mathcal{A}_{S, j, T}, \rho_{k,i} = r, Z_i = z\,\big) \notag \\
    &=& \frac{(N-k-1)!(k-1)!}{(N-1)!} \sum_{S,j,T}  \Pr \big(\, \sum_{l=k+1}^N U_l \leq m\,\big) \notag \\
    &=&  \Pr \big(\, \sum_{l=k+1}^{N-1} U_l \leq m\,\big)
\end{eqnarray}
and so $n_{x,i}-k$ have the same distribution as $\sum_{l=k+1}^{N-1} U_l$ given $Z_i = z$ and $\rho_{k,i} = r$. Here the mean of $U_l$ is given by:
\begin{eqnarray}
    \E [U_l] &=& \Pr\big(\,\|X_l - x\| < r\big|~\|Z_l - z\| > r\,\big) = \frac{\Pr\big(\,\|X_l - x\| < r, \|Z_l - z\| > r\,\big)}{\Pr\big(\,\|Z_l - z\| > r\,\big)} \notag \\
    &=& \frac{\int_{\|u-x\|<r} f_X(u) du - \iint_{\|(u,v)-(x,y)\| , r} f(u,v) dudv}{ 1 - \iint_{\|(u,v)-(x,y)\| , r} f(u,v) dudv}  .
\end{eqnarray}
Since $\|H(f_X)\| \leq C$ almost everywhere, if $r < r_N$ and $r_N$ decays as $N$ goes to infinity, for sufficiently large $N$, we have the following bound for $\E [U_l]$:
\begin{eqnarray}
    \E [U_l] &<& \int_{\|u-x\|<r} f_X(u) du \notag \\
    &=& \int_{\|u-x\| < r} f_X(x) + (u-x) \nabla f_X(x) + (u-x)^T H(f_X)(x) (u-x) + o(\|u-x\|^2) \notag \\
    &<& f_X(x) c_{d_x} r^{d_x} (1+Cr^2) \notag \\
    &<& f_X(x) c_{d_x} r^{d_x} (1+\varepsilon/8)\;,
\end{eqnarray}
and
\begin{eqnarray}
    \E [U_l] &>& \int_{\|u-x\|<r} f_X(u) du - \int_{\|(u,v)-(x,y)\| , r} f(u,v) dudv \notag\\
    &>& f_X(x) c_{d_x} r^{d_x} (1-Cr^2) - f(x,y) c_{d_x+d_y} r^{d_x+d_y} (1+Cr^2) \notag \\
    &>& f_X(x) c_{d_x} r^{d_x} (1-\varepsilon/8).
\end{eqnarray}

\subsubsection{Proof of Lemma~\ref{lemma:n_yi}}

Given that $Z_i = z = (x,y)$ and $\rho_{k,i} = r $. Define $\mathcal{A}_{S, j, T}$ as same as in Lemma ~\ref{lemma:n_xi}. Let $V_l = \frac{u(X_l)}{f_X(X_l)} \mathbb{I}\{\|Y_l-y\| < r \big| \|Z_l - z\| > r\}$. Condition on event $\mathcal{A}_{S,j,T}$, the cdf of $m^{(2)}_{y,i}$ is given by:,
\begin{eqnarray}
    &&\Pr\big(\,m^{(2)}_{y,i} \leq m\big|\mathcal{A}_{S, j, T}, \rho_{k,i} = r, Z_i = z\,\big) \notag \\
    &=& \Pr\big(\,\sum_{t \in T} \frac{u(X_t)}{f_X(X_t)} \mathbb{I}\{\|Y_t - y\| < \rho_{k,i}\} \leq m \big|~\|Z_s - z\| < r, \forall s \in S, \|Z_j - z\| = r, \|Z_t - z\| > r,\forall t \in T, Z_i = z\,\big) \notag \\
    &=& \Pr\big(\,\sum_{t \in T} \frac{u(X_t)}{f_X(X_t)} \mathbb{I}\{\|Y_t - y\| < \rho_{k,i}\} \leq m \big|~ \|Z_t - z\| > r,\forall t \in T\,\big) \notag \\
    &=& \Pr \big(\, \sum_{l=k+1}^{N-1} V_l \leq m\,\big).
\end{eqnarray}

Similarly we can drop the conditions of $Z_s$'s for $s \not\in T$. Therefore, given that $\|Z_t-z\| > r$ for all $t \in T$, the variables $\frac{u(X_t)}{f_X(X_t)}\mathbb{I}\{\|Y_t - y\| < r\}$ are i.i.d. and have the same distribution as $V_l$. Therefore, we have:
\begin{eqnarray}
    && \Pr\big(\,m^{(2)}_{y,i} \leq m\big|\rho_{k,i} = r, Z_i = z\,\big) \notag \\
    &=& \frac{(N-k-1)!(k-1)!}{(N-1)!} \sum_{S,j,T} \Pr\big(\,m^{(2)}_{y,i} \leq m\big|\mathcal{A}_{S, j, T}, \rho_{k,i} = r, Z_i = z\,\big) \notag \notag \\
    &=& \frac{(N-k-1)!(k-1)!}{(N-1)!} \sum_{S,j,T}\Pr \big(\, \sum_{l=k+1}^{N-1} V_l \leq m\,\big) \notag \\
    &=&  \Pr \big(\, \sum_{l=k+1}^{N-1} V_l \leq m\,\big)
\end{eqnarray}

so $m^{(2)}_{y,i}$ have the same distribution as $\sum_{l=k+1}^{N-1} V_l$ given $Z_i = z$ and $\rho_{k,i} = r$. Here $V_l \leq \sup_{x} \frac{u(x)}{f_X(x)} = 1/C_1$. The mean of $V_l$ is given by:
\begin{eqnarray}
    \E [V_l] &=& \E \big[\, \frac{u(X_l)}{f_X(X_l)} \mathbb{I}\{\|Y_l - y\| < r\} \big|~\|Z_l - z\| > r\,\big] \notag \\
    &=& \frac{\iint_{\|v-y\|<r} \frac{u(u)}{f_X(u)} f(u,v) du dv - \iint_{\|(u,v)-(x,y)\| < r} \frac{u(u)}{f_X(u)} f(u,v) dudv}{ 1 - \iint_{\|(u,v)-(x,y)\| , r} \frac{u(u)}{f_X(u)} f(u,v) dudv}  .
\end{eqnarray}
Since $\|H(f_U)\| \leq C$ almost everywhere, if $r < r_N$ and $r_N$ decays as $N$ goes to infinity, for sufficiently large $N$, we have the following bound for $\E [V_l]$:
\begin{eqnarray}
    \E [V_l] &<& \iint_{\|v-y\|<r} \frac{u(u)}{f_X(u)} f(u,v) du dv \notag \\
    &=& \int_{\|v-y\| < r} f_U(v) dy \notag \\
    &=& \int_{\|v-y\| < r} f_U(y) + (v-y) \nabla f_U(y) + (v-y)^T H(f_U)(y) (v-y) + o(\|v-y\|^2) \notag \\
    &<& f_U(y) c_{d_y} r^{d_y} (1+Cr^2) \notag \\
    &<& f_U(y) c_{d_y} r^{d_y} (1+\varepsilon/8)
\end{eqnarray}
and
\begin{eqnarray}
    \E [V_l] &>& \iint_{\|v-y\|<r} \frac{u(u)}{f_X(u)} f(u,v) du dv - \iint_{\|(u,v)-(x,y)\| , r} \frac{u(u)}{f_X(u)} f(u,v) dudv \notag\\
    &=& \int_{\|v-y\| < r} f_U(v) dy- \iint_{\|(u,v)-(x,y)\| , r} f_U(u,v) dudv \notag\\
    &>& f_U(y) c_{d_y} r^{d_y} (1-Cr^2) - f_U(x,y) c_{d_x+d_y} r^{d_x+d_y} (1+Cr^2) \notag \\
    &>& f_U(y) c_{d_y} r^{d_y} (1-\varepsilon/8).
\end{eqnarray}

\subsection{The case of discrete $\mathcal{X}$}

Under Assumption \ref{assumption:umi},
we prove a more general version of the theorem. Let $(X_1,Y_1), \dots ,(X_n,Y_n)$ be i.i.d. samples drawn from some unknown prior $p_X(x)$ anb let $q_X(x)$ be some known distribution on $\mathcal{X}$ such that $q_X(x)/p_X(x) \in [C_3, C_4]$ for all $x \in \mathcal{X}$. Then define
\begin{eqnarray}
w^{(q)}_{x} &\equiv& \frac{N q_X(x)}{n_x} \,,\\
n^{(q)}_{y,i} &\equiv& \sum_{j \neq i} w^{(q)}_{X_j} \mathbb{I} \{\|Y_j - Y_i\| < \rho_{k, i}\}.
\end{eqnarray}
The proposed estimator is:
\begin{eqnarray}
    \hI^{(q)}_{k,N} (X,Y) &\equiv & \frac1N \sum_{i=1}^N w^{(q)}_{X_i} \,\Big( \psi(k) + \log (N) - \big(\, \log(n_{X_i}) + \log (n^{(q)}_{y,i}) \,\big) \,\Big) .
    \label{def:qumi}
\end{eqnarray}
We claim that $\hI^{(q)}_{k,N}$ converges to the true value in probability, i.e.
\begin{eqnarray}
\lim_{N\to\infty} \Pr\big(\, \big|\hI^{(q)}_{k,N} (X,Y)-I^{(q)}(f_{Y|X})\big|>\varepsilon\,\big) = 0\,,
\end{eqnarray}
where
\begin{eqnarray}
I^{(q)}(f_{Y|X}) &\equiv& \sum_{x \in \mathcal{X}} q_X(x) \int f_{Y|X}(y|x) \log \frac{f_{Y|X}(y|x)}{f_q(y)} dy
\end{eqnarray}
and
\begin{eqnarray}
f_q(y) &\equiv& \sum_{x' \in \mathcal{X}} q_x(X) f_{Y|X}(y|x').
\end{eqnarray}
Notice that Theorem~1 is a special case when $q_X(x)$ is uniform.
Define
\begin{eqnarray}
g(X_i, Y_i) &\equiv& \psi(k) + \log(N) - \big( \log(n_{x,i}) + \log( n^{(q)}_{y,i}) \big)
\end{eqnarray}
such that $\hat{I}^{q}_{k,N}(X,Y) = \frac{1}{N} \sum_{i=1}^N \,w^{(q)}_{X_i} \,g(X_i, Y_i)$. Define each quantity with the true prior $p_X(x)$ as
\begin{eqnarray}
w'_{x} &\equiv& \frac{q_X(x)}{p_X(x)} \,,\\
n'_{y,i} &\equiv& \sum_{j \neq i} w'_{X_j} \mathbb{I} \{\|Y_j - Y_i\| < \rho_{k, i}\} \,,\\
g'(X_i, Y_i) &\equiv& \psi(k) + \log(N) - \big( \log(n_{X_i}) + \log(n'_{y,i}) \big) \,.
\end{eqnarray}
We apply triangular inequality,
and show that each term converges to zero in probability.
\begin{eqnarray}
    && \big| \hI^{(q)}_{k,N}(X,Y) - I^{(q)}(f_{Y|X}) \big| \notag \\
    & = & \Big| \frac{1}{N}\sum_{i=1}^N w^{(q)}_{X_i} g(X_i, Y_i) - \sum_{x \in \mathcal{X}} q_X(x)  \int f_{Y|X}(y|x) \log \frac{f_{Y|X}(y|x)}{\sum_{x' \in \mathcal{X}} q_X(x) f_{Y|X}(y|x')} dy \Big| \notag\\
    &　\leq & \frac{1}{N} \Big|\, \sum_{i=1}^N \big( w^{(q)}_{X_i} g(X_i,Y_i) - w'_{X_i} g'(X_i, Y_i) \big) \,\Big| \label{eq:m_kde_error}\\
    & + & \Big|\, \frac{1}{N} \sum_{i=1}^N w'_{X_i} g'(X_i, Y_i) - \sum_{x \in \mathcal{X}} q_X(x)  \int f_{Y|X}(y|x) \log \frac{f_{Y|X}(y|x)}{\sum_{x' \in \mathcal{X}} q_X(x) f_{Y|X}(y|x')} dy\,\Big| . \label{eq:m_knn_error}
\end{eqnarray}

The first term~(\ref{eq:m_kde_error}) captures the error in estimating $p_X(x)$. Similar as in~\eqref{eq:tail1}, the probability that it deviates from 0 is is upper bounded by:
\begin{eqnarray}
\Pr \Big(\, \frac{1}{N} \big|\, \sum_{i=1}^N \big( w^{(q)}_{X_i} g(X_i,Y_i) - w'_{X_i} g'(X_i, Y_i) \big) \,\big| > \varepsilon \,\Big) &\leq& \Pr \Big(\, \max_{x \in \mathcal{X}} |w^{(q)}_x - w'_x| > \varepsilon/(3\log{N})\,\Big) \label{eq:m_tail1} \,
\end{eqnarray}
for sufficiently large $N$. Recall that $w_x = Nq_X(x)/n_x$, ~\eqref{eq:m_tail1} is bounded by:
\begin{eqnarray}
&& \Pr \Big(\, \max_{x \in \mathcal{X}} |w^{(q)}_x - w'_x| > \varepsilon/(3\log{N})\,\Big) \,\notag\\
&=& \Pr \Big(\, \max_{x \in \mathcal{X}} \Big| \frac{Nq_X(x)}{n_x} - \frac{q_X(x)}{p_X(x)} \Big| > \varepsilon/(3\log{N})\,\Big) \,\notag\\
&=& \Pr \Big(\, \forall x \in \mathcal{X}, n_x \not\in [\frac{Nq_X(x)}{\frac{q_X(x)}{p_X(x)}+\frac{\varepsilon}{3\log{N}}}, \frac{Nq_X(x)}{\frac{q_X(x)}{p_X(x)}\frac{\varepsilon}{3\log{N}}}]\, \Big)\notag\\
&\leq& \Pr \Big(\, \forall x \in \mathcal{X}, n_x \not\in [N p_X(x) (1-\frac{\varepsilon p_X(x)}{6\log{N}q_X(x)}), Np_x(1+\frac{\varepsilon p_X(x)}{6\log{N}q_X(x)})]\, \Big)\label{eq:m_tail2}
\end{eqnarray}
for sufficiently large $N$ such that $\frac{\varepsilon p_x}{3\log{N}q_X(x)} < 1/3$. Recall that for each $x \in \mathcal{X}$, $n_x = \sum_{i=1}^N \mathbb{I}\{X_i = x\}$. Therefore, $n_x$ is a binomial random variable with parameter $(N,p_X(x))$. Therefore, by Hoeffding's inequality, for any $x \in \mathcal{X}$, we have:
\begin{eqnarray}
\Pr \big(\, |n_x-N p_X(x)| \leq \frac{N \varepsilon p^2_X(x)}{6\log{N}q_X(x)} \,\big) \leq 2\exp\{-\frac{1}{2N} (\frac{N \varepsilon p^2_X(x)}{6\log{N}q_X(x)})^2\} \leq 2\exp\{-\frac{N\varepsilon^2 C_1^2}{72|\mathcal{X}|^2 C_4^2 (\log{N})^2}\}
\end{eqnarray}
where the last inequality comes from the assumption that $p_X(x) > C_1/|\mathcal{X}|$ and $q_X(x)/p_X(x) < C_4$. Then by union bound, ~\eqref{eq:m_tail2} is upper bounded by:
\begin{eqnarray}
&& \Pr \Big(\, \forall x \in \mathcal{X}, n_x \not\in [N p_X(x) (1-\frac{\varepsilon p_X(x)}{6\log{N}q_X(x)}), Np_x(1+\frac{\varepsilon p_X(x)}{6\log{N}q_X(x)})]\, \Big) \,\notag\\
&\leq& |\mathcal{X}| \max_{x \in \mathcal{X}} \Pr \Big(\, |n_x-N p_X(x)| \leq \frac{N \varepsilon p^2_X(x)}{6\log{N}q_X(x)} \,\Big) \,\notag\\
&\leq& 2 |\mathcal{X}|\exp\{-\frac{N\varepsilon^2 C_1^2}{72|\mathcal{X}|^2 C_4^2 \log{N}}\}.
\end{eqnarray}
Combining with~\eqref{eq:m_tail1}, we know that~\eqref{eq:m_kde_error} converges to 0 in probability.

The second term in the error \eqref{eq:m_knn_error} comes from the sample noise in density estimation.
we decompose our estimator into three terms:
\begin{eqnarray*}
    \frac1N \sum_{i=1}^N w'_{X_i} g'(X_i, Y_i) &=& \hH^q_{k,N}(Y) - \hH^q_{k,N}(Y|X) - \sum_{i=1}^N \frac{w'_{X_i}}{N}(\log(N-1) - \log{N} + \log(n_{X_i}) - \psi(n_{X_i}))\;,
\end{eqnarray*}
where
\begin{eqnarray}
    \hH^q_{k,N}(Y|X) &\equiv& \sum_{i=1}^N \frac{w'_{X_i}}{N}  \big(\, -\psi(k) + \psi(n_{X_i}) + \log{c_{d_y}} + d_y \log \rho_{k,i} \,\big) \;, \label{def:m_hHUxy}\\
    \hH^q_{k,N}(Y) &\equiv& \sum_{i=1}^N \frac{w'_{X_i}}{N} \big(\, -\log {\tn_{y,i}} + \log(N-1) + \log c_{d_y} + d_y \log \rho_{k,i}  \, \big) .\label{def:m_hHUy}
\end{eqnarray}
Notice that $\sum_{i=1}^N \frac{w'_{X_i}}{N}(\log(N-1) - \log{N} + \log(n_{X_i}) - \psi(n_{X_i}))$ converges to 0 in probability as $N$ goes to infinity.
The desired claim follows directly from
the following two lemmas showing the convergence each entropy estimates to corresponding conditional entropy ${H}^q(Y|X)$ and entropy ${H}^q(Y)$. The desired claim immediately follows the two lemmas.

\begin{lemma}
    \label{lem:m_HUxy}
    Under the hypotheses of Theorem  \ref{thm:convergence_umi}, for all $\varepsilon > 0$
    \begin{eqnarray}
    \lim_{N \to \infty} \Pr\left(\, \Big|\hH^q_{k,N}(Y|X) - \big(\, -\sum_{x \in \mathcal{X}} q_X(x) \int f_{Y|X}(y|x) \log f_{Y|X}(y|x) dy \,\big) \Big| > \varepsilon\,\right) = 0 \;.
    \end{eqnarray}
\end{lemma}

\begin{lemma}
    \label{lem:m_HUx_HUy}
    Under the hypotheses of Theorem  \ref{thm:convergence_umi}, for all $\varepsilon > 0$
    \begin{eqnarray}
    \lim_{N \to \infty} \Pr\left(\, \Big|\hH^q_{k,N}(Y) - \big(\, -\sum_{x \in \mathcal{X}} q_X(x) \int f_{Y|X}(y|x) \log f_q(y)\big) dy\,\big)\Big| > \varepsilon\,\right) = 0 \;,
    \end{eqnarray}
    where $f_q(y)=\sum_{x \in \mathcal{X}} q_X(x) f_{Y|X}(y|x)$.
\end{lemma}

\subsubsection{Proof of Lemma~\ref{lem:m_HUxy}}

Define
\begin{eqnarray}
\hat{f}_{Y|X}(Y_i|X_i) = \frac{\exp\{\psi(k) - \psi(n_{X_i})\}}{c_{d_y} \rho_{k,i}^{d_y}}\;,
\end{eqnarray}
so that
\begin{eqnarray}
\hat{H}^q_{k,N}(Y|X) &=& -\sum_{i=1}^N \frac{w'_{X_i}}{N} \log \hat{f}_{Y|X}(X_i, Y_i) \,,
\end{eqnarray}
Notice that $\hat{f}_{Y|X}(Y_i|X_i)$ is just the $k$-nearest neighbour density estimator for the conditional pdf $f_{Y|X}(y|x)$. Therefore, by Theorem 8.~\cite{singh2003nearest}, we have
\begin{eqnarray}
\lim_{N \to \infty} \E \big[\log \hat{f}_{Y|X}(Y_i|X_i) \big| (X_i, Y_i) = (x,y) \big] &=& \log f_{Y|X}(y|x)\;,
\end{eqnarray}
Notice that $w'_{X_i} \log \hat{f}(Y_i|X_i)$ are identically distributed, therefore, we have
\begin{eqnarray}
&&\lim_{N \to \infty} \E \hat{H}^U_{k,N}(X,Y) \notag\\
&=& -\lim_{N \to \infty} \E [w'_{X_i} \log \hat{f}_{X,Y}(X_i,Y_i)] \notag\\
&=& - \lim_{N \to \infty} \sum_{x \in \mathcal{X}} \frac{q_X(x)}{p_X(x)} p_X(x) \Big(\, \int \E \big[\log \hat{f}_{X,Y}(X_i,Y_i) \big| (X_i, Y_i) = (x,y)\big] f_{Y|X}(y|x) dy \,\Big) \notag \\
&=& - \lim_{N \to \infty} \sum_{x \in \mathcal{X}} q_X(x) \Big(\, \int \E \big[\log \hat{f}_{X,Y}(X_i,Y_i) \big| (X_i, Y_i) = (x,y)\big] f_{Y|X}(y|x) dy \,\Big)
\end{eqnarray}
Use the same technique in the proof of Lemma~\ref{lem:HUxy} and Equation \eqref{eq:Fatou1}, we can switch the order of limit and integration. Therefore,
\begin{eqnarray}
&& \lim_{N \to \infty} \sum_{x \in \mathcal{X}} q_X(x) \Big(\, \int \E \big[\log \hat{f}_{X,Y}(X_i,Y_i) \big| (X_i, Y_i) = (x,y)\big] f_{Y|X}(y|x) dy \,\Big)  \notag \\
&=& \sum_{x \in \mathcal{X}} q_X(x) \Big(\, \int \lim_{N \to \infty}  \E \big[\log \hat{f}_{X,Y}(X_i,Y_i) \big| (X_i, Y_i) = (x,y)\big] f_{Y|X}(y|x) dy \,\Big)  \notag\\
&=& \sum_{x \in \mathcal{X}} q_X(x) \int \log_{Y|X}(y|x) f_{Y|X}(y|x) dy  \;, \label{eq:Fatou2}
\end{eqnarray}
Therefore,
\begin{eqnarray}
\lim_{N \to \infty} \E \hat{H}^q_{k,N}(Y|X) = -\sum_{x \in \mathcal{X}} q_X(x) \int f_{Y|X}(y|x) \log f_{Y|X}(y|x) dy. \label{eq:mean_m_HU}
\end{eqnarray}
Moreover, by Theorem11.~\cite{singh2003nearest}, we have:
\begin{eqnarray}
\lim_{N \to \infty} \V [\hat{f}_{Y|X}(Y_i|X_i)] &=& (\frac{\Gamma'(k)}{\Gamma(k)})' \, \V [\log f_{Y|X}(y|x)]< \infty \;,
\end{eqnarray}
and for any $j \neq i$:
\begin{eqnarray}
\lim_{N \to \infty} \Cov [\hat{f}_{Y|X}(Y_i|X_i), \hat{f}_{Y|X}(Y_j|Y_i)] = 0.
\end{eqnarray}
Since $w'_{x} \leq C_4$ for all $x$, similarly as in Lemma~\ref{lem:HUxy}, we obtain
\begin{eqnarray}
\lim_{N \to \infty} \V \big[ \hat{H}^q_{k,N}(Y|X) \big] = 0 \label{eq:var_m_HU}\;,
\end{eqnarray}
Combining~(\ref{eq:mean_m_HU}) and~(\ref{eq:var_m_HU}), we know $\hat{H}^q_{k,N}(Y|X)$ converges to its mean in $L^2$, hence in probability, i.e.,
\begin{eqnarray}
    \lim_{N \to \infty} \Pr\Big(\, \big|\hH^q_{k,N}(X,Y) - \big(\, -\sum_{x \in \mathcal{X}} q_X(x) \int \log_{Y|X}(y|x) f_{Y|X}(y|x) dy \,\big) \big| > \varepsilon\,\Big) = 0 .
    \end{eqnarray}

\subsubsection{Proof of Lemma~\ref{lem:m_HUx_HUy}}

Define
\begin{eqnarray}
\hat{f}_q(Y_i) &\equiv& \frac{\tn_{y,i}}{(N-1)c_{d_y}\rho_{k,i}^{d_y}} \;,
\end{eqnarray}
such that
\begin{eqnarray}
\hat{H}^q_{k,N}(Y) = -\sum_{i=1}^N \frac{w'_{X_i}}{N}  \log \hat{f}_q(Y_i).
\end{eqnarray}
By triangle inequality, we can write the formula in Lemma~\ref{lem:m_HUx_HUy} as:
\begin{eqnarray}
    && \big|\hH^q_{k,N}(Y) - \big(\, -\sum_{x \in \mathcal{X}} q_X(x) \int f_{Y|X}(y|x) \log f_q(y)\big) dy \,\big)\big| \notag \\
    & = & \big|\sum_{i=1}^N \frac{w'_{X_i}}{N} \log \hat{f}_q(Y_i) -\sum_{x \in \mathcal{X}} q_X(x) \int f_{Y|X}(y|x) \log f_q(y)\big) dy \big| \notag\\
    &　\leq & \big| \sum_{i=1}^N \frac{w'_{X_i}}{N} \log f_q(Y_i) -\sum_{x \in \mathcal{X}} q_X(x) \int f_{Y|X}(y|x) \log f_q(y)\big) dy \big| \label{eq:m_snis_error}\\
    &+&  \sum_{i=1}^N \frac{w'_i}{N} \Big| \log \hat{f}_q(Y_i) - \log f_q(Y_i)  \Big| . \label{eq:m_estimation_error}
\end{eqnarray}

The first term comes~(\ref{eq:m_snis_error}) from  sampling. Recall that $w'_{X_i} = q_X(X_i)/p_X(X_i)$. Therefore by strong law of large numbers,
\begin{eqnarray}
     \sum_{i=1}^N \frac{w'_{X_i}}{N} \log f_q(Y_i) \to \E \Big(\, \frac{q_X(x)}{p_X(x)} \log f_q(y) \,\Big)
\end{eqnarray}
almost surely. The mean is given by
\begin{eqnarray}
    \E \Big(\, \frac{q_X(x)}{p_X(x)} \log f_q(y) \,\Big) &=& \sum_{x \in \mathcal{X}} q_X(x) \int f_{Y|X}(y|x) \log f_q(y) dy.
\end{eqnarray}
Therefore,~\eqref{eq:m_snis_error} converges to 0 almost surely.
\\

The second term \eqref{eq:m_estimation_error} comes from density estimation. For any fixed $\varepsilon > 0$, by union bound, we obtain that
\begin{eqnarray}
    && \Pr\big(\,\sum_{i=1}^N \frac{w'_{X_i}}{N} \big| \log \hat{f}_q(Y_i) - \log f_q(Y_i)  \big| > \varepsilon\,\big) \notag\\
    & \leq & \Pr\big(\,\bigcup_{i=1}^N \big\{ \big| \log \hat{f}_q(Y_i) - \log f_q(Y_i)  \big| \Big| > \varepsilon/2 \big\} \,\big) + \Pr(\sum_{i=1}^N w'_{X_i} > 2N).
\end{eqnarray}
The second term converges to zero by the law of large numbers. The first term is bounded by:
\begin{eqnarray}
    && \Pr\big(\,\bigcup_{i=1}^N \big\{ \Big| \log \hat{f}_q(Y_i) - \log f_q(Y_i)  \big| \Big| > \varepsilon/2 \big\} \,\big) \\
    & \leq & N \cdot \Pr\big(\, \Big| \log \hat{f}_q(Y_i) - \log f_q(Y_i)  \big| \Big| > \varepsilon/2\,\big) \notag\\
    & = & N \sum_{x \in \mathcal{X}} p_X(x) \int \underbrace{\Pr\big(\, \big| \log \hat{f}_q(Y_i) - \log f_q(Y_i)  \big| \big| > \varepsilon/2 \Big|(X_i,Y_i) = (x,y)\,\big)}_{\leq I_1(x,y) + I_2(x,y) + I_3(x,y)} f_{Y|X}(y|x) dy
\end{eqnarray}
where
\begin{eqnarray}
     && I_1(x,y) = \Pr\big(\,\rho_{k,i} > (N^{1/2} p_X(x) f_{Y|X}(y|x) c_{d_y})^{-1/d_y} \big|X_i=x,Y_i=y\,\big) \label{eq:m_i1}\\
     && I_2(x,y) = \Pr\big(\,\rho_{k,i} < (\log{N})^{1+\delta/2} (N f_q(y) c_{d_y})^{-1/d_y} \big|X_i=x,Y_i=y\,\big) \label{eq:m_i2}\\
     && I_3(x,y) = \int_{r=r_2}^{r_1} \Pr\big(\,\big|\log \hat{f}_q(Y_i) - \log f_q(Y_i) \big| > \varepsilon/2 \Big| \rho_{k,i} = r, (X_i,Y_i)=(x,y) \,\big) f_{\rho_{k,i}}(r) dr \label{eq:m_i3}
\end{eqnarray}
where $f_{\rho_{k,i}}(r)$ is the pdf of $\rho_{k,i}$ given $X_i$ and $Y_i$. Here $r_1 =(N^{1/2} p_X(x) f_{Y|X}(y|x) c_{d_y})^{-\frac{1}{d_y}}$ and $r_2 = (\log{N})^{1+\delta/2} (N f_q(y) c_{d_y})^{-\frac{1}{d_y}}$. We will consider the three terms separately.
\\

\textit{$I_1(z)$:} Let $B(x, y, r) = \{(X,Y): \|Y-y\| < r, X = x\}$ be the $d_y$-dimensional ball centered at $y$ with radius $r$ with same $x$. Since the Hessian matrix of $H(f_{Y|X})$ exists and $\|H(f_{Y|X})\|_2 < C$ almost everywhere for any $x \in \mathcal{X}$, then for sufficiently small $r$, the probability mass within $B(x,y,r)$ is given by
\begin{eqnarray}
&& \Pr\big(\,(u,v) \in B(x,y,r)\,\big) = p_X(x) \int_{\|v-y\| \leq r} f_{Y|X}(v) dv \notag \\
& = & p_X(x) \int_{\|v-y\| \leq r} f_{Y|X}(y) + (v-y)^T \nabla f_{Y|X}(y) + (v-y)^T H(f_{Y|X})(y) (v-y) + o(\|v-y\|^2) du \notag \\
& \in & \big[\, p_X(x)f_{Y|X}(y|x)c_{d_y} r^{d_y}(1 - C r^2)), p_X(x)f_{Y|X}(y|x)c_{d_y} r^{d_y}(1 + C r^2))\,\big].
\end{eqnarray}
Then for sufficiently large $N$, the probability mass within $B(x,y,r_1)$ is lower bounded by
\begin{eqnarray}
    p_1 & \equiv & \Pr\big(\,(u,v) \in B(x,y,(N^{1/2} p_X(x)f_{Y|X}(y|x)c_{d_y})^{-1/d_y})\,\big) \notag \\
    & \geq & p_X(x) f_{Y|X}(y|x) c_{d_y} \big(\, (N^{1/2} p_X(x) f_{Y|X}(y|x) c_{d_y})^{-1/d_y} \,\big)^{d_y} \big(\, 1-C ((N^{1/2} p_X(x) f_{Y|X}(y|x) c_{d_y})^{-\frac{1}{d_y}})^2 \,\big) \notag \\
    & \geq & \frac{1}{2} N^{-1/2}.
\end{eqnarray}
$I_1(z)$ is the probability that at most $k$ samples fall in $B_Z(z, r_1)$, so it is upper bounded by
\begin{eqnarray}
I_1(z) & = & \Pr\big(\,\rho_{k,i} > r_1 \big| Z_i = z\,\big) \notag \\
& = &\sum_{m=0}^{k-1} {N-1 \choose m} p_1^m (1-p_1)^{N-1-m} \notag \\
&\leq & \sum_{m=0}^{k-1} N^m (1-p_1)^{N-1-m} \notag \\
&\leq & k N^{k-1} (1-\frac{1}{2\sqrt{N}})^{N-k-1} \notag \\
&\leq & k N^{k-1} \exp\{-\frac{N-k-1}{2\sqrt{N}}\} \label{eq:ub_m_i1}
\end{eqnarray}
for any $d_x, d_y \geq 1$.
\\

\textbf{$I_2$:} Let $r_2 = (\log{N})^{1+\delta/2} (N f_q(y) c_{d_y})^{-1/d_y}$. Then for sufficiently large $N$, the probability mass within $B(x,y,r_2)$ is given by:
\begin{eqnarray}
    p_2 & \equiv &\Pr\big(\,u \in B(x,y,(\log{N})^{1+\delta/2} (N f_q(y) c_{d_y})^{-1/d_y})\,\big) \notag \\
    & \leq & p_X(x) f_{Y|X}(y|x) c_{d_y} \big(\,(\log{N})^{1+\delta/2} (N f_q(y) c_{d_y})^{-1/d_y} \,\big)^{d_y} \big(\, 1+C ((\log{N})^{1+\delta/2} (N f_q(y) c_{d_y})^{-1/d_y})^2 \,\big) \notag \\
    & \leq & \frac{2p_X(x) f_{Y|X}(y|x)}{f_q(y)} (\log{N})^{(1+\delta/2)d_y} N^{-1} \notag \\
    & \leq & \frac{2p_X(x) f_{Y|X}(y|x)}{\sum_{x \in \mathcal{X}} q_X(x) f_{Y|X}(y|x)} (\log{N})^{(1+\delta/2)d_y} N^{-1} \notag \\
    & \leq & \frac{2}{C_3|\mathcal{X}|N} (\log{N})^{(1+\delta/2)d_y}
\end{eqnarray}
where the last equation comes from the assumption that $q_X(x)/p_X(x) > C_3$.
$I_2(z)$ is the probability that at least $k$ samples lying in $B(x,y,r_2)$. Therefore, it is upper bounded by
\begin{eqnarray}
I_2(z) &= & \Pr\big(\,\rho_{k,i} < r_2 \big | Z_i = z\,\big) \notag \\
& = &\sum_{m=k}^{N-1} {N-1 \choose m} p_2^m (1-p_2)^{N-1-m} \notag \\
&\leq &\sum_{m=k}^{N-1} \frac{N^m p_2^m}{m!} \notag \\
&\leq &\sum_{m=k}^{N-1} \frac{N^m p_2^m}{(m/e)^m} \notag \\
&\leq& \sum_{m=k}^{N-1} (\frac{Nep_2}{k})^m \notag \\
&\leq& \sum_{m=k}^{N-1} (\frac{2e}{C_3|\mathcal{X}|} (\log{N})^{(1+\delta/2)d_y}/k)^m.
\end{eqnarray}
Here we use the fact that $m! > (m/e)^m$ for all $m$. Since $k > (\log{N})^{(1+\delta)d_y}$ by assumption, $(\log{N})^{(1+\delta/2)d_y}/k$ is decreasing as $N$ increases. For sufficiently large $N$ such that $\frac{2e}{C_1|\mathcal{X}|} (\log{N})^{(1+\delta/2)d_y}/k < 1/2$, we obtain:
\begin{eqnarray}
I_2(z) &\leq& 2(\frac{2e}{C_3|\mathcal{X}|} (\log{N})^{(1+\delta/2)d_y}/k)^k \,\notag\\
&\leq& 2(\frac{2e}{C_3|\mathcal{X}|})^{(\log{N})^{(1+\delta)d_y}} (\log{N})^{-\delta (log{N})^{(1+\delta)d_y}/2} .\label{eq:ub_m_i2}
\end{eqnarray}
\\

$I_3$: Given that $(X_i,Y_i) = (x,y)$ and $\rho_{k,i} = r$. Recall that $\hat{f}_q(Y_i) = \frac{\tn_{y,i}}{(N-1)c_{d_y}r^{d_y}}$, then we have
\begin{eqnarray}
    && \Pr\big(\,\big|\log \hat{f}_q(Y_i) - \log f_q(Y_i)\big| > \varepsilon/2\big| \rho_{k,i} = r, (X_i,Y_i)=(x,y) \,\big) \notag \\
    &=&\Pr\big(\,\big|\log \tn_{y,i} - \log(N-1) - \log c_{d_y} - d_y \log \rho_{k,i} - \log f_q(y)\big| > \varepsilon/2\big| \rho_{k,i} = r, (X_i,Y_i)=(x,y)\,\big) \notag \\
    &=& \Pr\big(\,\big|\log \tn_{y,i} - \log(N-1)c_{d_y}r^{d_y}f_q(y)\big| > \varepsilon/2\big| \rho_{k,i} = r, (X_i,Y_i)=(x,y)\,\big) \notag \\
    &=& \Pr\big(\, \tn_{y,i} > (N-1)c_{d_y}r^{d_y}f_q(y)e^{\varepsilon/2}\big| \rho_{k,i} = r, (X_i,Y_i)=(x,y)\,\big) \notag\\
    &+& \Pr\big(\, \tn_{y,i} < (N-1)c_{d_y}r^{d_y}f_q(y)e^{-\varepsilon/2}\big| \rho_{k,i} = r, (X_i,Y_i)=(x,y)\,\big).
\end{eqnarray}
Following a similar technique as the analysis of $I_4$ in proof of Lemma~\ref{lem:HUx_HUy}, we obtain
\begin{eqnarray}
&&\Pr\big(\,\big|\log \hat{f}_q(Y_i) - \log f_q(Y_i)\big| > \varepsilon/2\big| \rho_{k,i} = r, (X_i,Y_i)=(x,y) \,\big) \,\notag\\
&\leq& 2\exp\{-\frac{C_3\varepsilon^2}{128(1+7\varepsilon/24)}(N-k-1) c_{d_y}r^{d_y}f_q(y)\}
\end{eqnarray}
where $C_3$ is the lower bound of $q_X(x)/p_X(x)$. Therefore, $I_3(x,y)$ is upper bounded by:
\begin{eqnarray}
    I_3(x,y)&=& \int_{r=r_2}^{r_1} \Pr\big(\,\big|\log \hat{f}_q(Y_i) - \log f_q(Y_i)\big| > \varepsilon/2\big| \rho_{k,i} = r, (X_i,Y_i)=(x,y) \,\big) f_{\rho_{k,i}}(r) dr \notag \\
    &\leq& \int_{r=(\log{N})^{1+\delta/2} (N f_q(y) c_{d_y})^{-1/d_y}}^{(N^{1/2} p_X(x) f_{Y|X}(y|x) c_{d_y})^{-1/d_y}} \Pr\big(\,\big|\log \hat{f}_q(Y_i) - \log f_q(Y_i)\big| > \varepsilon/2 \big| \rho_{k,i} = r, (X_i,Y_i)=(x,y) \,\big) f_{\rho_{k,i}}(r) dr \notag \\
    &\leq& \int_{r=(\log{N})^{1+\delta/2} (N f_q(y) c_{d_y})^{-1/d_y}}^{(N^{1/2} p_X(x) f_{Y|X}(y|x) c_{d_y})^{-1/d_y}} 2 \exp\{-\frac{C_3 \varepsilon^2}{128(1+7\varepsilon/24)}(N-k-1) c_{d_y}r^{d_y}f_q(y)\} f_{\rho_{k,i}}(r) dr \notag \\
    &\leq& 2\exp\{- \frac{C_3 \varepsilon^2}{256} N c_{d_y}f_q(y)((\log{N})^{1+\delta/2} (N f_q(y) c_{d_y})^{-\frac{1}{d_y}})^{d_y}\}  \notag \\
    &\leq& 2\exp\{-\frac{C_3 \varepsilon^2}{256} (\log{N})^{(1+\delta/2)d_y}\} \label{eq:ub_m_i3}
\end{eqnarray}
for sufficiently large $N$ such that $(N-k-1)/(1+7\varepsilon/24)) > N/2$.
\\

Now combine ~\eqref{eq:ub_m_i1}, ~\eqref{eq:ub_m_i2}, and~\eqref{eq:ub_m_i3}, and we obtain
\begin{eqnarray}
     && \Pr\big(\,\sum_{i=1}^N \frac{w'_i}{N} \big| \log \hat{f}_q(Y_i) - \log f_q(Y_i)  \big| > \varepsilon\,\big) \notag \\
    & \leq & N \sum_{x \in \mathcal{X}} p_X(x) \int (I_1(x,y) + I_2(x,y) + I_3(x,y))dy \notag\\
    &\leq& k N^k \exp\{-\frac{N-k-1}{2\sqrt{N}}\} + 2N\exp\{-\frac{C_3 \varepsilon^2}{256} (\log{N})^{(1+\delta/2)d_y}\} \,\notag\\
    &+& 2N(\frac{2e}{C_3|\mathcal{X}|})^{(\log{N})^{(1+\delta)d_y}} (\log{N})^{-\delta (log{N})^{(1+\delta)d_y}/2}.
\end{eqnarray}
One can easily see that the first and second terms converges to 0 as $N$ goes to infinity, given that $k < \sqrt{N}/(5\log{N})$. To see that the last term converges to 0, we will show that the logarithm goes to $-\infty$ as $N$ goes to infinity, which is
\begin{eqnarray}
&&\log (N(\frac{2e}{C_3|\mathcal{X}|})^{(\log{N})^{(1+\delta)d_y}} (\log{N})^{-\delta (log{N})^{(1+\delta)d_y}/2}) \,\notag\\
&=& \log{N} + \log(\frac{2e}{C_3|\mathcal{X}|}) (\log{N})^{(1+\delta)d_y} - \log{N} \delta (\log{N})^{(1+\delta)d_y}/2 \,\notag\\
&=& \log{N} + \log(\frac{2e}{C_3|\mathcal{X}|}) (\log{N})^{(1+\delta)d_y} - \frac{\delta}{2} (\log{N})^{(1+\delta)d_y+1}.
\end{eqnarray}
The negative term has the larger exponent, so the logarithm will goes to $-\infty$,
and we have
\begin{eqnarray}
\lim_{N \to \infty} \Pr\big(\,\sum_{i=1}^N \frac{w'_{X_i}}{N} \big| \log \hat{f}_q(Y_i)- \log f_q(Y_i) \big| > \varepsilon\,\big) = 0.
\end{eqnarray}
Therefore, by combining the convergence of error from sampling and error from density estimation, we obtain that $\hI^{(q)}_{k,N}(X,Y)$ converges to $I^{(q)}(f_{Y|X})$ in probability.

\section{Proof of the CMI estimator convergence}

\begin{assumption}
\label{assumption:mixed_cmi}
We make the following assumptions:
\begin{itemize}
  \item[$(a)$] $\int f_{Y|X}(y|x) \big|\log f_{Y|X}(y|x) \big| dy < \infty$, for all $x \in \mathcal{X}$.
  \item[$(b)$] $\int f_{Y|X}(y|x) \big(\, \log f_{Y|X}(y|x) \,\big)^2 dy < \infty$, for all $x \in \mathcal{X}$.
  \item[$(c)$] There exists a finite constant $C$ such that the Hessian matrix of $H(f_{Y|X})$ exists and $\|H(f_{Y|X})\|_2 < C$ almost everywhere, for all $x \in \mathcal{X}$.
   \item[$(d)$] There exists a finite constant $C'$ such that the conditional pdf $f_{Y|X}(y\big|x) < C'$  almost everywhere, for all $x \in \mathcal{X}$.
   \item[$(e)$] There exists finite constants $C_1 < C_3 < C_4 < C_2$ such that the ratio of the optimal prior $q^*$ of the maximizer in the definition of ${\cal C}(f_{Y|X})$ and the true prior satisfies that $q^*_X(x)/p_X(x) \in [C_3, C_4]$ for every $x \in \mathcal{X}$.
   \item[$(e)$] There exists finite constants $C_5 < C_6$ such that $p_X(x) > C_5/|\mathcal{X}|$ and $p_X(x) < C_6/|\mathcal{X}|$, for all $x \in \mathcal{X}$.
\end{itemize}
\end{assumption}

Define
\begin{eqnarray}
I(f_{Y|X})(q_X) &\equiv& \sum_{x \in \mathcal{X}} q_X(x) \int f_{Y|X}(y|x) \log \frac{f_{Y|X}(y|x)}{\sum_{x' \in \mathcal{X}} q_X(x') f_{Y|X}(y|x') } dy
\end{eqnarray}
and
\begin{eqnarray}
\hat{I}_{k,N}(X,Y)(w) &\equiv& \frac1N \sum_{i=1}^N w_{X_i} \,\Big( \psi(k) + \log (N) - \big(\, \log(n_{X_i}) + \log (n_{y,i}) \,\big) \,\Big)
\end{eqnarray}
such that $C(f_{Y|X}) = \max_{q_X \in Q} I(f_{Y|X})(q_X)$ and $\hat{C}^{\Delta}_{k,N}(X,Y) = \max_{w \in T_{\Delta}} \hat{I}_{k,N}(X,Y)(w)$. 
 First, consider the quantity:
\begin{eqnarray}
C^{\Delta}(f_{Y|X}) &\equiv& \max_{q_X \in T_{\Delta}(Q)} I(f_{Y|X})(q_X)
\end{eqnarray}
where the constraint set $T_{\Delta}(Q)$ is defined as:
\begin{eqnarray}
T_{\Delta}(Q) = \{q_X \in \mathbb{R}^{|\mathcal{X}|}: [ ( q_X(x)/p_X(x)) ] \in T_{\Delta} \textrm{ and }\sum_{x \in \mathcal{X}} q_X(x) \in [1- {|\cX|}\Delta,1+{|\cX|}\Delta]\}
\end{eqnarray}
we rewrite the error term in Theorem  \ref{eq:convergence_mixed_cmi} as
\begin{eqnarray}
\big|\hC^{\Delta}_{k,N} (X,Y)-C(f_{Y|X})\big| &\leq& \big|C^{\Delta}(f_{Y|X}) - C(f_{Y|X}) \big| + \big| \hC^{\Delta}_{k,N} - C^{\Delta}(f_{Y|X})\big|  . \label{eq:ub_cmi}
\end{eqnarray}

The first error comes from quantization. Let $q^*$ be the maximizer of $C(f_{Y|X})$. By assumption, $q^*(x)/p_X(x) \in [C_3, C_4] \subseteq [C_1, C_2]$, for all $x$. Since $T_{\Delta}(Q)$ is a quantization of the simplex $Q$, so there exists a $q_0 \in T_{\Delta}(Q)$ such that $|q_0(x) - q^*(x)| < \Delta \cdot p_X(x) < \Delta $ for all $x \in \mathcal{X}$. Now we will bound the difference of $I(f_{Y|X})(q_0)$ and $I(f_{Y|X})(q^*)$ by the following lemma:

\begin{lemma}
    \label{lem:cmi_cont}
    Under the assumptions of Theorem  \ref{eq:convergence_mixed_cmi},
    if $q(x)/p(x) \in [C_1,C_2] $ and $q'(x)/p(x) \in [C_1,C_2]$ for all $x \in \mathcal{X}$, then
    \begin{eqnarray}
    \big|\, I(f_{Y|X})(q) - I(f_{Y|X})(q') \,\big| \leq L \max_{x \in \mathcal{X}} |q(x)-q'(x)|\;,
    \end{eqnarray}
    for some positive constant $L$.
\end{lemma}

Then we have:
\begin{eqnarray}
C(f_{Y|X}) &=& I(f_{Y|X})(q^*) \,\notag\\
&\leq& I(f_{Y|X})(q_0) +  L \max_{x \in \mathcal{X}} |q_0(x)-q^*(x)| \,\notag\\
&\leq& \max_{q \in T_{\Delta}(Q)} I(f_{Y|X})(q) + L \Delta \,\notag\\
&=& C^{\Delta}(f_{Y|X}) + L \Delta.
\end{eqnarray}

Similarly, let $q^{**}$ be the maximizer of $C^{\Delta}(f_{Y|X})$, we can also find a $q_1 \in Q$ such that $|q_1(x) - q^{**}(x)| < \Delta $ for all $x \in \mathcal{X}$. Using Lemma~\ref{lem:cmi_cont} again, we will obtain $C^{\Delta}(f_{Y|X}) \leq C(f_{Y|X}) + L \Delta$. Therefore, the first term in~\eqref{eq:ub_cmi} is bounded by $O(\Delta)$.

Now consider the second term. Upper bound on the second term relies on the convergence of dicrete UMI estimation from  Theorem . Recall that in the proof of Theorem  \ref{thm:convergence_umi}, we have shown that under certain conditions,
\begin{eqnarray}
\Pr \big(\, \big| \hat{I}_{k,N}(X,Y)(w_q) - I(f_{Y|X})(q) \big| > \varepsilon/2 \,\big) \stackrel{N \to \infty }{\longrightarrow}  0
\end{eqnarray}
for any $q$ with bounded $q_X/p_X$. Here $(w_{q})_x= q(x)/p_X(x)$. Since the set $T_{\Delta}(Q)$ is finite, by union bound, we have:
\begin{eqnarray}
&&\lim_{N \to \infty} \Pr \big(\, \forall q \in T_{\Delta}(Q), \big| \hat{I}_{k,N}(X,Y)(w_q) - I(f_{Y|X})(q) \big| \leq \varepsilon/2 \,\big) \,\notag\\
&\geq& 1 - |T_{\Delta}(Q)| \lim_{N \to \infty} \Pr \big(\, \big| \hat{I}^{k,N}(X,Y)(w_q) - I(f_{Y|X})(q) \big| \leq \varepsilon/2 \,\big) = 1. \label{eq:cmi_cond1}
\end{eqnarray}
Also, by the strong law of large numbers, we have that
\begin{eqnarray}
\lim_{N \to \infty} \Pr \big(\,\forall x \in \mathcal{X}, |p_x(X) - n_X/N| <  \Delta/C_2|\mathcal{X}| \,\big) = 1. \label{eq:cmi_cond2}
\end{eqnarray}
We claim that if the events inside the probability in~\eqref{eq:cmi_cond1} and~\eqref{eq:cmi_cond2} happen simultaneously, then $\big| \hC^{\Delta}_{k,N} - C^{\Delta}(f_{Y|X})\big| < \varepsilon + O(\Delta)$, which implies the desired claim.

Let $w^* = \arg\max_{w \in T_{\Delta}} \hat{I}_{k,N}(X,Y)(w)$. Define $q_2(x) = w^*_x p_X(x)$. Since $[q_2(x)/p_X(x)] \in T_{\Delta}$ for all $x$ and
\begin{eqnarray}
\big|\, \sum_{x \in \mathcal{X}} q_2(x) - 1 \,\big| &=& \big|\, \sum_{x \in \mathcal{X}} w^*_x (p_X(x) - n_x/N)\, + (\Delta/2)|\cX| \big| \,\notag\\
	&\leq&  \,|\mathcal{X}| \,  ((\Delta/2)+C_2 \max_{x \in \mathcal{X}} \big|\, p_X(x) - n_x/N \,\big|) \nonumber \\
	&\leq& ({|\cX|}/2+1)  \Delta \;.
	\label{eq:q2}
\end{eqnarray}
Therefore, $q_2 \in T_{\Delta}(Q)$, so
\begin{eqnarray}
 \hC^{\Delta}_{k,N} &=& \hat{I}_{k,N}(X,Y)(w^*) \,\notag\\
 &\leq& I(f_{Y|X})(q_2) + \varepsilon \,\notag\\
 &\leq& C^{\Delta}(f_{Y|X}) + \varepsilon.
\end{eqnarray}

On the other hand, consider $q^{**} = \arg\max_{q_X \in T_{\Delta}(Q)} I(f_{Y|X})(q_X)$ again, and define $(w_0)_x = q^{**}(x)/p_X(x)$. We know that $(w_0)\in T_{\Delta}^{|\mathcal{X}|}$ but not necessarily $\sum_{i=1}^N (w_0)_{X_i} = N$. But we claim that the sum is closed to $N$ as follows
\begin{eqnarray}
\big|\, \sum_{i=1}^N (w_0)_{X_i} - N \,\big| &=& \big|\, \sum_{x \in \mathcal{X}} \frac{n_x q^{**}(x)}{p_X(x)} - N| \,\notag\\
&\leq& N \max_{x \in \mathcal{X}} \big\{\, \frac{q^{**}(x)}{p_X(x)} \big|\, \frac{n_x}{N} - p_X(x) \,\big| \,\big\}\,\notag\\
&\leq& N C_2 \frac{\Delta}{C_2|\mathcal{X}|} < N\Delta
\end{eqnarray}
so we can find a $(w_1) \in T_{\Delta}(W)$ such that $|(w_1)_{x} - (w_0)_{x}| \leq \Delta$ for all $x$. Let $q_4(x) = (w_1)_x p_X(x)$, similar as~\eqref{eq:q2}, we know that $q_4 \in T_{\Delta}(Q)$. Moreover, $\big|\, q_4(x) - q^{**}(x) \,\big| \leq p_X(x) \big|\, (w_1)_{x} - (w_0)_{x} \,\big| \leq \Delta$ for all $x$.
Then we have
\begin{eqnarray}
C^{\Delta}(f_{Y|X}) &=& I(f_{Y|X})(q^{**}) \,\notag\\
&\leq& I(f_{Y|X})(q_4) + L \max_{x \in \mathcal{X}} |q^{**}(x) - q_4(x)| \,\notag\\
&\leq& \hat{I}_{k,N}(X,Y)(w_1) + \varepsilon + L \Delta \,\notag\\
&=& \hat{C}^{\Delta}_{k,N} + \varepsilon + L \Delta.
\end{eqnarray}
Therefore, we have $|\hat{C}^{\Delta}_{k,N} - C(f_{Y|X})| < \varepsilon + O(\Delta)$, thus our proof is complete.

\subsection{Proof of Lemma~\ref{lem:cmi_cont}}
We will show that for any $x \in \mathcal{X}$, we have $|\frac{\partial}{\partial q_X(x)} I(f_{Y|X})(q) |\leq L/|\mathcal{X}|$ for some $L$. Therefore,
\begin{eqnarray}
\big|\, I(f_{Y|X})(q) - I(f_{Y|X})(q') \,\big| &\leq& \sum_{x \in \mathcal{X}} |\frac{\partial I(f_{Y|X})(q)}{\partial q_X(x)}| \, |q_X(x) - q'_X(x)| \,\notag\\
&\leq& L \max_{x \in \mathcal{X}} |q_X(x) - q'_X(x)|
\end{eqnarray}

Let $f_q(y) = \sum_{x \in \mathcal{X}} q_{X}(x) f_{Y|X}(y|x)$. Since $q_X(x) \in [C_1p_X(x), C_2p_X(x)] \subseteq [C_1C_5/|\mathcal{X}|,C_2C_6/|\mathcal{X}|]$ we know that
\begin{eqnarray}
f_q(y) \in [C_1C_5 \min_{x \in \mathcal{X}} f_{Y|X}(y|x), C_2C_6 \max_{x \in \mathcal{X}} f_{Y|X}(y|x)]
\end{eqnarray}
for all $x,y$. Therefore, the absolute value of the gradient is bounded by
\begin{eqnarray}
&&\Big|\, \frac{\partial I(f_{Y|X})(q)}{\partial q_X(x)} \,\Big|  \,\notag\\
&=&\Big| \frac{\partial}{\partial q_{X}(x)} \Big(\, \sum_{x \in \mathcal{X}} q_{X}(x) \int f_{Y|X}(y|x) \log \frac{f_{Y|X}(y|x)}{f_q(y)} dy \,\big) \Big| \,\notag\\
&=& \Big| \int f_{Y|X}(y|x) \log \frac{f_{Y|X}(y|x)}{f_q(y)} dy \Big| + \Big| \sum_{x' \in \mathcal{X}} q_X(x') \int \frac{f_{Y|X}(y|x')f_{Y|X}(y|x)}{f_q(y)} dy  \Big|  \,\notag\\
&\leq& \Big| \max_{y} \log \frac{f_{Y|X}(y|x)}{f_q(y)} \Big| + \Big| \max_{y} \frac{f_{Y|X}(y|x)}{f_q(y)} \Big| \,\notag\\
&\leq& \max \{|\log C_1C_5|, |\log C_2C_6|\} + 1/(C_1C_5)
\end{eqnarray}
where $L = |\mathcal{X}| \max\{|\log C_1C_5|, |\log C_2C_6|\}$.

%
%
%
%

\section{Proof of Lemma~\ref{lem:eqkde_error}}
\label{app:proof-of-kde_error}
  The term in Equation~\eqref{eq:kde_error}  is upper bounded by:
\begin{eqnarray}
&&\frac{1}{N} \Big|\, \sum_{i=1}^N \big( w_i g(X_i,Y_i) - w'_i g'(X_i, Y_i) \big) \,\Big| \,\notag \\
& \leq & \frac{1}{N} \sum_{i=1}^N \big| w_i g(X_i,Y_i) - w'_i g'(X_i, Y_i) \big| \, \notag \\
& \leq & \frac{1}{N} \sum_{i=1}^N \Big( | w_i - w'_i| \, |g'(X_i, Y_i)|  + w_i \, \big|g(X_i, Y_i) - g'(X_i, Y_i)\big| \Big) \, \notag \\
& = & \frac{1}{N} \sum_{i=1}^N \Big( | w_i - w'_i| \, |g'(X_i, Y_i)|  + w_i \, \big|\log(\tn_{y,i}) - \log(n'_{y,i})\big| \Big) \, \notag \\
& \leq & \frac{1}{N} \sum_{i=1}^N \Big( | w_i - w'_i| \, |g'(X_i, Y_i)|  + w_i \, \big|\tn_{y,i} - n'_{y,i}\big| (\frac{1}{2\tn_{y,i}} + \frac{1}{2n'_{y,i}})\Big) \,\notag \\
& \leq & \frac{1}{N} \sum_{i=1}^N | w_i - w'_i| \, |g'(X_i, Y_i)|  + \sum_{i=1}^N  \frac{w_i}{N}  \Big(\, \frac{\sum_{j \neq i} \mathbb{I} \{\|Y_j - Y_i| < \rho_{k,i}\} |w'_j - w_j|}{2 \sum_{j \neq i} \mathbb{I} \{\|Y_j - Y_i| < \rho_{k,i}\} w'_j} + \frac{\sum_{j \neq i} \mathbb{I} \{\|Y_j - Y_i| < \rho_{k,i}\} |w'_j - w_j|}{2 \sum_{j \neq i} \mathbb{I} \{\|Y_j - Y_i| < \rho_{k,i}\} w_j} \Big) \,\notag \\
& \leq & \frac{1}{N} \sum_{i=1}^N |w_i - w'_i| \, |g'(X_i, Y_i)| + \sum_{i=1}^N \frac{w_i}{N} \big(\, \frac{\max_{1 \leq j \leq N} |w'_j - w_j|}{2 \min_{1 \leq j \leq N} w'_j} +  \frac{\max_{1 \leq j \leq N} |w'_j - w_j|}{2 \min_{1 \leq j \leq N} w'_j} \,\big)\,\notag \\
& \leq & \max_{1 \leq i \leq N} |w_i - w'_i| \big(\max_{1 \leq i \leq N} |g'(X_i, Y_i)| + \frac{1}{2\min_{1 \leq j \leq N} w'_j} + \frac{1}{2\min_{1 \leq j \leq N} w_j})\;, \label{eq:ub_kde_error}
\end{eqnarray}
where the last inequality follows from the fact that $\sum_{i=1}^N w_i=N$.
We upper bound each term as follows. 
\begin{eqnarray}
w'_j &=& \frac{u(X_i)}{f_X(X_i)} \geq \frac{1/\mu(K)}{C_2/\mu(K)} = 1/C_2.
\end{eqnarray}
Similarly we have $w'_j \leq 1/C_1$.
This implies that $n_{y,i}'= \sum_{j\neq i} w'_j \mathbb{I}\{\|Y_i-Y_j\|\leq \rho_{k,i} \} \geq k/C_2$.
For finite $k$ and sufficiently large $N$, we have:
\begin{eqnarray}
g'(X_i, Y_i) &=& \psi(k) + \log(N)  + \log(\frac{c_{d_x}c_{d_y}}{c_{d_x+d_y}}) - \big( \log(n_{x,i}) + \log(\tn_{y,i}) \big) \,\notag \\
&\leq & \psi(k) + \log(N)  + \log(\frac{c_{d_x}c_{d_y}}{c_{d_x+d_y}}) - \big( \log(k) + \log(k/C_2) \big) \,\notag \\
&\leq & 2 \log{N}\;,
\end{eqnarray}
and similarly, using the fact that $\tn_{y,i} = \sum_{j\neq i} w'_j \mathbb{I}\{\|Y_i-Y_j\|\leq \rho_{k,i} \} \leq N/C_1$,
\begin{eqnarray}
g'(X_i, Y_i)
&\geq & \psi(k) + \log(N) +  \log(\frac{c_{d_x}c_{d_y}}{c_{d_x+d_y}}) - \big( \log(N) + \log(N/C_1) \big) \,\notag \\
&\geq & - 2 \log{N}.
\end{eqnarray}
We claim that for sufficiently large $N$such that $\log{N} > \max\{C_2\varepsilon/3, 3C_2/2\}$, if $|w_i - w'_i| < \varepsilon/(3\log{N})$ for all $i$, then~\eqref{eq:ub_kde_error} is upper bounded by $\varepsilon$.
\begin{eqnarray}
&&\max_{1 \leq i \leq N} |w_i - w'_i| \big(\max_{1 \leq i \leq N} |g'(X_i, Y_i)| + \frac{1}{2\min_{1 \leq j \leq N} w'_j} + \frac{1}{2\min_{1 \leq j \leq N} w_j}) \,\notag\\
&\leq& \frac{\varepsilon}{3\log{N}} \big(\, 2\log{N} + \frac{C_2}{2} + \frac{1}{2/C_2 - \frac{\varepsilon}{3\log{N}}}\big) \,\notag\\
&\leq& \frac{\varepsilon}{3\log{N}} \big(\, 2\log{N} + \frac{C_2}{2} + C_2\big) \leq \varepsilon .
\end{eqnarray}
Putting these bounds together,
we have, for any $\varepsilon > 0$ and sufficiently large $N$,
\begin{eqnarray}
\Pr \Big(\, \frac{1}{N} \big|\, \sum_{i=1}^N \big( w_i g(X_i,Y_i) - w'_i g'(X_i, Y_i) \big) \,\big| > \varepsilon \,\Big) &\leq& \Pr \Big(\, \max_{1 \leq i \leq N} |w_i - w'_i| > \varepsilon/(3\log{N})\,\Big). \label{eq:tail1}
\end{eqnarray}
Define
\begin{eqnarray}
w''_i = \frac{N/f_X(X_i)}{\sum_{j=1}^N \,\big( 1/f_X(X_j) \,\big)}
\end{eqnarray}
and applying  the triangle inequality and union bound for~(\ref{eq:tail1}), we have
\begin{eqnarray}
&& \Pr \Big(\, \max_{1 \leq i \leq N} |w_i - w'_i| > \frac{\varepsilon}{3\log{N}} \,\Big) \notag\\
&\leq& \Pr \Big(\, \max_{1 \leq i \leq N} |w'_i - w''_i|  + \max_{1 \leq i \leq N} |w''_i - w_i| > \frac{\varepsilon}{3\log{N}} \,\Big) \notag\\
&\leq& \Pr \Big(\, \max_{1 \leq i \leq N} |w'_i - w''_i| > \frac{\varepsilon}{6\log{N}} \,\Big) \label{eq:tail2} \\
&+& \Pr \Big(\, \max_{1 \leq i \leq N} |w''_i - w_i| > \frac{\varepsilon}{6\log{N}} \,\Big). \label{eq:tail3}
\end{eqnarray}
For~(\ref{eq:tail2}), recall that $w'_i = u(X_i)/f_X(X_i)$. Since  $u(X_i)/f_X(X_i) \in [1/C_2, 1/C_1]$ for all $i$. Therefore
\begin{eqnarray}
&& \Pr \Big(\, \max_{1 \leq i \leq N} |w'_i - w''_i| >\frac{\varepsilon}{6\log{N}}\,\Big) \notag\\
&=& \Pr \Big(\, \max_{1 \leq i \leq N} |\frac{u(X_i)}{f_X(X_i)} - \frac{N/f_X(X_i)}{\sum_{j=1}^N \,\big( 1/f_X(X_j) \,\big)}| > \frac{\varepsilon}{6\log{N}} \,\Big)\notag\\
&=& \Pr \Big(\, \big(\, \max_{1 \leq i \leq N} \frac{u(X_i)}{f_X(X_i)}  \big|\, 1 - \frac{N}{\sum_{j=1}^N \,\big( u(X_j)/f_X(X_j) \,\big)}\big|　\,\big) >\frac{\varepsilon}{6\log{N}}\,\Big) \notag\\
&\leq& \Pr \Big(\, \frac{1}{C_1} \big|\, \sum_{j=1}^N \frac{u(X_j)}{f_X(X_j)} - N \big| > \frac{\varepsilon }{6\log{N}} \sum_{j=1}^N \frac{u(X_j)}{f_X(X_j)} \,\big). \notag\\
&\leq& \Pr \Big(\,  \big|\, \sum_{j=1}^N \frac{u(X_j)}{f_X(X_j)} - N \big| > \frac{\varepsilon }{6\log{N}}\frac{NC_1}{C_2} \,\Big).
\end{eqnarray}
Note that $w'_j = \frac{u(X_j)}{f_X(X_j)}$ are i.i.d. random variables with $\E[w'_j] = \int f_X(x) \frac{u(x)}{f_X(x)} dx = 1$ and $w'_j \in [1/C_2, 1/C_1]$. Therefore, by Hoeffding's inequality, we obtain:
\begin{eqnarray}
&& \Pr \Big(\,  \big|\, \sum_{j=1}^N \frac{u(X_j)}{f_X(X_j)} - N \big| > \frac{\varepsilon NC_1 }{6\log{N}C_2}\,\Big) \,\notag\\
&\leq& 2\exp\left \{-\frac{2 \big(\,\frac{\varepsilon NC_1 }{6\log{N}C_2} \,\big)^2}{N(1/C_1-1/C_2)^2}\right\} \,\notag\\
&=& 2\exp\left\{-\frac{\varepsilon^2 NC_1^4}{18(\log{N})^2(C_2-C_1)^2}\right\} \label{eq:tail2bound}
\end{eqnarray}
which shows that the probability in~(\ref{eq:tail2}) goes to 0 as $N$ goes to infinity. For the probability in~(\ref{eq:tail3}), recall that for any $i$,
\begin{eqnarray}
|w''_i - w_i| &=& \big| \frac{N/f_X(X_i)}{\sum_{j=1}^N \,\big( 1/f_X(X_j) \,\big)} - \frac{N/\tilde{f}_X(X_i)}{\sum_{j=1}^N \,\big( 1/\tilde{f}_X(X_j) \,\big)}\big| .
\end{eqnarray}
We use the following lemma that shows an upper bound for the error of kernel density estimator.
\begin{eqnarray}
\tilde{f}_X(X_i) = \frac{1}{(N-1)h_N^{d_x}} \sum_{j \neq i} K(\frac{X_j - X_i}{h_N}).
\end{eqnarray}

\begin{lemma}
    \label{lem:KDE}
    Assume that $K(u) \leq A$ for all $u$, $\kappa_j(K) = \int_{\mathbb{R}^{d_x}} \|u\|^j K(u) du < +\infty$ for any positive integer $j \geq 1$ and $\int_{\mathbb{R}^{d_x}} u K(u) du = 0$. By choosing $h_N = \frac{1}{2}N^{-1/(2d_x+3)}$, we have for a given $i\in\{1,\ldots,N\}$,
    \begin{eqnarray}
    \Pr \Big(\, \big|\, \tilde{f}_X(X_i) - f_X(X_i) \,\big| > N^{-1/(2d_x+3)} \Big) \leq 2\exp\{-\frac{ N^{1/(2d_x+3)}}{16A^2}\}.
    \end{eqnarray}
\end{lemma}
Applying the union
bound we get that
with probability at least
$1-2N \exp\{-\frac{N^{1/(2d_x+3)}}{16A^2}\}$, we have
\begin{eqnarray}
    |\tilde{f}_X(X_i) - f_X(X_i)| &<& N^{-1/(2d_x+3)}\;,\label{eq:firsttermcondition}
    \end{eqnarray}
    for all $i$.
When this bound holds, we claim that the event inside of the probability in  \eqref{eq:tail3}  holds for sufficiently large $N$. Together with~(\ref{eq:tail2bound}), this proves the desired claim: for given $\varepsilon>0$ and large enough $N$,
\begin{eqnarray}
    \frac{1}{N} \big|\, \sum_{i=1}^N \big( w_i g(X_i,Y_i) - w'_i g'(X_i, Y_i) \big) \,\big| &\leq& \varepsilon
    \label{eq:firstterm}
\end{eqnarray}
with probability at least $1-2N \exp\{-\frac{N^{1/(2d_x+3)}}{16A^2}\} - 2\exp\{-\frac{\varepsilon^2 NC_1^4}{18(\log{N})^2 (C_2-C_1)^2}\}$.

Now, we are left to show that \eqref{eq:firsttermcondition}
implies event inside of the probability in  \eqref{eq:tail3}.
Given \eqref{eq:firsttermcondition}, we have
\begin{eqnarray}
    |\frac{\tilde{f}_X(X_i)}{f_X(X_i)} - 1| < \frac{N^{-1/(2d_x+3)}}{f_X(X_i)} < \frac{\mu(K) N^{-1/(2d_x+3)}}{C_1}
\end{eqnarray}
for all $i$. Therefore, for sufficiently large $N$, $w_i - w'_i$ is lower bounded by
\begin{eqnarray}
    w_i - w''_i &=& \frac{N/f_X(X_i)}{\sum_{j=1}^N \,\big( 1/f_X(X_j) \,\big)} - \frac{N/\tilde{f}_X(X_i)}{\sum_{j=1}^N \,\big( 1/\tilde{f}_X(X_j) \,\big)} \,\notag\\
    &\geq& \frac{N/f_X(X_i)}{\sum_{j=1}^N \,\big( 1/f_X(X_j) \,\big)} \big( 1 - \frac{1 + \frac{\mu(K) N^{-1/(2d_x+3)}}{C_1}}{1 - \frac{\mu(K) N^{-1/(2d_x+3)}}{C_1}}\big) \,\notag\\
    &\geq& \frac{N/f_X(X_i)}{\sum_{j=1}^N \,\big( 1/f_X(X_j) \,\big)} \big( 1- (1+\frac{3\mu(K)N^{-1/(2d_x+3)}}{C_1}) \big) \,\label{eq:lowerboundw1}\\
    &\geq& -\frac{3C_2\mu(K)N^{-1/(2d_x+3)}}{C^2_1} \,,\label{eq:lowerboundw2}
\end{eqnarray}
where \eqref{eq:lowerboundw1} follows from the fact that
$(1+a)/(1-a) \leq 1+3a$ for $a\in[0,1/3]$, and
\eqref{eq:lowerboundw2} follows from the fact that $C_1/\mu(K)\leq f_X(x)\leq C_2/\mu(K)$.
Similarly, it is upper bounded by
\begin{eqnarray}
    w_i - w''_i &=& \frac{N/f_X(X_i)}{\sum_{j=1}^N \,\big( 1/f_X(X_j) \,\big)} - \frac{N/\tilde{f}_X(X_i)}{\sum_{j=1}^N \,\big( 1/\tilde{f}_X(X_j) \,\big)} \,\notag\\
    &\leq& \frac{N/f_X(X_i)}{\sum_{j=1}^N \,\big( 1/f_X(X_j) \,\big)} \big( 1 - \frac{1 - \frac{\mu(K) N^{-1/(2d_x+3)}}{C_1}}{1 + \frac{\mu(K) N^{-1/(2d_x+3)}}{C_1}}\big) \,\notag\\
    &\leq& \frac{N/f_X(X_i)}{\sum_{j=1}^N \,\big( 1/f_X(X_j) \,\big)} \big( 1- (1-\frac{3\mu(K)N^{-1/(2d_x+3)}}{C_1}) \big) \,\label{eq:lowerbound3}\notag\\
    &\leq& \frac{3C_2\mu(K)N^{-1/(2d_x+3)}}{C^2_1}\;.
\end{eqnarray}
Here~\eqref{eq:lowerbound3} comes from the fact that $(1-a)/(1+a) \geq 1-3a$ for all $a \geq 0$. Therefore, $|w_i - w_i''| \leq {3C_2\mu(K)N^{-1/(2d_x+3)}}/{C^2_1}$.
For a given $\varepsilon>0$ and for sufficiently large $N$ such that
$\varepsilon/(6\log N) \geq 3C_2\mu(K)N^{-1/(2d_x+3)}/C_1^2$,
we have
\begin{eqnarray}
    && \Pr \Big(\, \max_{1 \leq i \leq N} |w''_i - w_i| > \frac{\varepsilon}{6 \log{N}}\,\Big) \,\notag\\
    &\leq & \Pr \Big(\, \max_{1 \leq i \leq N} |w_i - w''_i| > \frac{3C_2\mu(K)N^{-1/(2d_x+3)}}{C^2_1} \,\Big) \,\notag \\
    &\leq & \Pr \Big(\, \forall i \;, |\tilde{f}_X(X_i) - f_X(X_i)| < N^{-1/(2d_x+3)} \,\Big) \,\notag\;.
\end{eqnarray}
Together with \eqref{eq:tail1} and \eqref{eq:tail2bound}, this proves the desired convergence of the first term \eqref{eq:kde_error}.

\section{Proof of Proposition~\ref{prop:renyi}}
\label{sec:proof_renyi}
The proof steps are similar to that of Proposition~\ref{prop:cmi}, only requiring citations to properties of R\'{e}nyi divergence and asymmetric information.
\beit \item Clearly Axiom 0 holds -- it follows from a standard result that $D_\lambda =0$ if and only if $P = Q$ almost everywhere \cite{csiszar1995generalized}.
\item Axiom 1: Suppose $\textrm{CMI}_\lambda(P_{Z|X})$ is achieved with $P_X^{*}$. Consider the joint distribution $P_X^{*} P_{Y|X} P_{Z|Y}$. Utilizing the data-processing inequality for asymmetric mutual information (cf.\ Equation~(55) in \cite{polyanskiy2010arimoto}), we get
\beqa \textrm{CMI}_\lambda(P_{Y|X})  & = &  \max_{P_X} K_\lambda(P_X P_{Y|X})  \geq  K_\lambda(P_X^{*} P_{Y|X}) \nonumber\\
& \geq & K_{\lambda}(P_X^{*} P_{Z|X}) = \textrm{CMI}_\lambda(P_{Z|X}). \eeqa
Thus Axiom 1a is satisfied. Now consider Axiom 1b. With the same joint distribution, let $P_{Y}^{*}$ be the marginal of $Y$. Then  we have,
\beqa \textrm{CMI}_\lambda(P_{Z|Y})  & = &  \max_{P_Y} K_\lambda(P_Y P_{Z|Y})  \geq  K_\lambda(P_Y^{*} P_{Z|Y})\nonumber \\
& \geq & K_\lambda(P_X^{*} P_{Z|X}) = \textrm{CMI}_\lambda(P_{Z|X}). \eeqa

\item Axiom 2: The asymmetric mutual information has the same additivity property as traditional mutual  information, cf.\ Theorem 27 of \cite{van2014renyi}.   The corresponding additivity for $\textrm{CMI}_\lambda$ now follows.

\item Axiom 3a: The information-centroid representation for $\textrm{CMI}_\lambda$  states that (see \cite{csiszar1995generalized} or Equation~(44) of \cite{polyanskiy2010arimoto}):
\beqa \textrm{CMI}_\lambda(P_{Y|X})  & = & \min_{Q_Y}  \max_{x} D_\lambda(P_{Y|X=x} \| Q_Y) \label{eq:caP_alt_gen}.
\eeqa
This characterization allows us to make the  observation that $\textrm{CMI}_\lambda$ is a function only of the convex hull of the probability distributions $P_{Y|X=x}$, just as earlier: given a conditional probability distribution $P_{Y|X}$, we augment the input alphabet to have one more input symbol $x'$ such that $P_{Y|X=x'} = \sum_x \alpha_x P_{Y|X=x}$ is a convex combination of the other conditional distributions. We claim that the $\textrm{CMI}_\lambda$ of the new channel is unchanged: one direction is obvious, i.e., the new channel has capacity greater than or equal to the original channel, since adding a new symbol cannot decrease capacity. To show the other direction, we use \eqref{eq:caP_alt} and observe that, due to the quasi convexity of R\'{e}nyi divergence  in its arguments (cf.\ Theorem~13 in \cite{van2014renyi}), we get,
\beqa &&D_\lambda(P_{Y|X=x'} \| q_Y )   =  D_\lambda( \sum_x \alpha_x P_{Y|X=x} \| q_Y )  \leq \max_x D_\lambda(P_{Y|X=x} \| q_Y ). \nonumber\eeqa
%
Thus $\textrm{CMI}_\lambda$ is only a function of the convex hull of the range of the map $P_{Y|X}$, satisfying Axiom $3a$. This function is monotonic directly from  \eqref{eq:caP_alt_gen}, thus satisfying Axiom $3b$.

\item Axiom 4: For fixed output alphabet $\mc{Y}$, it is clear that $\max_{\mc{X}, P_{Y|X}} \textrm{CMI}_\lambda = \log |\mc{Y}|$ for each $\lambda$. Now suppose for some conditional distribution $P_{Y|X}$ we have  $\textrm{CMI}_\lambda(P_{Y|X}) = \log |\mc{Y}|$. This implies that, with the optimizing input distribution, $H_\lambda(Y) - H_\lambda(Y|X) = \log |\mc{Y}|$. This implies that $H_\lambda(Y) = \log |\mc{Y}|$ and $H_\lambda(Y|X) = 0$, thus $Y$ is a deterministic function of the essential support of $X$ and since $H_\lambda(Y) = \log |\mc{Y}|$, the Schur concavity of R\'{e}nyi entropy (cf.\ Theorem~1 of \cite{ho2015convexity}) implies that $P_Y = U_Y$, the uniform distribution and the deterministic function is onto.

\eeit

%% file: causal.bbl
\newcommand{\etalchar}[1]{$^{#1}$}
\begin{thebibliography}{BDGVdM97}

\bibitem[Ari72]{arimoto1972algorithm}
Suguru Arimoto.
\newblock An algorithm for computing the capacity of arbitrary discrete
  memoryless channels.
\newblock {\em Information Theory, IEEE Transactions on}, 18(1):14--20, 1972.

\bibitem[BDGVdM97]{BDG97}
Jan Beirlant, Edward~J Dudewicz, L{\'a}szl{\'o} Gy{\"o}rfi, and Edward~C
  Van~der Meulen.
\newblock Nonparametric entropy estimation: An overview.
\newblock {\em International Journal of Mathematical and Statistical Sciences},
  6(1):17--39, 1997.

\bibitem[Bla72]{blahut1972computation}
Richard~E Blahut.
\newblock Computation of channel capacity and rate-distortion functions.
\newblock {\em Information Theory, IEEE Transactions on}, 18(4):460--473, 1972.

\bibitem[CMMR12]{cornuet2012adaptive}
Jean Cornuet, Jean-Michel Marin, Antonietta Mira, and Christian~P Robert.
\newblock Adaptive multiple importance sampling.
\newblock {\em Scandinavian Journal of Statistics}, 39(4):798--812, 2012.

\bibitem[CS04]{csiszar2004information}
Imre Csisz{\'a}r and Paul~C Shields.
\newblock {\em Information theory and statistics: A tutorial}.
\newblock Now Publishers Inc, 2004.

\bibitem[Csi95]{csiszar1995generalized}
Imre Csisz{\'a}r.
\newblock Generalized cutoff rates and renyi's information measures.
\newblock {\em Information Theory, IEEE Transactions on}, 41(1):26--34, 1995.

\bibitem[Csi08]{csiszar2008axiomatic}
Imre Csisz{\'a}r.
\newblock Axiomatic characterizations of information measures.
\newblock {\em Entropy}, 10(3):261--273, 2008.

\bibitem[CT12]{CoverThomas}
Thomas~M Cover and Joy~A Thomas.
\newblock {\em Elements of information theory}.
\newblock John Wiley \& Sons, 2012.

\bibitem[DP84]{DP84}
Luc Devroye and Clark~S Penrod.
\newblock The consistency of automatic kernel density estimates.
\newblock {\em The Annals of Statistics}, pages 1231--1249, 1984.

\bibitem[GOV16]{gao2016demystifying}
Weihao Gao, Sewoong Oh, and Pramod Viswanath.
\newblock Demystifying fixed k-nearest neighbor information estimators.
\newblock {\em arXiv preprint arXiv:1604.03006}, 2016.

\bibitem[GSG14]{gao2014efficient}
Shuyang Gao, Greg~Ver Steeg, and Aram Galstyan.
\newblock Efficient estimation of mutual information for strongly dependent
  variables.
\newblock {\em arXiv preprint arXiv:1411.2003}, 2014.

\bibitem[GSG15]{gao2015estimating}
Shuyang Gao, Greg~Ver Steeg, and Aram Galstyan.
\newblock Estimating mutual information by local gaussian approximation.
\newblock {\em arXiv preprint arXiv:1508.00536}, 2015.

\bibitem[HJ96]{hjort1996locally}
Nils~Lid Hjort and MC~Jones.
\newblock Locally parametric nonparametric density estimation.
\newblock {\em The Annals of Statistics}, pages 1619--1647, 1996.

\bibitem[HV15]{ho2015convexity}
Siu-Wai Ho and Sergio Verd{\'u}.
\newblock Convexity/concavity of renyi entropy and $\alpha$-mutual information.
\newblock In {\em Information Theory (ISIT), 2015 IEEE International Symposium
  on}, pages 745--749. IEEE, 2015.

\bibitem[JBGW{\etalchar{+}}13]{Janzing13}
Dominik Janzing, David Balduzzi, Moritz Grosse-Wentrup, Bernhard Sch{\"o}lkopf,
  et~al.
\newblock Quantifying causal influences.
\newblock {\em The Annals of Statistics}, 41(5):2324--2358, 2013.

\bibitem[JSSS15]{JanSteShaSch15}
D.~Janzing, B.~Steudel, N.~Shajarisales, and B.~Sch{\"o}lkopf.
\newblock {\em Justifying Information-Geometric Causal Inference}, chapter~18,
  pages 253--265.
\newblock Springer International Publishing, 2015.

\bibitem[KBG{\etalchar{+}}07]{khan2007relative}
Shiraj Khan, Sharba Bandyopadhyay, Auroop~R Ganguly, Sunil Saigal, David~J
  Erickson~III, Vladimir Protopopescu, and George Ostrouchov.
\newblock Relative performance of mutual information estimation methods for
  quantifying the dependence among short and noisy data.
\newblock {\em Physical Review E}, 76(2):026209, 2007.

\bibitem[KKPW15]{kandasamy2015nonparametric}
Kirthevasan Kandasamy, Akshay Krishnamurthy, Barnabas Poczos, and Larry
  Wasserman.
\newblock Nonparametric von mises estimators for entropies, divergences and
  mutual informations.
\newblock In {\em Advances in Neural Information Processing Systems}, pages
  397--405, 2015.

\bibitem[KL87]{KL87}
LF~Kozachenko and Nikolai~N Leonenko.
\newblock Sample estimate of the entropy of a random vector.
\newblock {\em Problemy Peredachi Informatsii}, 23(2):9--16, 1987.

\bibitem[KSG04]{Kra04}
A.~Kraskov, H.~St{\"o}gbauer, and P.~Grassberger.
\newblock Estimating mutual information.
\newblock {\em Physical review E}, 69(6):066138, 2004.

\bibitem[KSM{\etalchar{+}}14]{Krishnaswamy14}
Smita Krishnaswamy, Matthew~H Spitzer, Michael Mingueneau, Sean~C Bendall, Oren
  Litvin, Erica Stone, Dana Pe'er, and Garry~P Nolan.
\newblock Conditional density-based analysis of t cell signaling in single-cell
  data.
\newblock {\em Science}, 346(6213):1250689, 2014.

\bibitem[L{\etalchar{+}}96]{loader1996local}
Clive~R Loader et~al.
\newblock Local likelihood density estimation.
\newblock {\em The Annals of Statistics}, 24(4):1602--1618, 1996.

\bibitem[LP16]{lombardi2016nonparametric}
Damiano Lombardi and Sanjay Pant.
\newblock Nonparametric k-nearest-neighbor entropy estimator.
\newblock {\em Physical Review E}, 93(1):013310, 2016.

\bibitem[MJG15]{mueller2015modeling}
Jonas Mueller, Tommi Jaakkola, and David Gifford.
\newblock Modeling trends in distributions.
\newblock {\em arXiv preprint arXiv:1511.04486}, 2015.

\bibitem[Mod89]{moddemeijer1989estimation}
Rudy Moddemeijer.
\newblock On estimation of entropy and mutual information of continuous
  distributions.
\newblock {\em Signal processing}, 16(3):233--248, 1989.

\bibitem[MPJ{\etalchar{+}}15]{Mooijetal2015}
J.M. Mooij, J.~Peters, D.~Janzing, J.~Zscheischler, and B.~Sch{\"o}lkopf.
\newblock Distinguishing cause from effect using observational data: methods
  and benchmarks.
\newblock {\em Journal of Machine Learning Research}, 2015.

\bibitem[Owe13]{mcbook}
Art~B. Owen.
\newblock {\em Monte Carlo theory, methods and examples}.
\newblock 2013.

\bibitem[Pan03]{paninski2003estimation}
Liam Paninski.
\newblock Estimation of entropy and mutual information.
\newblock {\em Neural computation}, 15(6):1191--1253, 2003.

\bibitem[Pea09]{Pearl}
Judea Pearl.
\newblock {\em Causality}.
\newblock Cambridge university press, 2009.

\bibitem[PPS10]{pal2010estimation}
D{\'a}vid P{\'a}l, Barnab{\'a}s P{\'o}czos, and Csaba Szepesv{\'a}ri.
\newblock Estimation of r{\'e}nyi entropy and mutual information based on
  generalized nearest-neighbor graphs.
\newblock In {\em Advances in Neural Information Processing Systems}, pages
  1849--1857, 2010.

\bibitem[PV10]{polyanskiy2010arimoto}
Yury Polyanskiy and Sergio Verd{\'u}.
\newblock Arimoto channel coding converse and r{\'e}nyi divergence.
\newblock In {\em Communication, Control, and Computing (Allerton), 2010 48th
  Annual Allerton Conference on}, pages 1327--1333. IEEE, 2010.

\bibitem[PXS12]{poczos2012nonparametric}
Barnab{\'a}s P{\'o}czos, Liang Xiong, and Jeff Schneider.
\newblock Nonparametric divergence estimation with applications to machine
  learning on distributions.
\newblock {\em arXiv preprint arXiv:1202.3758}, 2012.

\bibitem[RE15]{tutorial}
Robin~J Richardson and Thomas~S Evans.
\newblock Non-parametric causal models.
\newblock 2015.

\bibitem[R{\'e}n59]{renyi1959measures}
Alfr{\'e}d R{\'e}nyi.
\newblock On measures of dependence.
\newblock {\em Acta mathematica hungarica}, 10(3-4):441--451, 1959.

\bibitem[Sha48]{shannon}
C.E. Shannon.
\newblock A mathematical theory of communication.
\newblock {\em Bell System Tech. J.}, 27:379Ð423 and 623Ð656, 1948.

\bibitem[SJ91]{SJ91}
Simon~J Sheather and Michael~C Jones.
\newblock A reliable data-based bandwidth selection method for kernel density
  estimation.
\newblock {\em Journal of the Royal Statistical Society. Series B
  (Methodological)}, pages 683--690, 1991.

\bibitem[SJSB15]{ShaJanSchBes15}
N.~Shajarisales, D.~Janzing, B.~Sch{\"o}lkopf, and M.~Besserve.
\newblock Telling cause from effect in deterministic linear dynamical systems.
\newblock In {\em Proceedings of the 32nd International Conference on Machine
  Learning}, volume~37 of {\em JMLR Workshop and Conference Proceedings}, page
  285�294. JMLR, 2015.

\bibitem[SMH{\etalchar{+}}03]{singh2003nearest}
Harshinder Singh, Neeraj Misra, Vladimir Hnizdo, Adam Fedorowicz, and Eugene
  Demchuk.
\newblock Nearest neighbor estimates of entropy.
\newblock {\em American journal of mathematical and management sciences},
  23(3-4):301--321, 2003.

\bibitem[SRHI10]{sricharan2010empirical}
Kumar Sricharan, Raviv Raich, and Alfred~O Hero~III.
\newblock Empirical estimation of entropy functionals with confidence.
\newblock {\em arXiv preprint arXiv:1012.4188}, 2010.

\bibitem[TVdM96]{tsybakov1996root}
Alexandre~B Tsybakov and EC~Van~der Meulen.
\newblock Root-n consistent estimators of entropy for densities with unbounded
  support.
\newblock {\em Scandinavian Journal of Statistics}, pages 75--83, 1996.

\bibitem[VEH14]{van2014renyi}
Tim Van~Erven and Peter Harremo{\"e}s.
\newblock R{\'e}nyi divergence and kullback-leibler divergence.
\newblock {\em Information Theory, IEEE Transactions on}, 60(7):3797--3820,
  2014.

\bibitem[WKV09]{wang2009divergence}
Qing Wang, Sanjeev~R Kulkarni, and Sergio Verd{\'u}.
\newblock Divergence estimation for multidimensional densities
  via-nearest-neighbor distances.
\newblock {\em Information Theory, IEEE Transactions on}, 55(5):2392--2405,
  2009.

\end{thebibliography}
